\documentclass[12pt]{iopart}
\usepackage{epsfig,amssymb,bm}

\newcommand{\fd}{fluc\-tu\-a\-tion-dis\-si\-pa\-tion }
\newcommand{\be}{\begin{equation}}
\newcommand{\bea}{\begin{eqnarray}}
\newcommand{\ee}{\end{equation}}
\newcommand{\eea}{\end{eqnarray}}
\newcommand{\eq}{{\rm eq}}
\newcommand{\dpar}{\partial}
\renewcommand{\o}{{\mathcal O}}
\newcommand{\ot}{{\widetilde{\mathcal O}}}
\newcommand{\pht}{{\tilde\varphi}}
\newcommand{\p}{{\varphi}}
\newcommand{\scal}{{\rm scal}}
\newcommand{\can}{{\rm can}}

\def\en{\varepsilon}
\def\e{\epsilon}
\def\C{{\mathfrak{C}}}
\def\Fef{\mathfrak{F}}
\catcode`@=11 
\renewcommand\tableofcontents{%
  \section*{\contentsname}%
  \@starttoc{toc}%
}
\catcode`@=12

\begin{document}
\topical[Ageing Properties of Critical Systems]{Ageing Properties of
Critical Systems}

\author{Pasquale Calabrese$^1$ and Andrea Gambassi$^{2,3}$}
\address{$^1$\  Rudolf Peierls Centre for Theoretical Physics, 1 Keble
Road, Oxford OX1 3NP, United Kingdom}
\address{$^2$\ Max-Planck-Institut f\"ur Metallforschung,
Heisenbergstr. 3, D-70569 Stuttgart, Germany}
\address{$^3$\  Institut f\"ur Theoretische und Angewandte Physik,
Universit\"at Stuttgart,
Pfaffenwaldring 57, D-70569 Stuttgart, Germany}

\ead{calabres@thphys.ox.ac.uk, gambassi@mf.mpg.de}

\begin{abstract}

In the past few years systems with slow dynamics have attracted
considerable theoretical and experimental interest. 
Ageing phenomena are observed during this ever-lasting non-equilibrium evolution. 
A simple instance of such a behaviour is provided by the dynamics that takes place 
when a system is quenched from its high-temperature phase
to the critical point. The aim of this review is to summarize
the various numerical and analytical results that have been recently
obtained for this case. Particular emphasis is put to the
field-theoretical methods that can be used
to provide analytical predictions for the relevant dynamical
quantities.

Fluctuation-dissipation relations are discussed
and in particular the concept of fluctuation-dissipation ratio (FDR)
is reviewed, emphasizing its connection with the definition of
a possible effective temperature.

The Renormalization-Group approach to critical dynamics is
summarized and
the scaling forms of the time-dependent non-equilibrium
correlation and response functions of a generic observable are
discussed. From them the universality of the associated FDR follows
as an amplitude ratio. It is then possible to provide predictions for
ageing quantities in a variety of different models. In particular the
results for Model A, B and C dynamics of the $O(N)$ 
Ginzburg-Landau Hamiltonian,
and Model A dynamics of the weakly dilute Ising magnet
and of the $\p^3$ theory are reviewed
and compared with the available numerical results and exact solutions.   
The effect of a planar surface on the ageing behaviour of Model A dynamics is
also addressed within the mean-field approximation. 

\end{abstract}

\submitto{\JPA}


\maketitle

\tableofcontents

\section*{Nomenclature and notations}
\begin{description}
\item[A, B, C,\dots J] Dynamical universality classes
\item[FD] Fluctuation-Dissipation
\item[FDR] Fluctuation-Dissipation Ratio
\item[FDT] Fluctuation-Dissipation Theorem
\item[FT] Field Theory
\item[IR] Infra-Red
\item[MC] Monte Carlo (simulation)
\item[MF] Mean Field
\item[MS] Minimal Subtraction
\item[OT] Ordinary Transition (for surface critical behaviour)
\item[RG] Renormalization Group
\item[RIM] Random Ising Model
\item[$T_c$] Critical temperature
\item[$T_{\rm eff}$] Effective temperature
\item[SDE] Short-Distance Expansion
\item[SpT] Special Transition (for surface critical behaviour)
\item[$t,s$] Times with $t>s$, if not differently specified
\item[TTI] Time Translation Invariance
\item[TRS] Time Reversal Symmetry
\item[UV] Ultra-Violet
\item[$(\rmd q)$] $\rmd^d q/(2\pi)^d$
\end{description}


\section{Introduction}
\label{intr}

Understanding the non-equilibrium dynamics of physical systems is currently
one of the most challenging problems in statistical physics. 
Equilibrium statistical mechanics has been probably one of the most important
achievements during the last century. 
On the other hand, in nature equilibrium is more an
exception rather than a rule:  
real systems may persist out-of-equilibrium for several reasons, as 
external driving forces, very slow relaxation etc.
Many efforts are currently aiming at achieving a coherent theoretical picture 
of a prototype of such non-equilibrium system: Glassy systems, such as 
structural glasses and spin-glasses.   
One of the most striking features of glassy systems is a dramatic slowing down
of relaxational processes. 
After a perturbation (e.g., a sudden temperature change) 
the slow dynamics manifests itself in one-time quantities, such as the 
energy density, through some slow-decaying dependence on the time elapsed 
from the perturbation. However, even if these quantities get very 
close to an asymptotic value, this does not mean that the system is reaching 
some stationary or meta-stable state. 
This is elucidated by looking at two-time
quantities---mainly correlation and response functions--- that depend on the 
two times $s$ and $t>s$ not only through their difference, even for long times.
Furthermore the decay as a function of $t$ is slower for larger $s$.
This phenomenon is usually referred to as {\it ageing} \cite{struik}:
older samples respond more slowly.
The time $s$ is called the ``age'' of the system or also {\it waiting time}, 
being the time elapsed since  the preparation of the system. 
Useful quantities such as the temperature of the system can not be defined in
this genuine non-equilibrium regime. 

Ageing behaviour has been experimentally observed in a variety of glassy
systems (for a review see \cite{review}), and
it is expected whenever the equilibration time
exceeds any experimental observation time-window. 
This is the case for magnetic systems quenched from the high-temperature phase
to their critical point or below it, when the critical relaxation
and the phase ordering take place, respectively.
In the latter case both theory and experiment clearly display 
ageing behaviour as reviewed in \cite{bray}.

An important aspect of ageing systems is that the equilibrium \fd 
theorem (FDT) does not hold.
Such ``violations'' 
of the FDT have been the starting point for the introduction 
of fruitful ideas in the field of glassy systems
such as non-equilibrium \fd relations \cite{ck-93} and
effective temperatures \cite{ckp-97,cugl-99},
reviewed in \cite{cugl-02,cr-03,k-05}.
These intriguing theoretical ideas are currently under
experimental investigation in glassy \cite{gi-99,cgkv-99,bc-02,ho-02,ag-04}
and granular \cite{grani} materials.

However, as pointed out by Cugliandolo, Kurchan and Parisi \cite{ckp-94},
``violations'' of the FDT are not peculiar to glassy and disordered systems: 
Even a magnetic material quenched to its critical point
displays non-trivial fluctuation-dissipation (FD) relations.
Therefore, many aspects of ageing behaviour in classical ferromagnetic
models have been investigated.
Special attention has been devoted to the determination of the 
so-called Fluctuation-Dissipation Ratio (FDR), in particular after
the observation \cite{gl-00c,gl-02} that it is universal in the sense of
Renormalization-Group (RG) theory.
Due to universality in critical phenomena, this quantity, as well as
other aspects of ageing dynamics, is expected to be
the same for all the systems belonging to the same universality class,
irrespective of their specific realization and microscopic details.
In particular the predictions for universal quantities associated with
the dynamics of real systems and of the corresponding lattice models  can be 
obtained by means of suitable ``mesoscopic'' field-theoretical models. 
Understanding ageing behaviour in such simple instances might also provide 
insight into the phenomenology of glassy systems.

The aim of this paper is to give a comprehensive review of the rather large 
amount of theoretical ideas developed to describe ageing phenomena
at the critical point, since currently many results are scattered in the 
literature.
Particular emphasis is given to the perturbative field-theoretical approach 
as applied in \cite{cg-02a1,cg-02a2,cg-02rim,cg-03,cg-04} and to the 
comparison among the results of this analysis, the numerical estimates and 
the exact solutions that have been provided for the simplest models.
In  this paper we do not discuss the vast field of 
ageing in glassy systems, being well beyond our aims.
When required, we refer the reader to the comprehensive and updated
available reviews \cite{cugl-02,cr-03}.

The review is organized as follows. 

In section~\ref{FDTsec}, we recall some basic facts concerning the
description of dynamics in statistical systems. Then, 
after a simple derivation of the \fd theorem
(FDT) in equilibrium, 
we introduce the FDR in various forms and we remark its
importance in understanding the dynamics of slow-relaxing systems,
and in particular its connection with the idea of effective temperature.

In section~\ref{FT} we recall the basic steps which allow a  path-integral 
representation of the mesoscopic dynamics, given by a suitable Langevin
equation. 
After introducing the concept of dynamic universality classes, 
we briefly summarize the renormalization procedure for
the corresponding field theories.  
The scaling forms of the response and correlation functions 
of the order parameter and of other observables are obtained
from RG equations and  short-distance expansion, 
following the seminal work by Janssen {\it et al.} \cite{jss-89,jan-92}.

In section~\ref{modA} we consider the purely dissipative dynamics (Model
A in the notion 
of \cite{HH}) of the Landau-Ginzburg model with $O(N)$ symmetry. 
We review the numerical and exact results available for the relevant 
observables. We then consider in more detail
the model within the Gaussian approximation, where
the FDRs relative to different observables are equal.
Finally we review the calculation of two-loop response and correlation
functions of the order parameter and observables that are quadratic in
the order parameter. 
From the associated FDRs it has been shown that the Gaussian equivalence of 
all FDRs does not hold when interactions are taken into account.

In section \ref{secsurf} we investigate the effects of a spatial surface on 
the ageing properties of Model A dynamics. 
We report the scaling forms for correlation and
response functions and we provide a solution of the model in the Gaussian
approximation. 

In section~\ref{modB} the dynamics of a $O(N)$-symmetric conserved order 
parameter (Model B) is analyzed. 
We review numerical and analytical results. 
We derive the scaling forms for correlation and response functions solving
the corresponding RG equations. 
These findings agree with the solutions of particular models
and with some recent conjectures.
We solve the model exactly in the Gaussian approximation and in the limit 
of an infinite number of components of the order parameter.
Beyond the Gaussian approximation we show that 
the FDR is not affected by one-loop corrections. 

In section~\ref{modC} the case of a non-conserved 
order parameter coupled to a conserved density (Model C) is
considered. The results for two-point 
response and correlation functions are provided in a loop expansion up
to one loop.

In section~\ref{RIM} the purely dissipative dynamics (Model A) of a
weakly dilute Ising model is considered up to the first order in the loop
expansion.  

In section \ref{phi3} the purely dissipative dynamics (Model A) of a
$\p^3$ Landau-Ginzburg Hamiltonian is considered up to the first order 
in $\e=6-d$ expansion.  

Finally, in section \ref{disc}, we conclude the review presenting some 
issues that deserve further investigation.

\section{Dynamics, 
the Fluctuation-Dissipation Theorem and Fluctuation-Dissipation Ratios}
\label{FDTsec}

\subsection{Dynamics}
\label{dyn}

Let us consider a system in contact with a thermal bath at
a given temperature $T$. 
In principle the dynamics of the system is
specified by its {\it microscopic} Hamiltonian, either classical or
quantum, via the evolution equations for the density matrix and
phase-space density, respectively. However this fully microscopic
approach is rarely viable for real statistical systems. 

An alternative approach
makes use of the representation of the dynamics as a stochastic
process defined in the space of configurations of the system
\cite{vKbook,tauber}. At this description level, which
can still be termed microscopic, one has to assign the transition rates
between different configurations, on the base of
physical considerations.
Then the {\it master equation} rules the time evolution of the probability
of finding the system in a given state.
Denoting by $\C$ the generic configuration of the system and by $W[\C
\mapsto \C']$ the rate of the transition $\C \mapsto \C'$, the master
equation reads
\be
\dpar_t P[\C,t;\C_0] = 
\sum_{\C'} \{W[\C'\mapsto\C]P[\C',t;\C_0] - W[\C\mapsto\C']P[\C,t;\C_0]\}\,,
\label{ME}
\ee
where $P[\C,t;\C_0]$ is the probability of finding the system in the
configuration $\C$ at time $t$ starting from the configuration $\C_0$
at time $t=0$. 
It is possible to show that an equilibrium stationary distribution function 
$P_{\rm eq}[\C]$ (e.g., the Boltzmann-Gibbs or the microcanonical one) 
is reached for $t\rightarrow\infty$ if the 
so-called {\it detailed balance}
\be
W[\C\mapsto\C'] P_{\rm eq}[\C] =
W[\C'\mapsto \C]P_{\rm eq}[\C']\,,
\label{DB}
\ee 
holds (a proof of this statement can be
found in \cite{vKbook}).  
Equation (\ref{DB}) has to be fulfilled by the transition rates in the master
equation in order to describe a dynamics towards an 
equilibrium state. All the dynamical properties of the system can
therefore be computed once
$P[\C,t;\C_0]$ has been determined by solving the master equation with
the assigned transition rates.

Practically, solving the master equation (\ref{ME}) is rarely feasible.
A description of the dynamics in terms of 
{\it mesoscopic} variables is in some cases preferable, since it 
focuses directly on those quantities that are expected to determine
the dynamical properties at length and time scales (referred to as
{\it mesoscopic}) that are much larger
than the microscopic ones (at atomic or molecular level) but still
small compared to macroscopic one set by
the dimension of the sample. 
Mesoscopic variables (or observables), such as
the local magnetization density in magnetic systems or the local 
particle density in fluids, are obtained by averaging 
the corresponding microscopic quantities on mesoscopic length and time scales.
This average is usually referred to as coarse-graining.

A viable approach to dynamics, which was first successfully applied to
the Brownian motion, consists of a description
taking advantage of the separation between the typical time-scale of 
fast (microscopic) and slow (mesoscopic) dynamical processes,
clearly emerging in some cases.  
It is natural to assume that the dynamics of the mesoscopic observables
can be described as the result of an effective slow deterministic drift
towards a stationary state (equilibrium or not) and of a stochastic
force that sums up the effect of the fast microscopic fluctuations. Of
course this description fails to reproduce the dynamics 
taking place at microscopic time and length scales.
Hence, when the macroscopic physics is crucially related to some microscopic 
events (as claimed, for instance, in the case of rare intermittent events in 
some glasses \cite{sj-04}) such a mesoscopic description is not expected 
to capture the relevant physical mechanisms. 
We will strictly consider systems for which the mesoscopic 
description is feasible.

Let us assume that the mesoscopic properties of the system are described
by a set of variables $\phi_i$, labeled by an index $i$. 
In terms of $\{\phi_i\}$ the Hamiltonian is given by ${\cal H}[\phi]$. 
The previous heuristic considerations motivate the assumption 
that the dynamics of the system is described by the Langevin equation
\be
\dpar_t\phi_i(t) 
= - \sum_{j}{\mathbf D}_{ij}\frac{\delta {\cal H}[\phi]}{\delta \phi_j(t)} +
\zeta_i(t) \; .
\label{Eqlang}
\ee
The first term in the r.h.s. represents the deterministic evolution and 
${\mathbf D}_{ij}$ is called kinetic coefficient. 
The stochastic nature of the equation is embodied in the random 
noise $\zeta_i(t)$.
It is specified by its functional probability distribution function, 
or equivalently by the corresponding moments.
In the spirit of the central limit theorem, the first two moments are
sufficient to characterize the stochastic process, which hence is 
assumed to be Gaussian. 
Denoting with $\langle \cdot\rangle$ the mean over the possible 
realizations of the noise, we assume $\langle\zeta_i(t)\rangle=0$
and $\langle\zeta_i(t)\zeta_j(t')\rangle = 2 {\mathbf N}_{ij}\delta(t-t')$,
i.e., a so-called Gaussian white noise.

From the Langevin equation (\ref{Eqlang}) it is possible to derive
a Fokker-Planck partial differential equation for the probability distribution
function $P[\{\phi_i\},t]$ of $\phi_i$ at time $t$, given an initial
condition  \cite{zj,onuki,tauber}
\be
\label{eq:FP}
\dpar_t P[\{\phi_i\},t] = \frac{\delta}{\delta\phi_j}\left[
{\mathbf N}_{j k}\frac{\delta P}{\delta\phi_k} + 
{\mathbf D}_{j k}\frac{\delta{\mathcal H}}{\delta\phi_k}P\right] \;.
\ee
The stationary distribution $\dpar_t P[\{\phi_i\},t] = 0$ is the 
Boltzmann-Gibbs probability of equilibrium statistical mechanics
$P_{BG}[\{\phi_i\}]\propto e^{-\beta {\cal H}[\phi]}$ if and only if 
\be
\label{eq:EE}
{\mathbf D}=\beta{\mathbf N}\,.
\ee
This condition is the well-known Einstein's relation. 
If, given two arbitrary configurations $\{\phi_i\}$ and
$\{\phi'_i\}$, a realization of the noise $\zeta_i$ 
exists such that starting from
$\{\phi_i\}$ and evolving according to equation (\ref{Eqlang}), the
configuration $\{\phi'_i\}$ is reached in a finite time, the system 
is said to be ergodic.
If ergodicity holds the long-time limit of 
$P[\{\phi_i\},t]$ is $P_{BG}[\{\phi_i\}]$
independently of the initial condition.

The dynamics of the system can be studied by looking at time-dependent
quantities defined in terms of $\{\phi_i\}$ and averaged over the possible
realizations of the noise. 
The simplest quantity one might consider among those which are
measurable by looking at the system only at a given time $t$ (one-time
quantities) is $\langle\phi_i(t)\rangle$. In the long-time limit one
expects such a quantity to reach an asymptotic value and therefore it
is no longer possible to get information about the dynamics of the
system. In this sense one-time
quantities are not usually enough to characterize it. 
The next step is to consider quantities
that can be measured by looking at the configuration of the system at
two different times (two-time quantities). At variance with one-time
quantities, they also provide information about the dynamics of the system
in the long-time limit. 
Interesting two-time quantities are the
correlation function of the mesoscopic variables
$\{\phi_i\}$ and the response function. The former is defined as
$C_{ij}(t,s) \equiv \langle \phi_i(t)\phi_j(s)\rangle$ and is
related to the relaxation of spontaneous (thermal)
fluctuations of the variables $\{\phi_i\}$ whereas the latter expresses
the response of the system to an external perturbation.
In particular let us assume that this perturbation is due to an
external field $\{h_i\}$ that couples linearly to the variables
$\{\phi_i\}$ in the Hamiltonian ${\mathcal H}$ (i.e., $h_i$ is the field
conjugate to $\phi_i$): ${\mathcal H}_h[\phi] = {\mathcal H}[\phi]
-\sum_i h_i\phi_i$. The linear response function $R_{ij}(t,s)$ is
defined by
\be
R_{ij}(t,s) \equiv\left.\frac{\delta\langle\phi_i(t)\rangle_h}{\delta h_j(s)} 
\right|_{h=0} \;, 
\label{definitionR}
\ee
where $\langle\cdot\rangle_h$ indicates the average over the stochastic
dynamics generated by equation~(\ref{Eqlang}) with the 
Hamiltonian  ${\mathcal H}_h[\phi]$.
Due to causality, $\phi_i(t)$ does not depend on $h_j(s)$, whenever $t<s$. 
As a consequence $R_{ij}(t,s)\propto\theta(t-s)$, where $\theta(t) = 1$ 
for $t>0$ and $0$ otherwise. In the following we set $t>s$, so 
that $R(s,t)=0$.

To calculate the response function, we note that the presence of the external 
field $h$ is equivalent to a shift in the noise 
$\delta\zeta_i(t) = {\mathbf D}_{ij}\, h_j(t)$, so that 
equation~(\ref{definitionR}) may be written as
\be
R_{ij}(t,s) =  
{\mathbf D}_{j k}\langle \frac{\delta\phi_i(t)}{\delta \zeta_k(s)} \rangle 
= \frac{1}{2} 
({\mathbf D}{\mathbf N}^{-1})_{j k}\langle \phi_i(t)\zeta_k(s) \rangle = 
\frac{\beta}{2} \langle \phi_i(t)\zeta_j(s) \rangle\,,
\label{definitionRbis}
\ee
where we have used Einstein's relation and the algebraic identity
$\langle F[\zeta] \zeta_i(t)\rangle = 
2{\mathbf N}_{ij}\langle\frac{\delta F[\zeta]}{\delta\zeta_j(t)}\rangle$,
with $F[\zeta]$ an arbitrary function of the noise $\zeta$\footnote{%
This identity is easily derived taking into account that 
the probability distribution function $P_N[\{\zeta_i\}]$ of
the noise, being Gaussian, satisfies 
$-2 {\mathbf N}_{j k} \delta P_N/\delta\zeta_k(s) =
\zeta_j(s)P_N$. Multiplying both sides of this equation 
by an arbitrary $F[\zeta]$ and
performing the functional integration on $\{\zeta_i\}$ (by parts on
the l.h.s.) yields the result.
}.
A useful relation between $R_{ij}$ and $C_{ij}$ is obtained by considering
the time derivative of the correlation function 
$(\dpar_t-\dpar_s)C_{ij}(t,s) = \langle \dpar_t\phi_i(t)\phi_j(s)\rangle -
\langle\phi_i(t)\dpar_s\phi_j(s)\rangle$ as 
\be
\fl
(\dpar_t-\dpar_s)C_{ij}(t,s) =
-\langle{\mathbf D}_{i k}\frac{\delta {\cal H}[\phi]}{\delta \phi_k(t)}\phi_j(s)\rangle +
\langle\phi_i(t){\mathbf D}_{j k}\frac{\delta {\cal H}[\phi]}{\delta \phi_k(s)}\rangle -\langle\phi_i(t)\zeta_j(s) \rangle \;,
\ee
that can be cast in the form:
\be
2 T R_{ij}(t,s)=(\dpar_t-\dpar_s)C_{ij}(t,s)-A_{ij}(t,s)\,,
\label{FDTout}
\ee
where we defined the asymmetry
\be
A_{ij}(t,s)= 
\langle\phi_i(t){\mathbf D}_{j k}\frac{\delta {\cal H}[\phi]}{\delta \phi_k(s)}\rangle
- \langle{\mathbf D}_{i k}\frac{\delta {\cal H}[\phi]}{\delta \phi_k(t)}\phi_j(s)\rangle \,.
\ee 
Equation (\ref{FDTout}) relates correlation and response functions
by means of the asymmetry. A similar equation in terms of
microscopic degrees of freedom evolving according to 
a general class of master equations can be found in \cite{d-03}.

\subsection{The Fluctuation-Dissipation Theorem}

Here we present a simple derivation of the \fd theorem within the
approach to dynamics we are discussing. The same result can be
obtained in many different ways, with different
mathematical rigor (see, e.g., \cite{Ruelle,zj,cr-03,cugl-02,tauber}).

When the system equilibrates, equation (\ref{FDTout}) assumes a simpler form.
In fact the Time-Translational Invariance (TTI) implies that
the correlation and response functions satisfy $C_{ij}(t,s) = C_{ij}(t-s,0)$
and $R_{ij}(t,s) = R_{ij}(t-s,0)$. 
Therefore $(\dpar_t-\dpar_s)C_{ij}(t,s) = -2\dpar_s C_{ij}(t,s)$. 
Moreover, equilibrium is characterized by 
Time-Reversal Symmetry (TRS), and so the correlation function of two
observables $\o_1(t)$ and $\o_2(t)$ satisfies
$\langle \o_1(t)\o_2(s)\rangle = \langle\o_1(s) \o_2(t)\rangle$,
so that the asymmetry vanishes.
Taking into account these observations, one concludes that
\be
R_{ij}(t,s) = \beta\, \frac{\dpar C_{ij}(t,s)}{\dpar s} \;,
\label{ourFDT}
\ee
which is just one formulation of the \fd theorem.
Note that the FDT holds also if $C_{ij}$ is replaced by the connected 
correlation function 
$C_{ij}^{(c)}(t,s)=C_{ij}(t,s)-\langle\phi_i(t)\rangle\langle\phi_j(s)\rangle$,
since due to TTI $\langle\phi_i(t)\rangle$ is independent of time in 
equilibrium.

In the following, we are mainly concerned with the case of systems
whose physical properties close to a critical point are determined
by a suitable order parameter (see subsection~\ref{CPreminder}). 
In this case $\phi_i$ and $\phi_j$ stand for the values of 
the order parameter $\phi$ at different spatial points.
Using TTI, space-translational invariance and denoting by ${\bf x}$
the $d$-dimensional vector between the two points $i$ and $j$, 
equation (\ref{ourFDT}) can be written as
\be
R_{\bf x}(t-s)=\beta \dpar_s C_{\bf x}(t-s).
\label{FDTx}
\ee

Consider now a local observable $\o({\bf x},t)$ and define 
its correlation function (always with $t>s$) as 
$C_{{\bf x}-{\bf x}'}^\o(t,s)=\langle\o({\bf x},t)\o({\bf x}',s)\rangle$ and its response
function 
\be
R_{{\bf x}-{\bf x}'}^\o(t,s) \equiv\left.\frac{\delta\langle\o({\bf x},t)\rangle_{h_\o}}{\delta h_\o({\bf x}',s)} \right|_{h_\o=0} \;, 
\ee
where $h_\o({\bf x}',s)$ is the field conjugated to $\o({\bf x}',s)$. 
Equation~(\ref{FDTx}) can be generalized 
to $\o$ in the form \cite{zj,Ruelle}
\be
R_{\bf x}^\o(t-s)=\beta \dpar_s C^\o_{\bf x}(t-s),
\label{FDTO}
\ee
i.e., the FDT holds for any observable of the system.
This remarkable property allows one to define the temperature $T=\beta^{-1}$
through equation~(\ref{FDTO}), independently of the specific observable used.

Whenever the system does not reach thermal equilibrium one can not expect 
that the relations in equations~(\ref{ourFDT}), (\ref{FDTx}), 
and (\ref{FDTO}) hold. 
In this sense their validity may be considered as a signal of the fact that
the system in equilibrium.
Let us consider the following experiment. Prepare a system---a glass, a 
ferromagnet, etc.--- in an equilibrium state at a high temperature $T_0$, 
greater than the critical or glass transition temperature. 
At time $t_0=0$ quench the system to some temperature $T<T_0$ by taking it 
into contact with a thermal bath at temperature $T$ and let it evolve. 
On a general basis, one expects that the relaxation towards the equilibrium 
state corresponding to $T$ is characterized by two different regimes: 
(A) a transient one with non-equilibrium evolution,
for $t<t_{\rm \eq}(T)$, and (B) a stationary regime with equilibrium evolution
for $t\gg t_{\rm \eq}(T)$, where $t_{\rm eq}(T)$ is a characteristic 
equilibration time of the system.
During (A) one expects a dependence of the behaviour of the system on initial
conditions and both TRS and TTI are broken, whereas in (B) TTI and TRS are 
recovered: The dynamics of fluctuations is given by the ``equilibrium'' one, 
for which FDT holds.
In many systems regime (B) is not reached during  experimental or even 
geological times because $t_{\rm \eq} = \infty$ for all practical purposes. 
These systems always evolve out of equilibrium, even if no perturbation is 
acting on them.
In this case the standard concepts of equilibrium statistical mechanics do not
apply and in particular both TTI and TRS are broken. 
As a consequence two-time quantities, such as the response and correlation 
functions, depend separately on $s$ and $t$, even for large times.

\subsection{The Fluctuation-Dissipation Ratio}

A measure of the distance from equilibrium of an ageing system evolving in 
contact with a thermal bath at fixed temperature $T$ is the FDR \cite{ck-93}
\be
X_{\bf x}(t,s)=\frac{T\, R_{\bf x}(t,s)}{\dpar_s C_{\bf x}(t,s)} \; ,
\label{dx}
\ee
where we assume $t>s$.
When the age $s$ is greater than $t_{\rm eq}(T)$, TRS and TTI
hold, and the FDT yields $X_{\bf x}(t,s)=1$. 
This is not generically true in the ageing regime.

The asymptotic value of the FDR 
\be
X^\infty=\lim_{s\to\infty}\lim_{t\to\infty}X_{{\bf x}=0}(t,s)
\label{xinfdef}
\ee
turns out to be a very useful quantity in the description
of systems with slow dynamics, since  $X^\infty=1$ whenever 
$t_{\rm eq}(T)<\infty$, i.e., when the ageing evolution is interrupted and 
the system crosses over to equilibrium dynamics.
Accordingly, 
$X^\infty\neq 1$ is a signal of an asymptotic non-equilibrium dynamics.
Moreover $X^\infty$ can be used to define an effective temperature 
$T_{\rm eff}=T/X^\infty$, which might have some features of the 
temperature of an equilibrium system, e.g., controlling the direction of heat 
flows and acting as a criterion for thermalization \cite{ckp-97}.
This definition of $T_{\rm eff}$ is closely related to that introduced in the 
context of weak turbulence \cite{hs-89}.
Under some assumptions it has been shown that 
$X_{\bf x}(t,s)$ establishes a bridge between the dynamically inaccessible 
equilibrium state and the asymptotic dynamics for large times \cite{fmpp-98}.

In equation~(\ref{FDTx}) the FDR has been defined considering the
two-time correlation function $C_{\bf x}(t,s)$ and the two-time
response function $R_{\bf x}(t,s)$ of the order
parameter. Nevertheless it is possible to define a FDR for any observable
$\o$:
\be
X^\o_{\bf x}(t,s)=\frac{T\, R^\o_{\bf x}(t,s)}{\dpar_s C^\o_{\bf x}(t,s)} \, ,
\quad \mbox{ and } \quad
X^\infty_\o=\lim_{s\to\infty}\lim_{t\to\infty}X_{\bf x=0}^\o(t,s).
\label{dxO}
\ee
In \cite{ckp-97} it has been shown that a thermometer coupled to the 
observable $\o$ measures, on a proper time scale, the 
temperature $T_{\rm eff}^\o=T/X_\o^\infty$, although this point has been 
source of some controversy, since the measured temperature
seems to depend on the thermometer employed \cite{ep-00}.
It has been pointed out (see, e.g., \cite{sfm-02}) 
that the effective temperature can 
be of interest in order to devise some thermodynamics for the system
if its value is independent of $\o$,
i.e., $T_{\rm eff}^\o=T_{\rm eff}$ for all $\o$. 
This has been 
explicitly verified for infinite-range  
(mean-field) glass models \cite{ckp-97}.
Beyond these cases, the observable dependence of $T_{\rm eff}^\o$
has been investigated analytically in the case of the trap model \cite{fs-02}
and numerically for supercooled liquids \cite{bb-02} and 
in driven systems near a jamming transition \cite{jam}.
The case of ageing in critical systems will be reviewed in 
section \ref{modA}.

As already mentioned, these intriguing features are 
not only typical of glassy and disordered materials, but can be generally 
found to some extent whenever $t_{\rm eq}=\infty$, i.e., 
in slow-relaxing systems such as
ferromagnetic models in the low-temperature phase or at the critical
point, whose dynamics is characterized by
phase ordering and critical relaxation, respectively.
The phase-ordering dynamics has been mainly 
investigated by means of analytical solutions of exactly solvable
models \cite{ckp-94,nb-90,cd-95,zkh-00,cst-01,clz-02} and by
Monte Carlo simulations of more realistic 
systems \cite{h-89,barrat-98,bbk-99,prr-99,sc-99}. 
Due to these investigations the phenomenology of phase ordering is quite well
understood \cite{bray}, although some open questions remain about the scaling
of the response function \cite{clz-01,cclz-02,clz-04,hpgl-01,hpp-02,m-04}.
This issue goes beyond the scope of this review, therefore
the interested reader is referred to the literature. 
During phase ordering dynamics, the result $X^\infty=0$ 
(corresponding to an infinite effective 
temperature, as experimentally found in some colloidal glasses \cite{bc-02}) 
has been put on a firmer ground (see, e.g., \cite{gl-02}).

Particularly interesting is the case of a quench to the critical point.
In fact it is possible to get insight into this problem by means of the 
powerful tools of Renormalization Group \cite{zj}
and scale invariance \cite{cardyb} that have been developed during the last 
decades to investigate mainly equilibrium situations, but that 
can be applied to non-equilibrium relaxation as well. 
In particular, within this framework, 
it has been argued that $X^\infty$ and $X^\infty_\o$
are novel universal quantities of non-equilibrium critical 
dynamics \cite{gl-00c,gl-02,cg-02a1,cg-04}. This clearly emerges from the
discussion in sections \ref{secRenScal} and \ref{seccomp}.

\subsection{The Fluctuation-Dissipation Ratio in momentum space}
\label{secFDRms}

In the previous section we recalled the definition of 
the FDR for the response and correlation function of a space-dependent
observable considered in two points a distance ${\bf x}$ apart, and its 
long-time limit $X^\infty$ for ${\bf x}= {\bf 0}$. This is the
original form in which the FDR has been introduced in the literature on
glassy systems \cite{ck-93}, mainly with
the aim of investigating the {\it spatial} development of correlations.

However, within the field-theoretical approach to critical dynamics, 
it is more natural and computationally simpler to focus on the
behaviour of observables in {\it momentum} space. Accordingly
hereafter we mainly consider 
the momentum-dependent response $R_{\bf q}(t,s)$
and correlation $C_{\bf q}(t,s)$ functions, 
defined as the Fourier transform of the corresponding ones in real
space either on the lattice or in the continuum. 

To work in momentum space it is worth introducing a 
quantity that, just like $X_{\bf x}(t,s)$, ``gauges'' the distance
from equilibrium evolution. Its natural definition is~\cite{cg-02a1}
\be
{\cal X}_{\bf q}(t,s)=\frac{T R_{\bf q}(t,s)}{\dpar_s C_{\bf q}(t,s)} \;.
\label{Xq} 
\ee
Note that ${\cal X}_{\bf q}(t,s)$ is {\it not} the Fourier transform of 
$X_{\bf x}(t,s)$.
The long-time limit 
\be
{\cal X}^\infty_{{\bf q}=0}\equiv 
\lim_{s\rightarrow\infty}\lim_{t\rightarrow\infty} {\cal X}_{{\bf q}=0}(t,s)
\label{eq} \;
\ee
defines, in analogy to $X^\infty_{{\bf x} = {\bf 0}}$, 
an effective temperature as well.
It has been argued that these ``two'' effective temperatures in real and 
momentum space are equal \cite{cg-02a1}.
Hence, the effective temperature may still have a sound physical 
meaning. Let us outline this argument.
$X_{\bf x=0}$ can be written in terms of quantities in momentum space as
\be
X^{-1}_{{\bf x}=0} \equiv 
\frac{\int  (\rmd q) \, \dpar_s C_{\bf q}(t,s)}{T\int (\rmd q) R_{\bf q}(t,s)}= 
\frac{{\displaystyle \int} (\rmd q)\, R_{\bf q}(t,s) 
{\displaystyle \frac{\dpar_s C_{\bf q}(t,s)}{ T R_{\bf q}(t,s)}} }{\int (\rmd q)R_{\bf q}(t,s)}= 
{\langle {\cal X}^{-1}_{\bf q} \rangle}_{R_{\bf q}} \; .
\ee
where $(\rmd q)=\rmd^dq/(2\pi)^d$.
Thus $X^{-1}_{{\bf x}=0}$ is the average of ${\cal X}^{-1}_{\bf q}$, 
with weight $R_{\bf q}(t,s)$.
In the long-time limit $t\gg s\ \rightarrow \infty$, 
the main contribution to the integral comes from the small-${\bf q}$ region, 
provided that $R_{\bf q}(t,s)$ is peaked around ${\bf q}=0$ with a variance 
that vanishes for $t\gg s\rightarrow\infty$. This is the case for most
the critical
systems we are interested in.
As a consequence
\be
X^{\infty}_{{\bf x}={\bf 0}}={\cal X}_{{\bf q}={\bf 0}}^\infty\equiv X^\infty.
\label{equiv}
\ee
To our knowledge this argument is currently the only general one
available to support equation~(\ref{equiv}). 
Several results for specific models \cite{mbgs-03,s04,sl-04,sp-05}
confirm equation (\ref{equiv}).
As it stands, the argument is expected to be valid for
systems at criticality, whereas the extension to other cases might not be 
straightforward, being fundamental to know the form of $R_{\bf q}(t,s)$ 
for $t\gg s\rightarrow\infty$ 
(see also the discussion at the end of section~\ref{RIM}). 
Given the importance of equation (\ref{equiv}) for the existence of a proper, 
single-valued and well-defined effective temperature, 
a more rigorous proof and proper extensions of equation~(\ref{equiv}) 
would be very welcome.

We point out that the definition (\ref{Xq}) provides an alternative way to 
determine $X^\infty$ in numerical simulations, which is expected to increase 
the accuracy of the results, as shown in \cite{mbgs-03}.
The advantage of using  ${\cal X}_{{\bf q}={\bf 0}}$ is twofold.
On the one hand, ${\cal X}_{{\bf q}={\bf 0}}$ is defined by
using quantities with ${\bf q}={\bf 0}$ 
(referred to as {\it coherent} observables in \cite{mbgs-03}) which 
are the sums over the whole sample 
of the corresponding quantities in real space. 
For instance, when considering the magnetization of a system
of volume $V$, $X^\infty_{{\bf x} = {\bf 0}}$ and ${\cal X}_{{\bf q}={\bf 0}}$
are determined by measuring the correlation of the local magnetization 
$m_{\bf x}$ and of the total magnetization
$M=V^{-1}\sum_{{\bf x}\in V} m_{\bf x}$, respectively.
Even if in the latter case, one has to measure connected correlation 
functions (which are differences and hence more noisy than the single terms),
one expects that ${\cal X}_{{\bf q}={\bf 0}}$ 
(thus ${\cal X}_{{\bf q}={\bf 0}}^\infty$) has smaller statistical
fluctuations than $X_{{\bf x}={\bf 0}}$ 
(thus $X_{{\bf x}={\bf 0}}^\infty$).
However the key point is that coherent observables directly focus on the
non-equilibrated modes of the system, and their FD relations typically 
display very early
the crossover to $X^\infty$, when the correlation function is still 
comparable to its initial value. 
In local observables, 
on the other hand, the various relaxing
modes are mixed in, and so $X^\infty$ can only be measured at
very large times and small ratio $s/t$, corresponding to quite
small values of the correlation function.
We will show explicitly these differences when discussing 
the Gaussian model in section \ref{secgaux}.

\subsection{Quasi-equilibrium regime of $X_{\bf x}(t,s)$}
\label{quasi}

Throughout the review we will be mainly interested in the long-time
limit of $X_{{\bf x}=0}(t,s)$, i.e., in $X^\infty$. 
Nevertheless the function $X_{{\bf x}=0}(t,s)$ has also other interesting 
features in different time regimes.

In particular in the short-time regime,
defined as $\delta t \equiv t-s\ll s\rightarrow\infty$,
general arguments lead to the conclusion that the system 
is quasi-equilibrated and 
$\lim_{s\rightarrow \infty} X_{{\bf x}=0}(s^+,s)=1$.
This fact can be heuristically accounted for as follows. 
Let us consider the typical quench experiment in which the system is
initially in an uncorrelated state and then, at $t=0$, it is suddenly
quenched in the low-temperature phase~\cite{bray}. 
Thus the domain coarsening takes place. 
One expects that, at a given time, local quantities in 
two points that are inside the same 
domain display correlations typical of the thermodynamic phase of that domain. 
Deviations are expected when the two points belong to distinct domains. 
In the latter case the non-equilibrium behaviour is displayed.
Being the short-time evolution ($\delta t\ll s$) 
related to processes taking place at short spatial distances (at least for 
systems with short-range interactions), it is equilibrium-like, 
resulting in $X_{{\bf x}=0}(s+\delta t,s)=1$.
Conversely,  after a time $\delta t$ 
comparable with $s$, the behaviour at larger spatial distances starts
influencing the dynamics and the system falls out of equilibrium.

When the system is quenched at the critical point no domains are forming. 
Nevertheless, at a given time, the correlation and response functions are
equilibrium-like when referred to points within a given (growing) 
range, called dynamic correlation length, whereas they depart from 
equilibrium at larger distances. 
Thus, even in this case, $X_{{\bf x}=0}(t,s)$ exhibits a crossover from 
$1$ at $t\simeq s$ to $X^\infty$ for $\delta t\gg s$.

The previous heuristic description
can be made more rigorous considering specific
models. So far quasi-equilibrium behaviour has been observed 
in essentially all the studied cases (see the reviews \cite{cugl-02,cr-03}).

A general theorem \cite{cdk-97}, obtainable under quite reasonable assumptions
for system governed by a dissipative Langevin dynamics,  provides a bound 
for the quantity 
$V(t,s)=[1-X_{{\bf x}=0}(t,s)]\dpar_s C_{{\bf x}=0}(t,s)\,.$
This bound implies that $V(t,s)\rightarrow 0$ for fixed and finite $t-s$ 
and $s\rightarrow\infty$. Accordingly quasi-equilibrium is reached 
whenever  $\dpar_s C_{{\bf x}=0}(t,s)$ does not vanish.
However, for the one-dimensional Ising model, an observable ${\cal O}$ 
has been found such that $X^\o_{{\bf x}=0}(s,s)=3/4$ for 
$s\rightarrow\infty$ \cite{ms-04b}, while $\dpar_s C^\o_{{\bf x}=0}(t,s)$
vanishes in the same limit. 
This puts forward the idea that quasi-equilibrium can be defined only on a 
certain class of observables, termed {\it neutral} in 
reference \cite{ms-04b}.

Nevertheless, the notion of quasi-equilibrium is still very important: 
It was used in \cite{sdc-03} to define a nominal thermodynamic temperature
for models where the dynamics does not satisfy detailed balance and hence
$T_{\rm bath}$, the temperature of the thermal bath, does not exist a priori.
In any case, this can not be done in general non-equilibrium systems 
that do not satisfy detailed balance. In fact there are several examples of 
models where, due to a different scaling in time of the response and of the 
correlation functions, a non-trivial short-time $X_{{\bf x}=0}(t,s)$ 
can not be defined. This is the case of 
the contact processes \cite{contact} and of  
the deterministic non-linear evolution (i.e., equation (\ref{lang}) with 
$\zeta=0$) \cite{cardy-92}.

Let us remark that the quasi-equilibrium regime is not expected to be
detectable via ${\cal X}_{{\bf q}=0}(t,s)$. 
Indeed the two-point correlation and response functions involved in its
definition get non-equilibrium contributions (through the Fourier transform) 
from the corresponding functions in real space taken at points whose distance
is bigger than the domain size or than the dynamic correlation length.

\section{Field-Theoretical Approach to Non-equilibrium Dynamics}
\label{FT}

In this section we briefly review the Renormalization-Group approach
to the dynamics following a quench from a
high-temperature state to the critical point. 
In particular we take advantage of various
well-known methods to determine the scaling forms of 
the non-equilibrium response and correlation functions 
at criticality, computing the associated universal quantities in
perturbation theory.

\subsection{Critical Phenomena: A brief reminder}
\label{CPreminder}

The macroscopic properties of statistical systems 
(e.g., fluids, magnetic materials, etc.) are generally expected 
to depend strongly on the usually quite complex 
interactions among their microscopic constituents. 

Theoretical insight can be gained
only by studying analytically or numerically rather simplified models
of real systems and the predicted behaviour depends on
the parameters of the specific model considered. Accordingly,
it is important to have a clear understanding of the connection between
the model parameters and the microscopic properties of the actual
system  in order to provide predictions comparable
with experiments. Due to this difficulty only in few cases a 
quantitative comparison can be done. 
Nevertheless
there are circumstances in which a collective
behaviour emerges which is largely independent of the microscopic
details of the actual system and, as a consequence, also 
of the particular model used to describe it. This property is known as
{\it universality} and it naturally characterizes the physical
behaviour upon approaching a critical point,
where the system undergoes a continuous phase transition.

The onset of a collective behaviour is clearly revealed by
the {\it correlation length} $\xi$, defined as 
the typical distance over which the microscopic variables are correlated.
Far away from a critical point $\xi$ is typically of the order of the range of
microscopic interactions, whereas it diverges upon approaching the
critical point. Accordingly, close enough to the transition point,
$\xi$ becomes mesoscopic and indeed it provides the {\it only}
relevant length-scale of a critical system.

In view of the universality of critical properties, which is
justified within the renormalization-group theory 
(see subsection~\ref{RGandallthat}) and
supported by experimental data \cite{PV-r},
it is possible to study this collective behaviour in
terms of suitable mesoscopic 
field-theoretical models, in a formal development of
the Landau approach to phase transitions \cite{landau}. 
Indeed, as long as one is interested in the behaviour at mesoscopic 
length and time scales, an effective Hamiltonian 
which reflects the internal symmetries of the underlying
microscopic system can be used.
Such Hamiltonian depends only on the {\it order parameter} and
potentially on a few other slow modes, whose actual nature is determined 
specifically by the system. For instance the order parameter can be
identified with the magnetization in magnetic materials close to the
Curie temperature, with the particle density in fluids etc.
The previous considerations motivate the assumption that 
the static critical properties of a system with a
$N$-component vector order parameter $\p$, short-range interactions 
and $O(N)$ symmetry are captured for large distances by the
Landau-Ginzburg-Wilson Hamiltonian 
\be
{\cal H}[\p] = \int \rmd^d x \left[
\frac{1}{2} (\nabla \p )^2 +
\frac{1}{2} r_0 \p^2 +\frac{1}{4!} g_0 \left(\p^2\right)^2 \right] \, ,
\label{lgwA}
\ee
where $r_0\propto T$ is a parameter that has to be tuned to the
value $r_{0,{\rm crit}}$ in order to approach the critical point for
$T=T_c$ and $g_0 > 0$ is the coupling constant of the theory. 
From now on we absorb the factor $\beta = 1/T$ in the Hamiltonian, so
that ${\mathcal H}$ henceforth denotes the reduced Hamiltonian. 
Statistical averages of quantities $\o$ depending on the order parameter field
$\p$ are then properly computed, in the equilibrium canonical ensemble, as
\be
\langle \o[\p] \rangle = \frac{\int[\rmd \p]\o[\p] e^{-{\mathcal H}[\p]}}{\int
[\rmd \p]e^{-{\mathcal H}[\p]}}\;.
\ee
In a first approximation ({\it mean-field} -- MF) the integral
over the fields is determined by the
configuration $\p_{\rm MF}$ which minimizes ${\mathcal H}$, i.e., 
satisfying $\delta {\mathcal H}[\p]/\delta\p|_{\p = \p_{\rm
MF}}=0$. Corrections coming from fluctuations around $\p_{\rm MF}$ can
be accounted for by considering 
the successive terms of the formal expansion of ${\mathcal H}[\p]$
around $\p_{\rm MF}$. The term which is quadratic in $\p-\p_{\rm MF}$
defines the so-called {\it Gaussian approximation} of ${\mathcal H}$. 
Note
that the free energy $\Fef =
-\ln\int[\rmd \p]e^{-{\mathcal H}[\p]}$ is given, within the mean-field
approximation, by $\Fef_{\rm MF} = {\mathcal H}[\p_{\rm MF}]$. In this
sense ${\mathcal H}[\p]$ is sometimes referred to as Landau 
free-energy functional.

By means of field-theoretical techniques it is possible to determine the
non-analytic behaviour observed in various
thermodynamic quantities and structure factors upon approaching the
critical point. Such non-analyticities, parameterized by the
standard critical exponents, some associated amplitude ratios and
scaling functions turn out to be universal quantities (see, e.g.,
\cite{PV-r,pha}). 
The values of universal quantities and scaling functions 
characterize the so-called {\it universality class} of the model.

Upon approaching a critical point the typical time scale of dynamics 
of the fluctuations around the equilibrium state
diverges as $\sim \xi^z$ ({\it critical slowing down}), where $z$ is
the dynamic critical exponent. This  provides the
natural separation between the relevant slow evolution due to
the developing collective behaviour
and the fast one related to microscopic processes. 
This separation makes the
mesoscopic description of the dynamics 
a particularly viable approach to the problem, as explained in
section~\ref{dyn}.
Indeed it allows one to compute systematically 
the non-analytic behaviours observed in dynamical quantities, e.g., in the
low-frequency limit of the dynamic structure factor. In turn the
associated universal quantities define the {\it dynamic universality
class}.
One finds that each static universality class consists of several
dynamic sub-universality classes which differ, e.g., by different
conserved quantities, but nonetheless exhibit the same static
universal properties.
In subsection~\ref{duc} we provide some examples of physical systems
belonging to the same static universality class but to different
dynamical ones.

\subsection{Non-equilibrium critical dynamics}

In the following sections we will focus on a particular instance of
non-equilibrium behaviour: The one due to a sudden thermal quench to the 
critical point. 
From the theoretical point of view
this behaviour is induced by the initial conditions of the
dynamics and is not generic. Indeed as soon as the quench
is done at a temperature slightly above the critical one the system
thermalizes in a finite time $t_{\rm eq}\sim \xi^z$ 
and reaches an equilibrium state
characterized by the canonical distribution function proper to the
mesoscopic Hamiltonian ${\mathcal H}$ of the system.
At the critical point the effects of the initial
conditions persist forever and give rise to a non-equilibrium critical
behaviour with some universal features. 

It is possible to study theoretically this cases by simply accounting
for the initial conditions in the mesoscopic evolution equation
usually used to describe the dynamics of fluctuations
around the equilibrium state. In particular we consider only the case
of a disordered (high-temperature) initial condition.
The field-theoretical approach to this problem
was developed in a seminal paper by Janssen, Schaub and 
Schmittmann \cite{jss-89}, who
focus on the early stage of the relaxation process after the quench, 
i.e., on the short-time scaling.
The scaling forms obtained there for the two-time correlation and
response functions $C_{\bf q}(t,s)$ and $R_{\bf q}(t,s)$ ($t>s$), 
in the limit of long times, turn out to depend
on the ratio $s/t$.
As a consequence, the short-time regime investigated in \cite{jss-89}
defined by the limit $s\rightarrow0$ with fixed $t$ is equivalent to 
$t\rightarrow\infty$ with fixed $s$, that is the ageing limit of 
equations (\ref{xinfdef}), (\ref{dxO}) and (\ref{eq}). Accordingly we
can take advantage of the analysis presented in \cite{jss-89} to study
ageing behaviour at the critical point.

The main results obtained from the RG analysis of this problem 
are the scaling forms 
of the response and correlation functions, in equation (\ref{RXX}) 
and (\ref{CXX}), respectively.
From them one deduces that:
\begin{enumerate}
\item The critical 
exponent $z$ appearing in the equilibrium and non-equilibrium 
scaling forms is the same \cite{jss-89}, although its physical meaning is 
different.
\item Only {\it one} new independent 
critical exponent $\theta$ has to be introduced 
to describe the non-equilibrium dynamics \cite{jss-89}, compared to those
required for the equilibrium one.
\item The FDR $X^\infty$ [see equations (\ref{dx}) and (\ref{xinfdef})] 
is a  universal amplitude-ratio (this conclusion 
was first drawn in \cite{gl-00c} on the sole basis of scaling arguments).
\end{enumerate}

In the next subsections we assume the reader to be familiar with the
field-theoretical approach to critical phenomena 
as explained in many introductory textbooks \cite{zj,tauber,cardyb}
and we summarize here only the most relevant points,
mainly concerning dynamics.

\subsection{Path-Integral Representation of Dynamics}
\label{sec:PathInt}

The field-theoretical approach to dynamic critical phenomena relies 
on a path-integral description of stochastic processes.  
Here we recall the basic steps which allow the construction of the
path-integral associated with a given Langevin equation,
as developed in~\cite{MSR73,bjw-76} 
and reviewed in \cite{Janssen79,jan-92,tauber,zj}. 

Consider the following equation for the field $\p$ (here 
$\p$ can be either a single field, e.g., the order parameter as in Models
A and B or a set of slow and conserved modes in a
more complex case, e.g., 
$\p\mapsto (\p,\en)$ in Model C -- cf. subsection~\ref{duc})
\be
\dpar_t\p({\bf x},t) = {\cal F}[\p({\bf x},t)] 
+ \zeta({\bf x},t) \; .
\label{Langgeneral}
\ee
In this equation ${\cal F}$ is a local functional of $\p({\bf x},t)$
and $\zeta({\bf x},t)$ is a zero-mean Gaussian white noise with 
correlation
\be
\langle\zeta({\bf x},t)\zeta({\bf x'},t') \rangle =2
 {\mathcal N}
\delta(t-t')\delta({\bf x}-{\bf x'})\;,
\ee
where ${\mathcal N}$ can either be a constant (as it is the case in
Model A) or a differential operator acting on ${\bf x}$ (as in the
case of Model B). 
In all the models we are interested in  ${\cal F}$ has the form
\be
{\cal F}[\p({\bf x},t)] = - 
{\cal D}\frac{\delta {\cal H}[\p]}{\delta \p({\bf
x},t)} \;,
\ee
where ${\mathcal H}$  is a functional playing the role of a reduced
Hamiltonian and 
${\mathcal D}$ a constant or a differential operator.
To have TRS and thus a convergence towards the equilibrium
distribution function for the field 
$P[\p]\propto e^{-{\cal H[\p]}}$, the condition 
${\mathcal D}={\mathcal N}$ has to be fulfilled, 
which is the analogue of equation (\ref{eq:EE}).

Given an initial condition $\p({\bf x},t_0)$ the expectation value
of a generic observable $\o[\p]$ over all possible realizations of the noise $\zeta$
can be written as
\be
\langle {\mathcal O} \rangle \equiv  \int \![\rmd \zeta]\,\o[\p_\zeta] 
P_G[\zeta] = 
\int \![\rmd \p]\,\o[\p] \,
\left\{\int\![\rmd \zeta]\delta(\p-\p_\zeta) P_G[\zeta] \right\}\;. 
\ee
$P_G[\zeta]$ is the Gaussian functional probability distribution function
of the noise and $\p_\zeta$ is the solution of equation~(\ref{Langgeneral})
for a given realization of the noise, with the specified 
$\p({\bf x},t_0)$ as initial condition. Taking into account that
\be
\delta(\p-\p_\zeta) = 
\delta(\dpar_t\p - {\cal F}[\p] - \zeta) 
\det \left[\dpar_t-\frac{\delta{\cal F}}{\delta\p}\right]\;,
\label{changedelta}
\ee
it is possible to express the functional $\delta$-function as an exponential
by introducing a complex auxiliary field $\pht$: $\delta(\psi) =
\int\![\rmd \pht]\exp\{\int{\rmd t}{\rmd^dx}\,\pht({\bf x},t)\psi({\bf x},t)\}$. Then the average
over the Gaussian noise is straightforward and leads to
\be
\label{func-av}
\langle {\mathcal O} \rangle = 
\int \![\rmd \p \rmd\pht]\,\o \,
e^{-S_{t_0}[\p,\pht]} \;,
\ee
where
\be
S_{t_0}[\p,\pht] = 
\int_{t_0}^\infty \rmd t \int \rmd^d x 
\left\{\pht[\dpar_t\p - {\cal F}[\p]]
-\pht {\mathcal N}\pht\right\} \;.
\label{gendynfunc}
\ee
The functional $S_{t_0}[\p,\pht]$, usually referred to as the dynamic
functional,  
is the starting point for the field-theoretical approach to dynamics. 
Note that $\tilde{\p}({\bf x},t)$ has a clear physical meaning. 
Indeed, given an external field $h$ coupled linearly 
to $\p$, one has
${\cal H}[\p,h] = {\cal H}[\p] - \beta \int\!\rmd^d x\, h \, \p$.
This implies, following equation~(\ref{gendynfunc}), that 
$\pht$ is conjugated in $S_{t_0}[\p,\pht]$ to the external field $h$,
i.e., $S_{t_0}[\p,\pht,h] = S_{t_0}[\p,\pht] - \beta \int_{t_0}^\infty \rmd
t \int \rmd^d x \pht{\mathcal D} h$.
As a consequence, the linear response of an observable
$\o$ to the field $h$ is given by
\be
{\delta \langle \o \rangle \over \delta h({\bf x},s)} = 
\beta \langle \pht({\bf x},s){\mathcal D} \o\rangle \ .
\ee
For this reason $\tilde{\p}({\bf x},s)$ is termed response field.
In particular the response function (of the order parameter) reads 
\be
R_{{\bf x}-{\bf x}'}(t,s)\equiv \left.\frac{\delta\langle \p({\bf
x},t)\rangle_h}{\delta h({\bf x}',s)}\right|_{h=0} =
 \beta \langle\pht({\bf x}',s) {\mathcal D}\p({\bf x},t)\rangle\,.
\label{generalresp}
\ee
From now on we absorb the factor $T = \beta^{-1}$ in the definition of the
response function.

We remark that in equation~(\ref{gendynfunc}) the term
corresponding to the determinant in equation~(\ref{changedelta}) is missing.
To be properly evaluated, it requires a discretization of the
Langevin equation and eventually its expression depends on the
chosen discretization~\cite{tauber,lrt-79,jan-92}. 
Nevertheless the result of the computation of averages such as
equation~(\ref{func-av}) is actually independent of the particular
choice~\cite{Janssen79} which can be made in such a way to render the
determinant equal to one. In turn, this implies that $R_{{\bf x}-{\bf x}'}(t,t)
\propto \langle \p({\bf x},t)\zeta({\bf x}',t)\rangle = 0$,
corresponding to the so-called \^Ito prescription in 
stochastic calculus (see, e.g., \cite{vKbook,vK-81}). Accordingly, in the
perturbative expansion of averages as~(\ref{func-av}) all the diagrams
with at least one loop of the response function do not
contribute.

\subsection{Dynamic Universality Classes}
\label{duc}

A classification of several dynamic universality 
classes was done in the early
seventies and is reviewed in the classical paper by Hohenberg and Halperin
\cite{HH}. These universality classes have been named with 
capital letters, from A to J (some new classes have been 
added to the original classification of \cite{HH}).
A rather complete set of two-loop results for the equilibrium
dynamics has been only recently reached (see \cite{fm-02} for a brief review).

In the following we introduce the dynamic universality classes that we will 
consider in this review.

\subsubsection{Purely dissipative relaxation: Model A.}

The purely dissipative dynamics of a $N$-component field $\p_i$ ($i=1\dots N$)
can be specified in terms of the stochastic Langevin equation
\be
\dpar_t \p_i ({\bf x},t)=-\Omega 
\frac{\delta \cal{H}[\p]}{\delta \p_i({\bf x},t)}+\zeta_i({\bf x},t) \; ,
\label{lang}
\ee
where $\cal{H}[\p]$ is the static reduced 
Hamiltonian of the model, $\Omega$ is
a kinetic coefficient and $\zeta_i({\bf x},t)$ a zero-mean 
Gaussian white noise with correlations
\be
\langle \zeta_i({\bf x},t) \zeta_j({\bf x}',t')\rangle= 2 \Omega \, \delta({\bf x}-{\bf x}') \delta (t-t')\delta_{ij}\;,
\label{corrnoise}
\ee
where $T$ is the bath temperature. Equation (\ref{lang}) is a special
case of the more general one introduced in section~\ref{sec:PathInt},
with ${\mathcal D}=\Omega$, ${\mathcal N}=\Omega $  and it
fulfills Einstein's relation ${\mathcal D}={\mathcal N}$. 

The critical dynamics of some anisotropic magnets and alloys \cite{HH}
are described by equations~(\ref{lang}) and~(\ref{corrnoise}) 
with ${\mathcal H}$ given by equation~(\ref{lgwA}) with $N=1$, i.e.,
by the effective Hamiltonian of the Ising universality class. 
Model A describes also the critical dynamics of kinetic spin models on
the lattice and spin-flip sampling, i.e., 
in which the elementary step amounts to an arbitrary change in the 
orientation of the spin in a given site, performed with proper rates (e.g., 
the well-known Glauber dynamics \cite{glau}). 
In the case of $N$-component spin models
on a regular lattice and $O(N)$-symmetric 
short-range interactions, the critical
dynamics is described by equation~(\ref{lang}) where ${\mathcal H}$ 
is given by equation~(\ref{lgwA}).

The dynamical properties of Model A 
may be worked out by representing the Langevin 
equation~(\ref{lang}) as a dynamical functional, following the method
outlined in the previous section. 
The resulting action is 
\be
S[\p,\tilde{\p}]= \int_{t_0}^\infty \rmd t \int \rmd^dx 
\left[\tilde{\p} \dpar_t\p +
\Omega \tilde{\p} \frac{\delta \mathcal{H}[\p]}{\delta \p}-
\tilde{\p} \Omega \tilde{\p}\right].\label{mrsh}
\ee

To specify completely the dynamics one has to provide the initial condition
for the field $\p({\bf x},t)$: $\p({\bf x},t_0)=\p_0({\bf x})$.
More generally one can assign a probability distribution function for the 
initial condition, in the form $\exp\{-{\mathcal H}_0[\p_0]\}$ and then
average over the initial field $\p_0({\bf x})$. 
If the system is already in thermal equilibrium
at time $t_0$, then ${\mathcal H}_0[\p_0]={\mathcal H}[\p_0]$ 
and one can equivalently extend the time integration in 
$S[\p,\tilde{\p}]$, from $-\infty$ to $\infty$ \cite{jss-89}. 
This is possible since, assuming ergodicity, independently of
the initial condition in the far past, the same stationary 
order parameter distribution is reached at time $t_0$. 
The resulting theory is translational invariant both in space and time, and 
given that the Einstein's relation is fulfilled, equal-time correlation 
functions can be computed directly using the functional distribution 
$e^{-{\mathcal H}[\p]}$. 

\subsubsection{Conserved order parameter: Model B.}

In some cases Model A is not suited to
describe the dynamics of physical systems. For instance, when the order
parameter is related to the density of particles in a fluid, as it is the case
when studying  the liquid-gas critical point, one expects the
continuity equation to be obeyed. 
The simplest model in which such a local conservation
law is implemented is known as Model B, according to the
classification  of \cite{HH}.
In particular, the dynamics of the scalar order parameter $\p$ is
given by 
\be
\dpar_t \p({\bf x},t) + \nabla\cdot{\bf J}({\bf x},t) = 0\;, 
\label{contB}
\ee
where ${\bf J}({\bf x},t)$ is a fluctuating driving force. Thinking of
$\p$ as a particle density one expects the deterministic driving force
to be related to the gradient of some sort of local chemical
potential. Indeed it is generally assumed that
\be
{\bf J}({\bf x},t) = - \sigma \nabla_{\bf x}
\frac{\delta {\mathcal H}[\p]}{\delta
\p({\bf x},t)} + {\bf J}_\zeta({\bf x},t),
\label{currB}
\ee
where $\sigma$ is a kinetic coefficient and ${\bf J}_\zeta$ a Gaussian
zero-mean random current with correlations
\be
\langle J_{\zeta,i}({\bf x},t) J_{\zeta,j}({\bf x}',t')\rangle= 2
\sigma \, \delta({\bf x}-{\bf x}') \delta (t-t') \delta_{ij}\;.
\label{corrnoiseB}
\ee
In the example given above, $\sigma$ is related to the mobility
whereas $\bar \mu({\bf x},t) = \delta {\mathcal H}[\p]/\delta
\p({\bf x},t)$ represents a sort of chemical potential whose inhomogeneities
drive the diffusion of the particles.

It is not difficult to recast the dynamic model specified by the
previous equations in the form of the Langevin equation (\ref{lang}),
with the noise correlation (\ref{corrnoise}) [where $\zeta =
-\nabla\cdot{\bf J}_\zeta$], $N=1$ and 
$\Omega\rightarrow -\sigma \nabla^2_{\bf x}$ \cite{HH,zj}.

Model B describes the critical dynamic properties of some uniaxial
ferromagnets \cite{HH} which belong to the Ising universality class,
i.e., the effective Hamiltonian ${\mathcal H}$  in
equation~(\ref{currB}) is given by equation~(\ref{lgwA}) with $N=1$.
Moreover Model B describes the critical behaviour of lattice spin models 
with spin-exchange sampling (also known as Kawasaki
dynamics \cite{k66}), characterized by an elementary step which amounts to 
an exchange between the spins in two neighbouring sites.

The dynamical functional associated with Model B is given by
\be
S[\p,\tilde{\p}]= \int \rmd t \int \rmd^dx 
\left[\tilde{\p} \dpar_t\p -
\sigma \pht \nabla^2_{\bf x}\frac{\delta \mathcal{H}[\p]}{\delta \p}+
\tilde{\p} \sigma \nabla_{\bf x}^2 \tilde{\p}\right].
\label{mrshB}
\ee
The initial 
condition can be accounted for as described in the case of Model A.

\subsubsection{Coupling of a conserved density to the non-conserved order 
parameter: Model C.}
\label{UCmodelC}

The models of dynamics considered in the previous subsections involve
only the time evolution of the order parameter $\p$, subjected to
the thermal fluctuations due to the coupling to the thermal bath.
These models, although quite rich and complex in their phenomenology, 
are usually too oversimplified to be able to capture the relevant 
features of real systems, in particular those with a vector order 
parameter. 
In many cases the order
parameter is not the only relevant slow variable that has to be taken
into account. For instance, in the
case of a one-component fluid close to its critical point, 
the conserved order parameter 
interacts with three slow hydrodynamic modes \cite{onuki}. 

The simplest model with two interacting fields, called Model C,
consists of a non-conserved $N$-component 
vector order parameter $\p({\bf x},t)$ coupled 
to a non-critical conserved density $\en({\bf x},t)$.

The dynamics is specified by the following 
coupled stochastic Langevin equations
\bea
\label{langC}
\dpar_t \p ({\bf x},t)&=&-\Omega 
\frac{\delta \cal{H}[\p,\en]}{\delta \p({\bf x},t)}+\zeta_{\p}({\bf x},t) \; , \\
\dpar_t \en({\bf x},t)&=& \Omega\rho\nabla^2_{\bf x} 
\frac{\delta \cal{H}[\p,\en]}{\delta \en({\bf x},t)}+\zeta_\en({\bf x},t) \; ,
\eea
where $\Omega$ and $\rho$ are the kinetic coefficients,
$\cal{H}[\p,\en]$ the reduced Hamiltonian of the system and 
$\zeta_\p({\bf x},t)$, $\zeta_\en({\bf x},t)$  zero-mean Gaussian noises with
\bea
\langle \zeta_\p({\bf x},t) \zeta_\p({\bf x}',t')\rangle&=& 2 \Omega \, \delta({\bf x}-{\bf x}') \delta (t-t') ,\\ 
\langle \zeta_\en({\bf x},t) \zeta_\en({\bf x}',t')\rangle &=& - 2 \rho \,\Omega\, \nabla^2_{\bf x}  \delta({\bf x}-{\bf x}') \delta (t-t')\; .
\eea  
Note that Einstein's relation is fulfilled by this system of coupled
Langevin equations, therefore the equilibrium distribution function of
the fields $\en$ and $\p$ is given by $\exp\{-{\mathcal H}[\p,\en]\}$, 
suitably normalized.

In the case of a system with short-range $O(N)$-symmetric interactions, 
the Hamiltonian ${\mathcal H}[\p,\en]$ in equation~(\ref{langC}) 
is given by 
\be
{\cal H}[\p,\en] = {\cal H}[\p] + 
\int \rmd^d x \left[ \frac{1}{2} \en^2 +  \frac{1}{2}\gamma_0 \en\p^2 \right] ,\label{lgwC}
\ee
where ${\mathcal H}[\p]$ is the $O(N)$-symmetric 
Landau-Ginzburg Hamiltonian reported
in equation (\ref{lgwA}) and $\gamma_0$ the coupling constant between
$\p({\bf x},t)$ and  $\en({\bf x},t)$. In particular this coupling  does not 
change the critical static properties of the field $\p$ 
as it can be seen by computing (through a Gaussian integration)  the 
effective Hamiltonian
${\mathcal H}_{\rm eff}[\p]$ defined by
$\exp\{-{\mathcal H}_{\rm eff}[\p]\} = \int[\rmd\en] \exp\{-{\mathcal
H}[\p,\en]\}$. Indeed it turns out that ${\mathcal H}_{\rm eff}[\p] =
{\mathcal H}[\p]$ (see equation~(\ref{lgwA})) where $g_0\mapsto
g_0'\equiv g_0 - 12 \gamma_0^2$. 
On the same footing one can show that 
$\en$-field equilibrium correlation functions are related to 
$\p^2$-field correlation functions \cite{zj}. For this reason 
$\en$ is also referred to as energy density, given that $\p^2$ is
conjugate to the temperature ($\propto r_0$) in equation~(\ref{lgwA})
and therefore the derivatives of the free-energy with respect to the
temperature give rise to $\p^2$($\en$)-correlation functions.

Some lattice models belonging to 
this universality class are discussed in \cite{latC}.  Among 
real systems whose dynamic critical properties are described by Model C
with Hamiltonian (\ref{lgwC}) and $N=1$
we mention the case of structural (displacive) transitions in
crystalline materials and some phase transition in 
uniaxial antiferromagnets \cite{HH}.

Dynamical correlation functions, generated by the Langevin 
equations~(\ref{langC}) and averaged over the noises $\zeta_\p$ and 
$\zeta_\en$, may be obtained by means of the field-theoretical action 
\be
\fl
S = \int \rmd^dx \,\rmd t
\left[ \tilde{\p} \dpar_t\p +
\Omega \tilde{\p} \frac{\delta \mathcal{H}[\p,\en]}{\delta \p}-
\tilde{\p} \Omega \tilde{\p} +\tilde{\en} \dpar_t\en -
\rho\Omega \tilde{\en} \nabla_{\bf x}^2\frac{\delta \mathcal{H}[\p,\en]}{\delta \en} +
\tilde{\en} \rho \Omega \nabla_{\bf x} ^2\tilde{\en} \right],
\label{mrshC}
\ee
where two response fields $\tilde{\p}({\bf x},t)$ and  
$\tilde{\en}({\bf x},t)$ associated with $\p({\bf x},t)$ and 
$\en({\bf x},t)$ have been introduced. 
It is easy to read from equation~(\ref{mrshC}) 
and~(\ref{lgwC}) the interaction vertices, given by 
$-\Omega g_0 \tilde{\p} \p^3/3!$ (as in the case of Model~A),
$ - \Omega \gamma_0 \en \tilde{\p}\p$ and $\rho \Omega \gamma_0 \,\p^2\nabla^2\tilde{\en}/2$.

To take into account the effect of the initial condition 
one has also to average over the possible 
initial configurations of both the order parameter 
$\p_0({\bf x})=\p({\bf x},t=0)$ and the conserved density 
$\en_0({\bf x})=\en({\bf x},t=0)$
with a probability distribution  $e^{-H_0[\p_0,\en_0]}$ given 
by~\cite{oj-93}
\be
H_0[\p_0]=\int\! \rmd^d x\, \left\{ \frac{\tau_0}{2}[\p_0({\bf x})-u({\bf x})]^2 + \frac{1}{2 c_0}[\en_0({\bf x})- v({\bf x})]^2\right\}.
\ee
This specifies an initial state $u({\bf x})$ for $\p({\bf x},t)$ and 
$v({\bf x})$ for $\en ({\bf x},t)$ with correlations proportional to 
$\tau_0^{-1}$ and $c_0$, respectively.

The model with a conserved order parameter coupled to a conserved density
is called Model D \cite{HH}.

\subsubsection{Other dynamic universality classes.}
\label{UCother}

The models of critical dynamics introduced so far are in many cases 
still too simple to describe the critical dynamic properties of a 
variety of actual materials. 

For instance let us briefly discuss the case of isotropic
ferromagnets, microscopically described by the so-called Heisenberg
model for the three-component 
spin variables (see, e.g., \cite{tauber}). 
In the absence of symmetry-breaking fields the model
has a $O(3)$ symmetry.
At a mesoscopic level the system is described by a
three-component order parameter ${\bf S}({\bf x},t)$, representing
the local magnetization, which turns out to be a dynamically conserved
quantity. In view of the underlying $O(3)$ symmetry, the universal
aspects of the equilibrium critical properties of the model are
correctly captured by the Hamiltonian (\ref{lgwA}) with
$N=3$~\cite{PV-r} ($\p_i \mapsto S_i$).
The mesoscopic dynamics of order parameter ${\bf
S}({\bf x},t)$  has to account both for the conservation law
mentioned above and for the expected Larmor's precession of 
${\bf S}({\bf x},t)$
in the local magnetic field $\delta {\mathcal H}[{\bf S}]/\delta{\bf
S}({\bf x},t)$ generated by neighbouring spins.  
The Langevin equation accounting for these features and the associated
universality class are known as Model J \cite{HH}:
\be
\dpar_t{\bf S}({\bf x},t) = - g_L {\bf S}({\bf x},t)\times\frac{\delta {\mathcal H}[{\bf S}]}{\delta{\bf
S}({\bf x},t)} + \sigma \nabla^2_{\bf x} \frac{\delta {\mathcal H}[{\bf S}]}{\delta{\bf
S}({\bf x},t)} + {\bm \zeta}({\bf x},t)\,,
\ee
where $g_L$ is the coupling constant of the precession term, $\sigma$
the kinetic coefficient (analogous to that of Model B) and ${\bm
\zeta}$ a Gaussian random current with correlations given by equation
(\ref{corrnoise}) where $\Omega \mapsto -\sigma\nabla_{\bf
x}^2$. 
Apart from the precession term (which is an example of the general
class of {\it reversible mode-couplings}, see~\cite{tauber,zj} for details), 
the dynamics of each single component $S_i$ is given by a Model B
and satisfies Einstein's relation. The precession term, on the other
hand, does not influence the equilibrium distribution which is still given 
by $\exp\{-{\mathcal H}[{\bf S}]\}$, properly normalized \cite{tauber,zj}.
We do not discuss the
properties characterizing this universality class but we mention only that
they agrees with what has been observed in experiments \cite{HH}.
The critical properties of an isotropic antiferromagnet close to the
N\'eel temperature are instead described by a non-conserved
three-component order
parameter ${\bf m}_s({\bf x},t)$ (the local staggered magnetization) 
which is dynamically coupled, via precession-like terms, to a three-component 
conserved density representing the local magnetization 
${\bf m}({\bf x},t)$. The resulting dynamic universality class is known as
Model G and its predictions agree with the experimental results on
actual magnetic systems. We remark that the 
equilibrium properties of the model are given by 
the Hamiltonian (\ref{lgwA}) (with $N=3$) 
for the order parameter ${\bf m}_s$~\cite{PV-r}.
Accordingly, models G and J provide a non-trivial example of two different
dynamic universality classes of real magnetic systems 
that reduce to the same static universality class.
Other examples are provided by the
dynamic universality classes describing a planar magnet and the superfluid
$ ^4$He, known as Model E and 
F, respectively \cite{HH}, which belong to the same static
universality class as
the Hamiltonian (\ref{lgwA}) with two-component order
parameter (i.e., $O(2)$ symmetry).

Since we do not consider these models in what follows, we do not
discuss any further details of the associated dynamic universality
classes and refer the interested reader to the available review
\cite{HH} and textbooks \cite{onuki,tauber} on the subject.

\subsection{Renormalization and scaling of a purely dissipative critical 
system}

As we discussed in subsection~\ref{dyn} the quantities one is interested
in when 
studying dynamical processes are the response and correlation
functions. In the following we will generically refer to them as
correlation functions. In subsection~\ref{sec:PathInt} we described the method
that allows the computation of such functions as averages of the
form~(\ref{func-av}) where the specific expression of the functional
$S_{t_0}$ depends on the considered dynamics. 
Field-theoretical methods can be applied to compute such averages. As
usual, when going beyond the Gaussian approximation (i.e., considering
the anharmonic terms in the expansion of $S_{t_0}$)
one faces the problem of ultraviolet (UV) divergences that
can be regularized by standard methods 
and removed by means of a standard renormalization
procedure. 
The existence of a well-defined 
renormalized theory allows the derivation of RG equations for the
correlation functions. In particular their solutions highlight the
scaling properties of these functions in the large-distance long-time
limit [the so-called infrared (IR) behaviour] which can be computed  
by means of the renormalized perturbation theory.

To illustrate the general procedure, we consider as a specific example
the Model A dynamics of a $\p^4$ field theory. 
All the following basic steps can be carried over to more complex cases.

\subsubsection{RG, $\e$ expansion and Minimal Subtraction.}
\label{RGandallthat}

The RG idea is to construct a scale-dependent effective action 
$S_\varkappa[\p,\pht]$ (with $\varkappa\ge 1$) which has
connected correlation functions 
$G_{n,\tilde n} = \langle [\p]^n[\pht]^{\tilde n}\rangle_c$ satisfying
\be
G^{(\varkappa)}_{n,\tilde n}(\{{\bf x},t\})
={\mathcal Z}^{-n/2}(\varkappa)\tilde {\mathcal Z}^{-\tilde n/2}(\varkappa)
G_{n,\tilde n}(\{\varkappa{\bf x},\varkappa^z t\})+\dots,
\label{RGidea}
\ee
where $G^{(\varkappa)}_{n,\tilde n}$ is computed with
the effective action $S_\varkappa$, whereas $G_{n,\tilde n}$ is
obtained with the original action $S$. One can arbitrarily
fix $S_{\varkappa=1}\equiv S$. 
The ellipsis in equation~(\ref{RGidea}) stand for functions decreasing
faster than any power of $\varkappa$ for 
$\varkappa\rightarrow\infty$, $\{{\bf x},t\}$ stands for the collection of 
the $n+\tilde n$ space-time points, and we allow an anisotropic scaling
between space and time through the exponent $z$.
The mapping $S[\p,\pht]\rightarrow S_\varkappa[\p,\pht]$ 
is called RG transformation and
$\{S_\varkappa[\p,\pht]\}_{\varkappa\ge 1}$ is called the trajectory
of $S[\p,\pht]$ under the RG flow.
Various RG transformations differ by the form of ${\mathcal Z}(\varkappa)$,
$\tilde {\mathcal Z }(\varkappa)$, and of $S_\varkappa$.
In explicit constructions the transformations 
are generated by integrations over the large-momentum fluctuation
modes of the fields.

The coupling constants appearing in $S_\varkappa[\p,\pht]$ 
are explicit function of 
$\varkappa$. If for $\varkappa\rightarrow\infty$, 
the action  $S_\varkappa[\p,\pht]$ has a limit
$S^*[\p,\pht]$, called the fixed-point action, 
the correlation functions behave like
\be
G_{n,\tilde n}(\{\varkappa {\bf x}, \varkappa^z t\})
\sim
{\mathcal Z}^{n/2}_*(\varkappa) \tilde {\mathcal Z}^{\tilde n/2}_*(\varkappa)
G_{n,\tilde n}^*(\{{\bf x},t\})\,\quad \mbox{for}\quad
\varkappa\rightarrow\infty \,,
\ee
where $*$ is used to denote fixed-point quantities and $G^*_{n,\tilde
n}$ is computed with the fixed-point action $S^*$. It is possible to
show that under some reasonable assumptions~\cite{zj}
${\mathcal Z}_*(\varkappa)={\varkappa}^{-2 d_\p}$ and 
$\tilde {\mathcal Z}_*(\varkappa)=\varkappa^{-2 d_\pht}$.
Accordingly, correlation functions have a scaling behaviour at 
large distances and long times:
\be
G_{n,\tilde n}(\{\varkappa{\bf x},\varkappa^z t\})\sim
\varkappa^{-n d_\p-n d_\pht} G_{n,\tilde n}^*(\{{\bf x},t\})\,\quad
\mbox{for}\quad\varkappa\rightarrow\infty \, ,
\ee
where the exponents
$d_\p$, $d_\pht$ (the so-called scaling dimensions of the fields), 
and $z$ are properties of the fixed point, i.e., they depend only on 
$S^*[\p,\pht]$.
Correlation functions of all those actions that flow under RG 
transformations into the same fixed-point action $S^*$ display
the same critical 
scaling behaviour: This provides the basis for the universality
observed in critical phenomena. Within this framework  
a given universality class is characterized by
the values of the exponents $d_\p$, $d_\pht$ and $z$ and by the scaling 
function $G_{n,\tilde n}^*$. 
There are several ways and approximations that 
allow to calculate such quantities in physical dimensions $d=2,3$.

Renormalization-Group equations for the correlation functions 
can be conveniently 
derived within the field-theoretical approach. 
They eventually
lead to a relation of the form~(\ref{RGidea}) for correlation
functions.
The first step is to regularize the theory in such a way to render
finite those Feynman diagrams of the perturbative expansion that would be
otherwise UV-divergent. To this end we will assume dimensional
regularization \cite{zj}. 
The original UV divergences of the four-dimensional theory 
appear in the correlation functions as poles when $\e\rightarrow0$,
i.e., when the regularization is formally removed.
Let us consider the Model A dynamics with Hamiltonian
given by equation~(\ref{lgwA}). In this specific case one introduces
the {\it renormalized parameters} as:
\be
\eqalign{{\stackrel\circ\p}= Z^{1/2}\p,\quad 
{\stackrel\circ \pht}= \tilde Z^{1/2} \pht,\\
{\stackrel\circ\Omega}=Z_\omega \Omega, \quad
r_0\equiv{\stackrel\circ r}= Z^{-1}Z_r r, \quad\mbox{and}
\quad  g_0\equiv{\stackrel\circ g}= Z_g\mu^{4-d} g.}
\label{renormMA}
\ee
Here the parameters with the superscript $\circ$ are those originally
appearing in equation (\ref{mrsh}) ({\it bare parameters})
whereas the others represent the renormalized ones. 
Note that the renormalized coupling constant $g$ is
defined in such a way to be dimensionless.
$\mu$ is an arbitrary scale with the dimension of a mass 
that in general enters in the 
renormalization of the parameters that are not dimensionless.
The $Z$ factors are called renormalization constants and can be
computed in many different ways. In the following we will use the
so-called Minimal Subtraction (MS) scheme \cite{zj} which allows one to
study directly the critical theory, hence we
fix $r=0$ in what follows. Within MS the renormalization constants
take the form $Z = 1+\sum \alpha_n(g)/\e^n$, where the coefficient
$\alpha_n(g)=O(g^n)$ are computed order by order in perturbation theory.
It is possible to show that the FDT yields the relation
$Z_\omega=(Z/\tilde Z)^{1/2}$ \cite{zj}.
Once the correlation functions of the fields are expressed in terms of the 
renormalized parameters only, they turn out to be regular functions for
$\e\rightarrow 0$.

To renormalize $G_{n,\tilde n}$ one uses the $Z$-factors in 
equations~(\ref{renormMA}) obtaining
\be
{\stackrel\circ G}_{n,\tilde n} = Z^{n/2}\tilde Z^{\tilde n/2}
G_{n,\tilde n} \; ,
\ee
where the r.h.s. is expressed in terms of renormalized quantities.
The RG equations may be derived by exploiting the fact that the bare
correlation functions are independent of the mass scale $\mu$ introduced
to define the renormalized theory
\be
\left .\mu \dpar_\mu \stackrel{\circ}G_{n,\tilde n}
\right|_0 = 0 \; ,
\label{RGE}
\ee 
where the notation $|_0$ reminds that the derivative is taken with fixed 
bare parameters. As we show below, this equation defines a RG flow in
the parameter space $\{g,\mu,\Omega\}$.

In terms of renormalized quantities, equation (\ref{RGE}) reads 
\be
\fl
\left[\mu\frac{\partial}{\partial \mu}+
\frac{\tilde n}{2} \tilde\gamma +
\frac{n}{2} \gamma +
\gamma_\omega \Omega  \frac{\partial}{\partial \Omega}+
\beta_g \frac{\partial}{\partial g}\right]
G_{n,\tilde n}(\{{\bf x},t\};g,\mu,\Omega)=0\,,
\label{RG2}
\ee
where we the so-called Wilson's functions are defined by
\be
\fl
\gamma(g) = \mu\dpar_\mu \ln Z|_0\,, \quad
\tilde\gamma(g) = \mu\dpar_\mu \ln \tilde Z|_0\,,\quad 
\gamma_\omega(g) = \mu\dpar_\mu \ln \Omega|_0\,,
\label{gammavarie}
\ee
and 
\be 
\beta(g) = \mu\dpar_\mu g|_0\,,
\label{beta}
\ee
which are regular for $\e=0$ \cite{zj}.

To solve RG equation (\ref{RG2}), the method of characteristic is
usually employed.
One introduces a continuous real scale parameter $l$, 
in terms of which the flowing mass scale $\mu$ reads 
$\mu(l)=\mu l$ and $\mu$ plays the role of an initial condition for $l=1$. 
The IR behaviour is obtained for $l\rightarrow0$.
Then one defines effective $l$-dependent parameters by solving the
ordinary differential equations
\be
l \frac{d\Omega(l)}{dl}=\Omega(l)\gamma_\omega(l), \quad
l \frac{d g(l)}{dl}= \beta(l)\,, \quad {\rm etc.}\,,
\ee
with $\Omega(l=1)=\Omega$, $g(l=1)=g$ etc., where the RG functions on
the r.h.s. depend on $l$ through $g(l)$, 
e.g., $\gamma(l)=\gamma(g(l))$.
The RG equation for the renormalized correlation functions becomes an ordinary 
differential equation 
\be
\left[
\frac{\tilde n}{2} \tilde\gamma(l) +
\frac{n}{2} \gamma(l) +
l\frac{\rmd}{\rmd l}\right]
G_{n,\tilde n}(\{{\bf x},t\};g(l),\mu(l),\Omega(l))=0\,,
\label{RG3}
\ee
whose solution is 
\be 
\fl
G_{n,\tilde n}(\{{\bf x},t\};g(l),\mu(l),\Omega(l))=
G_{n,\tilde n}(\{{\bf x},t\};g,\mu,\Omega)
\exp\left\{\int_l^1\frac{\rmd l'}{2l'}[n \gamma(l')
+\tilde n  \tilde\gamma(l')]\right\}\,.
\label{RGsol}
\ee

Physically, varying the parameter $l$, one changes the scale at which the 
model is explored. The infrared regime is reached for $l\rightarrow0$.
The flowing parameters $\Omega(l)$
and $g(l)$ can be interpreted as 
effective values of the original ones on different length/time scales
set by $\mu l$.
They evolve under scale transformation $\mu\rightarrow\mu l$ according to 
the flow equations, that are determined by the required renormalizations
through the functions $\gamma$, $\tilde \gamma$, $\gamma_\omega$ and $\beta$
(see equations (\ref{gammavarie}) and (\ref{beta})).

Under a scale transformation, the dimensionless renormalized coupling $g$ 
changes according to equation (\ref{beta}). Therefore $g(l)$
increases for $l\rightarrow 0$ when 
the $\beta$ function is negative and decreases in the opposite case.
Fixed points for the flow of $g$ correspond to zeroes $\bar g$ of the
$\beta$ function, i.e., $\beta(\bar g)=0$. Those where 
$\beta'(\bar g) < 0$ are not stable in the limit $l\rightarrow 0$ 
(IR behaviour) we
are interested in: The effective coupling
moves away from them. Conversely, when $\beta'(\bar g) > 0$, the zero
is stable. Indeed linearizing the flow equation for $g$ close to $\bar
g$ one finds that $g(l) - \bar g = (g - \bar g) l^{\beta'(\bar g)}$
for $l \ll 1$. 
Let us assume we find such an IR-stable fixed point for $g=g^*$. 
We denote the fixed-point values of the 
RG functions as $\eta=\gamma(g^*)$, $\tilde\eta=\tilde\gamma (g^*)$, 
$\eta_\omega=\gamma_\omega(g^*)=(\tilde\eta-\eta)/2$ (this last
equality follows from the FDT),
and $z= 2+\eta_\omega$. Accordingly, $\Omega(l\rightarrow 0) =
\Omega\, l^{z-2}$.

The leading IR behaviour is obtained from equation (\ref{RGsol})
\be
G_{n,\tilde n}(\{{\bf x},t\};g,\mu,\Omega)\simeq_{l\rightarrow0}
l^{\eta n/2+\tilde \eta\tilde n/2} G_{n,\tilde n}(\{{\bf x},t\};g(l),\mu(l),\Omega(l))\,.
\label{RGasy}
\ee
From the dimensional analysis one finds that 
\be\fl
G_{n,\tilde n}(\{{\bf x},t\};g,\mu,\Omega)=
\ell^{-n(d-2)/2-\tilde n(d+2)/2} 
G_{n,\tilde n}(\{{\bf x}/\ell,t/\tau\};g,\mu\ell,\Omega \tau/\ell^2)\,,
\label{diman}
\ee
where $\ell$ and $\tau$ are arbitrary units of measure of length and
time, respectively. 
Applying this equation to the r.h.s. of equation (\ref{RGasy}) with $\ell
= (\mu l)^{-1}$ and $\tau = \Omega^{-1} \ell^2l^{2-z}$, one finally
arrives at the leading IR scaling behaviour of the correlation functions 
\be
\fl
G_{n,\tilde n}(\{{\bf x},t\};g,\mu,\Omega) = 
l^{\delta(n,\tilde n)} \mu^{n(d-2)/2+\tilde n(d+2)/2} 
G_{n,\tilde n}(\{l(\mu {\bf x}),l^z (\Omega \mu^2\,t)\}; g^*,1,1) \; ,
\label{genscalingeq}
\ee
where
\be
\delta(n,\tilde n)= n \frac{d-2+\eta}{2} +\tilde n 
\frac{d+2+\tilde\eta}{2} \;.
\label{genexpeq}
\ee
In the following, to simplify the notation, we will set the constants
$\mu$ and $\Omega$ to 1.

Let us consider in more detail the scaling forms of the two-point 
critical correlation function $C_{\bf q}(t,s)$ and 
response function $R_{\bf q}(t,s)$, with $t>s$.
From equation~(\ref{genscalingeq}) and equation~(\ref{genexpeq}), using 
TTI, one has 
\bea
C_{\bf q}(t,s) &\equiv& \int \!\!\rmd^d x \,e^{\rmi {\bf q x}}
\,G_{2,0}=l^{\eta-2} f_C(l^{-1}{\bf q},l^z (t- s))\;\nonumber ,\\
R_{\bf q}(t,s) &\equiv& \Omega \int \!\!\rmd^d x \,e^{\rmi {\bf q x}}
\,G_{1,1}=l^{\eta+z-2} f_R(l^{-1}{\bf q},l^z(t-s))\; .
\label{scalzeroeq}
\eea
If we choose $l=(t-s)^{-1/z}$, with $t-s \gg 1$, these scaling forms may be 
rewritten as ($q=|{\bf q}|$)
\bea
C_{\bf q}(t,s) &=& (t-s)^{(2-\eta)/z}F^{\rm eq}_C (q (t-s)^{1/z})\; ,
\nonumber\\
R_{\bf q}(t,s) &=& (t-s)^{(2-\eta-z)/z}F^{\rm eq}_R(q(t-s)^{1/z})\; ,
\label{scaloneeq}
\eea
where $F^{\rm eq}_R(x)$ and $F^{\rm eq}_C(x)$ are regular functions for small
argument. 
The FDT provides a differential relation between $F^{\rm eq}_R(x)$ and
$F^{\rm eq}_C(x)$.

By means of RG, one reduces the problem of calculating universal quantities
to the evaluation of RG functions at the IR-stable fixed point.
In order to find such a fixed point, a  useful tool is the $\e$ 
expansion. It is based on the observation that $d=4$ is a special dimension 
for  the model we are considering, in fact using the definition of the 
$\beta$ function we get
\be
\beta(g)=\mu\partial_\mu g|_0=-(4-d) 
\left(\frac{d\log g Z_g}{d g}\right)^{-1}\,.
\ee
Being $Z_g=1+O(g)$, the fixed point $g^*=0$ is IR stable for 
$d>4$ and unstable in the opposite case. Thus, for $d>4$ the critical 
behaviour is governed by the fixed-point $g^*=0$ (Gaussian fixed
point), corresponding to the
Gaussian model.
For $d<4$ this is no longer true.
However, adiabatically decreasing the dimension from four dimensions, the 
new IR stable fixed point is expected to be close to the Gaussian one,
i.e., $g^*=O(\e)$, where $\e=4-d$. 
As originally pointed out 
in a seminal paper by Wilson and 
Fisher \cite{wf-72}, one can perform a double expansion of RG functions
in terms of $g$ and $\e$ and find the zero(es) of the $\beta$ function as 
series in $\e$. Finally the critical exponents and other universal quantities
are obtained expanding in $\e$ the corresponding RG functions evaluated at 
the stable fixed point. 

This procedure allows one to obtain the critical quantities as series in $\e$.
The analytical continuation of such series at physical dimensions $d=3,2$ 
(i.e., $\e=1,2$) may seem not so straightforward. 
However, the validity of this continuation  to $\e=1,2$ 
is corroborated by the very good agreement of the $\e$-expansion
estimates (obtained applying resummation techniques based on Borel
resummation \cite{zj}) with other theoretical and experimental 
values \cite{PV-r,zj}.

\subsubsection{Non-equilibrium renormalization and scaling.}
\label{secRenScal}

Let us consider the case with a non-equilibrium initial condition.
As explained in section \ref{duc}, it can be studied by averaging the
correlation functions with the probability distribution function of
the initial condition, specified by $\exp\{-{\mathcal H}_0\}$
\cite{jss-89}. In turn this implies that expectation values in
equation~(\ref{func-av}) have to be computed with a total dynamical
functional given by $S[\p,\pht] + {\mathcal H}_0$ (see
equation~(\ref{mrsh})).
As long as ${\mathcal H}_0$ has the same form as the static Hamiltonian 
${\mathcal H}$, but with different bare couplings, the usual
renormalizations are enough to render the theory finite~\cite{jss-89,jan-92}. 
This is no longer true if, for instance, the initial state is an
uncritical one like a high-temperature state with short-range correlations
and a small initially prepared magnetization
$\langle \p_0({\bf x})\rangle = a({\bf x})$, i.e. 
\be
\langle [\p_0({\bf x})-a({\bf x})]
[\p_0({\bf x}')-a({\bf x}')]\rangle = 
\tau_0^{-1}\delta({\bf x}-{\bf x}') \;.
\ee
The corresponding ${\mathcal H}_0[\p_0]$ 
is Gaussian
\be
{\mathcal H}_0[\p_0]=\int\! \rmd^d x\, \frac{\tau_0}{2}
[\p_0({\bf x})-a({\bf x})]^2 \; .
\label{initialH}
\ee
The canonical mass dimension of $\tau_0$ is $[\tau_0]_\can = 2$, thus
the possible fixed points of $\tau_0$ can be $\pm \infty$ and $0$.
Since $\tau_0=0$ and $-\infty$ yield non-normalizable distributions,
the fixed point of physical interest is $\tau_0=\infty$ \cite{jss-89}.
This corresponds in the language of surface critical behaviour to the 
ordinary transition \cite{diehl-86}. Corrections due to finite value 
of $\tau_0$ will be irrelevant in the RG sense, thus one can set 
$\tau_0^{-1}=0$ from the very beginning of the calculation,
as long as the leading contribution does not vanish with
$\tau_0^{-1}=0$.

Following standard methods, the response and correlation functions may
be obtained by a perturbative expansion of the functional weight
$\exp\{-S[\p,\tilde{\p}]-{\cal H}_0[\p_0]\}$ in terms of the
coupling constant $g_0$.
From the technical point of view, the breaking of TTI does not allow 
the factorization of connected correlation functions in terms of one-particle 
irreducible ones, as usually done when TTI is not broken. As a consequence
all the perturbative computations must be done in terms of connected 
functions only \cite{diehl-86,jss-89}.

The addition of ${\mathcal H}_0[\p_0]$ gives rise to new
divergences in perturbation theory whenever $s$ approaches
the ``time surface'' located at $t_0\equiv0$. 
On the other hand, the bulk renormalization functions defined by equations 
(\ref{renormMA}) are not changed by the presence of ${\mathcal H}_0[\p_0]$,
if MS is employed to renormalize the theory.
It is possible to remove these new singularities by counterterms
``located'' at that time surface, as it has been shown in 
\cite{diehl-86} for the case of surface critical phenomena. 
Moreover, being the canonical mass dimension of time variables $-2$, 
the degree of divergence of a generic correlation function 
decreases by $2$ for each vanishing time argument. As a consequence the new
renormalizations are required only in the case of two-point 
functions.
A detailed analysis shows that once equilibrium theory has been renormalized
according to equation (\ref{renormMA}), only one new 
renormalization constant is required to render finite both the correlation 
and the response functions. 
In the case of non-equilibrium Model A dynamics the new renormalization is
\be
\pht_0({\bf x})\mapsto (Z_0 \tilde Z)^{1/2} \pht_0({\bf x}),
\label{renormB}
\ee
and $Z_0$ is the new renormalization constant.

The scaling properties of connected correlation functions 
\be
G_{n,\tilde n}^{\tilde n_0} = \langle [\p]^n[\pht]^{\tilde n}
[\pht_0]^{\tilde n_0} \rangle
\label{GFR}
\ee
may be exploited by using RG equations as in equilibrium. 
To renormalize $G_{n,\tilde n}^{\tilde n_0}$ one uses all the $Z$-factors 
previously introduced (equations~(\ref{renormMA}) and (\ref{renormB})), obtaining
\be
G_{n,\tilde n}^{\tilde n_0} \mapsto 
{\stackrel\circ G}_{n,\tilde n}^{\tilde n_0} = Z^{n/2}\tilde Z^{\tilde n/2}
(\tilde Z Z_0)^{\tilde n_0/2} G_{n,\tilde n}^{\tilde n_0} \; ,
\ee
where again the r.h.s. is expressed in terms of renormalized quantities.
The RG equations are derived from  
$\left .\mu \dpar_\mu \stackrel{\circ}G_{n,\tilde n}^{\tilde n_0}
\right|_0 = 0$, in analogy with equation (\ref{RGE}). 
To solve it, a new Wilson's function $\gamma_0 = \mu\dpar_\mu \ln Z_0|_0$ 
has to be introduced.
The leading scaling behaviour of the correlation functions (\ref{GFR})
is determined as in equilibrium
\be
G_{n,\tilde n}^{\tilde n_0}(\{{\bf x},t\}) = 
l^{\delta(n,\tilde n,\tilde n_0)} 
G_{n,\tilde n}^{\tilde n_0}(\{l{\bf x},l^z t\}) \; ,
\label{genscaling}
\ee
where now
\be
\delta(n,\tilde n,\tilde n_0)= n \frac{d-2+\eta}{2} +\tilde n 
\frac{d+2+\tilde\eta}{2} + \tilde n_0 \frac{d+2+\tilde\eta +\eta_0}{2} \;, 
\label{genexp}
\ee
with the novel exponent $\eta_0 = \gamma_0(g^*)$, in terms
of which the so-called initial-slip 
exponent $\theta = -\eta_0/(2 z)$ is defined~\cite{jss-89}.
The scaling dimensions of the surface 
field is $[\pht_0]_\scal = (d+2+\tilde\eta +\eta_0)/2$.

According to the same procedure, the analysis has been done for
various models. 
Model C dynamics (see sections~\ref{UCmodelC} and~\ref{modC}) has been
studied in \cite{oj-93},
whereas Models E and G [see section~(\ref{UCother})]
have been studied in \cite{oj-93b}. Model A dynamics of a tricritical 
point is investigated in \cite{oj-94}. 
In \cite{kissner-92,oj-95} the weakly dilute Ising Model 
has been considered in the case of uncorrelated impurities.
The case of extended random defects has been also addressed \cite{f-04}.
Other interesting universality classes have been studied \cite{cinesi}.

Let us consider in more details the scaling forms of the two-point 
critical correlation function $C_{\bf q}(t,s)$ and 
response function $R_{\bf q}(t,s)$, with $t>s>0$ (i.e., with $t$ and $s$ 
in the ``bulk'').
From equation~(\ref{genscaling}) and equation~(\ref{genexp}) one 
has (we set $\Omega =1$) 
\bea
C_{\bf q}(t,s) &\equiv& G^0_{2,0}=l^{\eta-2} C_{l^{-1}{\bf q}}(l^z t,l^z s)\;\nonumber ,\\
R_{\bf q}(t,s) &\equiv& G^0_{1,1}=l^{\eta+z-2} R_{l^{-1}{\bf q}}(l^z t,l^z s)\; .
\label{scalzero}
\eea
Considering $l=(t-s)^{-1/z}$, these scaling forms may be rewritten as 
\bea
C_{\bf q}(t,s) &=& (t-s)^{(2-\eta)/z}\tilde{\mathcal F}_C (q (t-s)^{1/z},s/t)\; ,
\nonumber\\
R_{\bf q}(t,s) &=& (t-s)^{(2-\eta-z)/z}\tilde{\mathcal F}_R(q(t-s)^{1/z},s/t)\; ,
\label{scalone}
\eea
where, as a fundamental difference to equilibrium ones, they depend 
on the ratio $s/t$.

We observe that the scaling functions $\tilde{\mathcal F}_C$ and
$\tilde{\mathcal F}_R$ 
just introduced are not expected to be regular when $s$ approaches
the time surface, i.e., for $s\rightarrow 0$.
To determine the functional form of the correlation 
functions when their
arguments approach exceptional points, the short-distance expansion (SDE)
must be used (see \cite{zj,cardyb} for general reference 
and \cite{diehl-86} for applications to surface critical phenomena). 
The starting point is a formal expansion of the fields $\p({\bf x},s)$ and
$\pht({\bf x},s)$ around $s=0$. First of all one notes that, 
when inserted in correlation functions with bulk fields, the following
relations hold~\cite{jss-89,jan-92}
\be
\p({\bf x},0) = \p_0({\bf x}) = 0\quad \mbox{and}\quad 
\dpar_s\p({\bf x},s)|_{s=0} \equiv \dot{\p}_0({\bf x})
= 2\Omega \pht_0({\bf x})\; .
\label{relationSDE}
\ee
As a consequence, for small $s$, one can formally expand the fields as
\bea
\p({\bf x},s) &\sim& \phi(s)\pht_0({\bf x}) + \mbox{h.o.c.f.}\; ,
\label{SDE1}\\
\pht({\bf x},s) &\sim& \tilde\phi(s)\pht_0({\bf x}) 
+ \mbox{h.o.c.f.}\; ,
\label{SDE2}
\eea
where h.o.c.f stands for higher-order composite fields which are neglected if
one is interested only in the leading contributions in the 
limit $s/t\rightarrow0$. 
Inserting the relations~(\ref{SDE1}) and (\ref{SDE2}) 
into the correlation functions and taking
into account the scaling behaviour equation~(\ref{genscaling}), one deduces 
that, at criticality~\cite{jss-89,jan-92},
\bea
\phi(s) & = &  a_C s^{1-\theta} \; ,\nonumber\\
\tilde\phi(s) & = & a_R s^{-\theta}\;,
\eea
(i.e., $[\phi]_\scal = [\p]_\scal - [\pht_0]_\scal = - z(1-\theta)$ and
$[\tilde{\phi}]_\scal = [\pht]_\scal - [\pht_0]_\scal = z\theta$, with
$[{\rm time}]_\scal = - z$)
where $a_C$ and $a_R$ are two non-vanishing constants.
Accordingly for small $s$ one has
\bea
C_{\bf q}(t,s)   & =&
 \phi(s) \langle \p({\bf q},t)\pht_0(-{\bf q})\rangle \; \nonumber,\\
R_{\bf q}(t,s)  &= & 
\tilde\phi(s) \langle \p({\bf q},t)\pht_0(-{\bf q})\rangle \; .
\eea
From equation~(\ref{genscaling}) the scaling form of  
$\langle\p({\bf q},t)\pht_0(-{\bf q})\rangle 
\equiv G_{1,0}^{1}(\{{\bf q},t\};r)$ 
can be determined and at criticality 
\be
G_{1,0}^{1}(\{{\bf q},t\};0) = 
l^{-\theta z-(2-z-\eta)} G_{1,0}^{1}(\{l^{-1}{\bf q},l^z t\};0) \; .
\label{scalinginiz}
\ee
Taking into account the previous three relations, the following 
conclusion may be drawn ($q = |{\bf q}|$)
\bea
C_{\bf q}(t,s) &=& a_C t^{(2-\eta)/z}(t/s)^{\theta-1} f_C(qt^{1/z}) \; ,\nonumber\\
R_{\bf q}(t,s) &=& a_R t^{(2-z-\eta)/z} (t/s)^\theta f_R(qt^{1/z})\; .
\label{scaltwo}
\eea
Comparing these forms with equation~(\ref{scalone}) one concludes that
for $y\rightarrow 0$
\bea
\tilde{\mathcal F}_C (x,y) & \sim & a_C y^{-\theta+1} f_C(x)\; ,\nonumber\\
\tilde{\mathcal F}_R (x,y) & \sim & a_R y^{-\theta} f_R(x)\; .
\label{scalthree}
\eea
It is possible to rewrite equation~(\ref{scalone}) in terms of scaling
functions $\tilde F_C(x,y)$ and  $\tilde F_R(x,y)$ with a regular
behaviour for $y\rightarrow 0$, i.e.,
\bea
C_{\bf q}(t,s) &=& (t-s)^{(2-\eta)/z}(t/s)^{\theta-1} 
{\tilde F}_C (q(t-s)^{1/z},s/t)\;, \label{scalfin1}\\
R_{\bf q}(t,s) &=& (t-s)^{(2-\eta-z)/z}(t/s)^{\theta} 
{\tilde F}_R(q(t-s)^{1/z},s/t)\; .
\label{scalfin2}
\eea
These results refer to systems in the infinite-volume limit. 
Finite-size effects on the non-equilibrium
dynamics have been studied in \cite{rd}.

For ${\bf q}=0$ the scaling forms (\ref{scalfin1}) and (\ref{scalfin2}) may be written as 
\bea
R_{{\bf q}=0}(t,s) &=& A_R\, (t-s)^a(t/s)^{\theta} F_R(s/t)\; ,\label{scalR}\\
C_{{\bf q}=0}(t,s) &=& A_C\,s(t-s)^a(t/s)^{\theta} F_C (s/t)\; ,
\label{scalC}
\eea
where $a = (2-\eta-z)/z$. We singled out explicitly 
the non-universal amplitudes $A_{R,C}$ by fixing
$F_{R,C}(0) = 1$. With this normalization, $F_{R,C}$ are universal
scaling functions. From the previous scaling forms one deduces that
\be
\dpar_s C_{{\bf q}=0}(t,s) = A_{\dpar C}\, 
(t-s)^a(t/s)^{\theta} F_{\dpar C} (s/t)\; ,
\label{scaldC}
\ee
where the non-universal amplitude $A_{\dpar C}$ has been defined in 
such a way that $F_{\dpar C} (0) = 1$. The relation between $F_{\dpar
C}(x)$ and $F_C(x)$ can be easily worked out. 
The amplitudes are related by $A_{\dpar C} = A_C (1-\theta)$.

Using equation~(\ref{scalR}) and (\ref{scaldC}) one finds that
\be
{\cal X}_{{\bf q}=0}(t,s)\equiv\frac{R_{{\bf q}=0}(t,s)}{\dpar_s C_{{\bf q}=0}(t,s)}=
\frac{A_R F_R(s/t)}{A_{\dpar C} F_{\dpar C}(s/t)}\,
\label{Xdissut}
\ee
is a universal scaling function, 
being the ratio of two quantities ($R_{{\bf q}=0}(t,s)$ and $\dpar_s
C_{{\bf q}=0}(t,s)$) that have the same scaling dimensions. 
Furthermore it is a function only of $s/t$ and not of $s$
and $t$ separately.

From equation (\ref{Xdissut}) and using the results of 
section~\ref{secFDRms}, one easily finds that
\be
X^\infty=
\lim_{s\rightarrow\infty}\lim_{t\rightarrow\infty} 
\frac{R_{{\bf q}=0}(t,s)}{\dpar_s C_{{\bf q}=0}(t,s)}=
\lim_{s/t\rightarrow0}{\cal X}_{{\bf q}=0}(s/t)=
\frac{A_R}{A_{\dpar C}}=\frac{A_R}{A_C(1-\theta)}\;,
\label{Xinfamplitratio}
\ee
is a universal, dimensionless amplitude-ratio in the sense of \cite{pha}.

Fourier transforming equations (\ref{scalfin1}) and (\ref{scalfin2}) we 
obtain the scaling forms of the autoresponse and autocorrelation functions 
\bea
R_{{\bf x}=0}(t,s) &=& {\cal A}_R\, (t-s)^{a-d/z}(t/s)^{\theta} {\cal F}_R(s/t)\; \label{RXX},\\
C_{{\bf x}=0}(t,s) &=& {\cal A}_C\,s(t-s)^{a-d/z}(t/s)^{\theta} {\cal F}_C (s/t)\; ,
\label{CXX}
\eea
with ${\cal A}_R$, ${\cal A}_C$, ${\cal F}_R(y)$ and  ${\cal F}_C(y)$
that are (in analogy with their counterpart in the real space) non-universal 
and universal, respectively.

Note that even $X_{{\bf x}=0}(t,s)$ is a function of $s/t$ only.
This is an important difference compared to mean-field glassy models, 
where
$X_{{\bf x}=0}$ can be written as a 
function of $C_{{\bf x}=0}(t,s)$ (see, e.g., \cite{cugl-02}).

\subsection{The effect of the initial condition and the irrelevance of $\tau_0^{-1}$.}
\label{sectau0}

In the previous section, by means of dimensional analysis, we concluded that
$\tau_0^{-1}$ is an irrelevant variable and so it can be neglected as long as 
one is interested in the leading long-time behaviour. 
However, this is not possible when considering 
quantities that vanish for $\tau_0^{-1}=0$.
This is actually the case of the correlation function when the smaller
time $s$ goes to zero, as it is clear from equations (\ref{scalC}) and (\ref{CXX}).
These forms have been obtained assuming large $s$, but 
it is simple to show that due to the Dirichlet boundary condition in 
time (implicit in the assumption $\tau_0^{-1}=0$) the correlation functions
with the insertion of one $\p({\bf x},0)$ vanish identically, 
see equation (\ref{relationSDE}). 
In these cases the contribution depending on $\tau_0^{-1}$ is the leading one 
and so the scaling forms can not be calculated setting $\tau_0^{-1}=0$. 
It was shown that the insertion of a boundary field is equivalent to 
the insertion of $\tau_0^{-1} \pht_0$ \cite{jan-92}:
\be
 \p({\bf x},0)=\tau_0^{-1}  \pht_0({\bf x})\,,
\ee
when inserted in a correlation function with bulk fields. Therefore,
using equations (\ref{genscaling}) and (\ref{genexp}),
the scaling of the autocorrelation function for $s=0$ and large $t$ 
turns out to be (neglecting higher-order corrections due to initial
correlations, i.e., to finite $\tau_0$):
\be
A(t)=C_{{\bf x}=0}(t,0)=
\tau_0^{-1} \langle \p ({\bf x},t)  \pht_0({\bf x})\rangle
\simeq \tau_0^{-1} t^{-\delta(1,0,1)/z}= \tau_0^{-1} t^{\theta'-d/z},
\label{at}
\ee
where we introduced the so-called magnetization initial-slip exponent, 
given by $\theta'=[d-\delta(1,0,1)]/z$ [see equation~(\ref{genexp})]. 
For Model A dynamics it can be written as  
\be
\theta'=\theta+a=\theta+(2-\eta-z)/z\,.
\label{scalthetap}
\ee
Equations (\ref{at}) and (\ref{CXX}) imply that for
$t\gg s$, $C_{{\bf x}=0}(t,0)\sim \tau_0^{-1}t^{-d/z+\theta'}$ whereas 
$C_{{\bf x}=0}(t,s)\sim s^{1-\theta}t^{-d/z+\theta'}$, 
at the leading order in $\tau_0^{-1}$. In both the cases the
contributions decrease as $t^{-d/z+\theta'}$,
allowing 
for the identification of the autocorrelation exponent, defined as \cite{h-89}
\be
C_{{\bf x}=0}(t,s)\simeq t^{-\lambda/z}\; \quad {\rm for} \quad t\gg s \,,
\label{autocorrexp}
\ee
so that
\be
\lambda= d-\theta' z= d-\theta z-2+\eta+z\,.
\label{scaltheta}
\ee 
Let us point out that the large-$t$ behaviour of 
$C_{{\bf x}=0}(t,0)$ and $C_{{\bf x}=0}(t,s)$ can, a priori, be
different, giving rise to a temporal cross-over for finite $s$,
as in the case of Model B dynamics considered in 
section \ref{modB}. 

For sake of completeness we comment that also autoresponse exponent is
usually defined in the literature \cite{ph-02}, through
\be
R_{{\bf x}=0}(t,s)\sim t^{-\lambda_R/z}\; \quad {\rm for}\quad t\gg s\,.
\label{autorespexp}
\ee
For Model A dynamics $\lambda_R$ equals $\lambda$. 
This is not the case in the more general case, see section \ref{modB}.

\subsection{Comments}

Some remarks are now in order concerning the particular choices of the noise 
and of the initial conditions.

If in equation (\ref{lang}) we fix $\zeta({\bf x},t)=0$, we obtain 
the deterministic evolution of an order parameter under a diffusion-like 
equation. 
In turn, the associated dynamical action has no quadratic term in $\pht$.
Physically this can be used to describe the evolution of 
a system quenched from a fully disordered initial state to a critical point,
at which the thermal noise is negligible. This should be the case of 
systems with $T_c=0$. However, it is not clear under which 
conditions the microscopic models defined via a master equation (and hence
stochastic) are in the same universality classes as the corresponding 
deterministic mesoscopic equations.
Examples of different resulting universality classes have been 
provided \cite{detlg}. 
The RG study and $\e$ expansion of this deterministic model 
has been worked out in \cite{cardy-92}. For our discussion, a relevant
finding is that response and correlation functions scale in
time with $\lambda\neq\lambda_R$, resulting in a trivial FDR. For this reason
we will not further consider the deterministic evolution.

We have also assumed the noise to be Gaussian and white. A non-constant
spectral density of the noise may also occur (in this case the noise is 
called ``coloured'').
To our knowledge, the non-equilibrium behaviour under the effects of a
coloured noise has been considered only for the random 
walk \cite{po-03}, corresponding to the Gaussian approximation of the 
${\bf q}=0$ mode in equation (\ref{lang}).

So far, we considered only the case of short-range correlations in the
initial state. Long-range ones (for instance decaying as 
$\varsigma_0 |{\bf x}|^{-\rho}$ for large ${\bf x}$) 
may change the dynamic non-equilibrium
universality class and in particular the initial-slip exponent
$\theta$ and $X^\infty$. 
Decreasing the exponent $\rho$ of the long-range
correlations, the canonical dimension of $\varsigma_0$, that is negative 
for large enough $\rho$, may become positive. 
In this case $\varsigma_0$ turns to a relevant parameter in the
RG sense and one expects that the stable fixed 
point and consequently the critical properties get modified. 
Exact calculations for the spherical model \cite{ph-02} and for the 
one-dimensional Ising model \cite{hs-04} substantiate this picture.

Finally a scaling correlation function of the form of equation
(\ref{CXX}) was also found in models of self-organized 
criticality \cite{soc}.

\subsection{Scaling of high-order observables}
\label{seccomp}

In this section we briefly discuss the scaling behaviour of 
observables that are powers of the order parameter at the same 
space-time point (composite operators in FT language). 
More details can be found in reference \cite{cg-04}. 

Let us consider an observable $\o$ with
$h_\o$ as a conjugate field (e.g., $\o$ is the energy density and $h_\o$ the 
temperature), that couples to ${\mathcal H}$ according to
${\mathcal H} \mapsto {\mathcal H} - \beta\,h_\o \o$.
As a consequence, the dynamical functional $S$ changes according to 
$S \mapsto S -\beta\, h_\o \ot$, 
where the associated operator $\ot$ is given by
\be
\ot=\int \rmd t \rmd^d x \; \pht({\bf x},t) {\mathcal D} \frac{\delta\o}{\delta\p({\bf x},t)} \;.
\label{defot}
\ee
The linear response of an observable ${\mathcal A}$ to a variation in the
field $h_\o$ conjugated to $\o$ can be expressed as
\be
\left .\frac{\delta \langle
{\mathcal A} \rangle_{h_\o}}{\delta h_\o}\right|_{h_\o=0} 
= \beta\, \langle{\mathcal A} \ot \rangle\,,
\label{defrespO}
\ee
where $\langle \cdot \rangle_{h_\o}$ stands for the average over the
dynamics associated with the dynamical functional in the presence of $h_\o$.

We are interested in the scaling properties of the correlation functions
of the form $\langle\o \ot\rangle$ and $\langle\o \o\rangle$. 
Using RG equations as in the previous sections one obtains \cite{cg-04}
\bea
\langle \o({\bf q},t)\o({\bf - q},s)\rangle & =& 
(t-s)^{a_\o+1}\hat{F}_C(q^z(t-s),s/t)\,, \nonumber \\
\langle \o({\bf q},t)\ot({\bf - q},s)\rangle & =&
(t-s)^{a_\o}\hat{F}_R(q^z(t-s),s/t)\,,
\label{scalformR}
\eea
where $a_\o=(2-\eta_\o-z)/z$ and $\eta_\o$ is the scaling dimension of 
the observable $\o$.

In the presence of a time surface the functions $\hat{F}_C(q^z(t-s),s/t)$
and $\hat{F}_R(q^z(t-s),s/t)$ are not regular in the limit $s/t\rightarrow0$, 
as in the case of $\o=\p$ of the previous section.
To extract the leading singularity in the ageing regime a proper 
short-distance expansion should be considered. 
This can be easily worked out for $\o^{(m)}(t)= \p^m(t)$ (hence 
$\ot^{(m)}(t) \sim \pht(t)\p^{m-1}(t)$). 
In this case the analysis is straightforward and the final scaling forms 
are \cite{cg-04}
\bea
\fl\langle \o^{(m)}({\bf q},t)\o^{(m)}({\bf - q},s)\rangle & =& s
(t-s)^{a_\o}(t/s)^{-(m-1)+m\theta}  F^\o_C(q^z(t-s),s/t)
\nonumber\\ 
\fl\langle \o^{(m)}({\bf q},t)\ot^{(m)}({\bf - q},s)\rangle & =&
(t-s)^{a_\o}(t/s)^{-(m-1)+m\theta}  F^\o_R(q^z(t-s),s/t)
\label{scalformRfin}
\eea
where the functions $F^\o_C$ and $F^\o_R$ are regular for
$s/t\rightarrow 0$. Furthermore they are also universal once we fix the 
normalization for small arguments.

Let us focus on the scaling properties of the correlation and response
functions of the two zero-momentum observables with $m=2$:
\bea
E(t) &\equiv& \sum_{i=1}^N \int \!\rmd^d x\; \p_i^2({\bf x},t) = 
\sum_{i=1}^N \int \!
(\rmd q)\; \p_i({\bf q},t)\p_i({\bf -q},t)\,, \label{defE}\\
T_{i\neq j}(t) &\equiv& \int \!\rmd^d x\; \p_i({\bf x},t)\p_j({\bf x},t)= 
\int \! (\rmd q)\; \p_i({\bf q},t)\p_j({\bf -q},t) \;.
\label{defT}
\eea
$E$ and $T_{i\neq j}$ are usually referred to as the ``energy'' and 
the quadratic tensor operator, respectively \cite{cg-04}.
The corresponding response operators are 
\bea
\widetilde E(t) &=& \sum_{i=1}^N \int \!
(\rmd q) \; 2 \pht_i({\bf q},t)\p_i({\bf -q},t) \nonumber\\
\widetilde T_{i\neq j}(t) &=& 
\int \! (\rmd q) \; [\pht_i({\bf q},t)\p_j({\bf -q},t) +
\p_i({\bf q},t)\pht_j({\bf -q},t)] \;.
\eea
In the following we will generically refer to them as $\o(t)$.
In this case, being the scaling dimension of $\o(t)$ well-known,
the exponent $a_\o$ can be written 
in terms of the usual critical exponents \cite{cg-04}.
The corresponding scaling forms for the correlation and response 
functions (the latter defined as in equation~(\ref{defrespO}) with ${\mathcal
A}={\mathcal O}$) are 
\bea
C^\o(t,s) &=& \langle \o(t)\o(s)\rangle = A^\o_C\,s
(t-s)^{a_\o}(t/s)^{-1+2\theta} F^\o_C(s/t)\;,\label{scalOfinC}\\
R^\o(t,s) &=& \langle \o(t)\ot(s)\rangle =
A^\o_R \,(t-s)^{a_\o}(t/s)^{-1+2\theta} F^\o_R(s/t)\;,
\label{scalOfin}
\eea
where the non-universal amplitudes $A^\o_{C,R}$ are chosen in such a way 
that $F^\o_{C,R}(0) =1$. (As previously done, we have absorbed the
factor $\beta^{-1}$ in the definition of $R^\o(t,s)$.)
In terms of these quantities we can write the (zero-momentum) FDR as
\be
{\mathcal X}_\o(t,s) \equiv \frac{R^\o(t,s)}{\dpar_s C^\o(t,s)}
\ee
that, as its analogous (\ref{Xdissut}), depends only on the ratio $s/t$,
and it is a universal function. 
In particular its long-time limit
\be
{\mathcal X}^\infty_\o = 
\lim_{s\rightarrow \infty}\lim_{t\rightarrow \infty} {\mathcal X}_\o(t,s)=
\frac{1}{2}\frac{ A^\o_R}{(1-\theta)A^\o_C}  \;,
\ee
is universal (analogously to equation~(\ref{Xinfamplitratio})).
These considerations hold in general for all the FDRs that can be obtained 
from equations (\ref{scalformRfin}).

\section{Purely dissipative dynamics of an O(N) model (Model A)}
\label{modA}

One of the simplest non-trivial models displaying ageing after a quench to
the critical point is a lattice spin model in $d$ dimensions 
with $O(N)$ symmetry,
evolving according to a purely dissipative dynamics (no conservation laws). 
In the simplest instance its 
Hamiltonian is given on the lattice by 
\be
{\cal H}= - \sum_{\langle {\bf ij}\rangle} {\bf s_i}\cdot{\bf s_j} \;,
\label{HON}
\ee
where ${\bf s_i}$ is a $N$-component spin located at the lattice site
${\bf i}$, with ${\bf s_i}^2 = 1$. Here and in the following the symbol 
$\langle{\bf ij}\rangle$ means that the sum runs on all
nearest-neighbour pairs  of lattice sites. 
When $N=1$ the Hamiltonian (\ref{HON}) describes the Ising model,
whereas for $N=2$ the XY or planar model and for $N=3$ the 
(isotropic) Heisenberg model.
A purely dissipative dynamics for the lattice model (\ref{HON}) proceeds by
elementary moves that amount to random changes in the direction of the
spin ${\bf s}_{\bf i}$ (spin-flip sampling). 
The transition rates can be arbitrarily chosen provided that 
the detailed-balance condition is satisfied. 
For analytical studies the most suited is the
Glauber dynamics \cite{glau}, which allows some exact solutions. 
Given its relative simplicity, this lattice model 
is the most studied and best understood. 
In the following we review analytical and numerical results that have to be 
compared with the field-theoretical (FT) ones presented in the next sections.
In fact the coarse-grained continuum dynamics described by the Langevin
equations (\ref{lang}) is expected to be in the same universality class
as lattice models with $O(N)$ symmetry, short-range interactions and 
spin-flip dynamics \cite{HH}.
All the reviewed determinations of the FDRs are summarized in 
table \ref{tab} (Magnetization FDR) and table \ref{tab2} 
(other observable FDRs) together with the FT estimates.

{\it One-dimensional Ising model}: 
In this case there is no finite temperature phase transitions, since $d=1$
is the lower critical dimensionality of the Ising model, therefore $T_c=0$. 
Analytical solutions for the critical response and correlation 
functions have been provided \cite{lz-00,gl-00i,fins-01,ms-04,mbgs-03}.
From them the critical exponents read $z=2$ and $\theta=1/2$
(as first shown in reference \cite{b-89}). 
$X^\infty=1/2$ has also been found \cite{lz-00,gl-00i} 
(actually this result was indirectly obtained several
years before in \cite{kh-89}). 
Correlation and response functions of more complex observables have
been studied \cite{sfm-02,mbgs-03,ms-04} and 
in particular the FDR for the energy  has been found to be
$X^\infty_E = 0$ at $T=0$. 
The fact that $X^\infty\neq X^\infty_E$ is interpreted
in \cite{mbgs-03} as an interplay between criticality and coarsening,
a peculiarity of those models with $T_c=0$. For many other observables, 
instead, the equality $X^\infty_\o=X^\infty$
between the critical FDRs holds \cite{sfm-02,mbgs-03,ms-04}. 
In this context it must be recalled (as stated in previous sections) that 
the correspondence between the Glauber-Ising model and the Landau-Ginzburg 
dynamics is not obvious and they probably belong to two different 
universality classes (in fact they display different non-equilibrium exponent 
$\theta$ \cite{detlg}).

{\it d-dimensional spherical and $O(\infty)$ models}:
The static critical behaviour of the $O(N)$ model in the limit 
$N\rightarrow\infty$ is equivalent to that of the spherical
model, defined by the Hamiltonian
\be
{\cal H}=\frac{1}{2}\sum_{\langle {\bf ij}\rangle}(s_{\bf i}-s_{\bf j})^2\,,
\ee
where $s_{\bf i}$ are real numbers 
subjected to the constraint $\sum_{\bf i}s_{\bf i}^2=L^d$ ($L$ being the 
linear dimension of the $d$-dimensional lattice, assumed for simplicity to 
be hypercubic).
Indeed Stanley proved \cite{stanley-68} that the free energies of the 
two models are exactly the same. From this equality it 
follows that critical exponents, universal scaling functions etc. are equal.
The same equivalence
also holds for equilibrium critical dynamics defined in the spherical model  
by means of a Langevin equation properly 
modified in order to prescribe a dynamics that is
compatible with the spherical constraint.
As far as we are aware, a proof of this equivalence has not yet been carried 
over to the 
non-equilibrium critical dynamics we are interested in.
Let us recall that the free energies of the 
models in the presence of a spatial boundary (whose corresponding field theory
looks very similar to the non-equilibrium one that we are considering) 
are not equal \cite{sfbound}.

The spherical model has attracted a lot of attention, since its essentially 
Gaussian Hamiltonian makes it exactly solvable, even though the resulting 
critical behaviour is not mean-field-like in $2<d<4$, because of the 
spherical constraint. 
Even in the $O(\infty)$ model several quantities are exactly calculable
because the fourth order interaction term in the Hamiltonian (\ref{lgwA}) 
can be self-consistently decoupled according to
$g/N (\p^2)^2\rightarrow g C_{{\bf x}=0}(t) \p^2$ and the resulting
Langevin equation (\ref{lang}) is linear and thus solvable \cite{jss-89}.
The dynamical critical exponents for both $O(\infty)$ \cite{jss-89} 
and spherical models \cite{gl-00c} are  $z=2$ and $\theta=1-d/4$, for $2<d<4$.
The spherical FDR of the magnetization is $X_M^\infty=1-2/d$ \cite{gl-00c}. 
This result has been found in agreement with the direct calculation for the 
$O(\infty)$ model \cite{cg-04}.
Recently Sollich \cite{s04} has determined several FDRs of 
quadratic operators (in the spin $s_{\bf i}$) that could be 
compared with FT results. He considered the bond energy observable 
$B_{\bf i}=\frac{1}{2}(s_{\bf i}-s_{\bf j})^2$, the product observable 
$P_{\bf i}=s_{\bf i} s_{\bf j}$ (with ${\bf i},{\bf j}$ 
nearest neighbours)  both in the real and in the momentum space, 
and the total energy $E$. The {\it exact} results of his analysis are that
$X^\infty_P=X^\infty_B=X^\infty_M\neq X^\infty_E$. 
The scaling forms for these observables have been also derived and they agree
with FT expectations equations (\ref{scalOfinC}) and (\ref{scalOfin}) with 
$a_B=-1-d/2$, $a_P=1-d/2$ and $a_E=d/2-3$ in $d<4$ (and 
$a_E=a_P=a_B+2=1-d/2$ in $d>4$). 
The calculation of response and correlation functions of high-order
observables is rather cumbersome and it has not yet been carried out
completely (see for details \cite{cg-04}).

{\it Two-dimensional Ising model}:
The physically relevant cases with finite $N$ and $d=2,3$ are not 
analytically solvable and have been intensively investigated by
means of Monte Carlo (MC) simulations. 
For the two-dimensional Ising model the dynamical critical exponents are 
known with very high precision from lattice techniques (we recall that 
the static model is solvable, see, e.g., \cite{baxter}).
The most accurate determination of the critical exponent $z$ is 
$z=2.1667(5)$ \cite{nb-00} (this number is obtained by means of equilibrium
MC, since the exponent $z$ is the same in- and out-of-equilibrium).
This value is in agreement with other less precise 
simulations \cite{2dimz,g-95,2dimt,ws-95},
but it is somehow bigger than the most precise FT estimates, i.e., the 
three-loop $\e$ expansion \cite{av-84}, four-loop fixed dimension
\cite{o-95b,pif-97} and the interpolation of two-loop result between one 
and four dimensions \cite{bdjz-81}.
The most accurate value of the non-equilibrium critical exponent $\theta'$ has been 
obtained in reference \cite{g-95}, i.e., $\theta'=0.191(3)$
that, using the scaling relation (\ref{scalthetap}) leads to 
$\theta=0.383(3)$, compatible with other MC results \cite{h-89,2dimt,ws-95} 
and again slightly bigger than two-loop FT estimate \cite{jss-89}.

Concerning the FDR, earlier investigations indicate 
$X^\infty=0.26(1)$ \cite{gl-00c}. Recent simulations provide results of
increasing accuracy thanks to the improved methods that 
have been introduced to measures the quantities of interest.
From the analysis of zero-momentum observables 
the very accurate estimate $X^\infty=0.340(5)$ has been 
obtained \cite{mbgs-03} (the method employed in \cite{mbgs-03} has been 
questioned, see for details \cite{p-03,mbgs-03r}). 
Recently a new algorithm has been proposed by Chatelain~\cite{c-03} 
and subsequently revised in \cite{rt-03}. 
In standard simulations some integrated form of the response
function is measured after the application of a magnetic field.  
The new algorithm instead allows one to measure the integrated linear 
response function without applying any magnetic field to the system, 
avoiding in such a way an eventual crossover towards the non-linear regime. 
By applying this algorithm, the result $X^\infty=0.33(2)$ has been 
found \cite{c-03}. The study of a more general model that belongs (apart one 
exceptional point in its phase diagram) to the Glauber-Ising universality 
class, even if it 
does not satisfy detailed balance, gives $X^\infty=0.33(1)$ \cite{sdc-03}. 
A few attempts have been made to measure the critical FDR for the energy 
$X^\infty_E$. 
In \cite{sdc-03} the standard algorithm produced a signal 
too noisy to have any reliable result. By considering, instead,
coherent observables $X^\infty_E=0.33(2)$ has been obtained \cite{mbgs-03}. 
This result led the authors of \cite{mbgs-03} to put forward the
conjecture that $X^\infty$ is the same for all the observables,
according to the idea of the existence of a  unique effective
temperature, as it is the case for mean-field glassy models \cite{ckp-97}. 
We comment that along the same lines of references \cite{c-03,rt-03}, an 
alternative algorithm measuring directly the response function has been 
proposed \cite{lcz-04}, but it has only been tested in one-dimensional 
systems.

{\it Three-dimensional Ising universality class}:
In this case the numerical simulations are obviously less accurate than 
in two dimensions. 
The values of the static critical exponents are reviewed in reference
\cite{PV-r}
and will be not reported here, neither will for the $O(N)$ models with $N>1$.
The best determinations of the equilibrium dynamic critical exponent are
$z=2.032(4)$ \cite{g-95} and $z=2.055(10)$ \cite{ihoo-00}. 
They are not in agreement within three standard deviations, but they are 
anyway almost compatible with the result of less accurate simulations \cite{3do,ws-95}. 
Again the numerical values are a bit higher than the best FT estimates giving 
$z\simeq 2.02$ \cite{av-84,o-95b,pif-97,bdjz-81}.
For the initial slip exponent the best value is $\theta'=0.104(3)$ 
\cite{g-95} in agreement with less precise estimates \cite{h-89,3do} and 
with the two-loop $\e$ expansion \cite{jss-89,jan-92}.
Through the scaling relation (\ref{scalthetap}) and 
using the safe value $z=2.04(2)$, this value of $\theta'$ leads to 
$\theta=0.14(1)$.
These values confirm that  the $\e$ expansion converges better in 
$d=3$ than in $d=2$.

Regarding the FDR, only the preliminary analysis of reference \cite{gl-00c} 
is currently available, giving $X^\infty\sim0.40$.

{\it Two-dimensional XY universality class}: 
The non-equilibrium dynamics of this universality class has been 
studied in \cite{ckp-94,ktvari,ooi-03,bbj-00,bhs-01,ak-04a}. 
The model is quite peculiar since, despite the absence of a spontaneous
magnetization, it is critical for all temperatures below the so-called
Kosterlitz-Thouless temperature $T_{KT}$ 
(i.e., in the RG language, it has a line of fixed points)
with a continuously varying critical exponent $\eta(T)$ that
ranges from $0$ to $1/4$ as $T$ changes from $0$ to $T_{KT}$ \cite{kt}.
Regarding the exponent $z$, all the approaches give $z=2$, but with
logarithmic corrections to the scaling depending upon the presence 
of vortices in the initial state \cite{bbj-00}. 
For a quench from a high-temperature state to one of the critical points, 
numerical simulations have been so far inconclusive in determining 
non-equilibrium exponents.
Indeed in \cite{ooi-03} the value of the critical exponent $\theta'$
was found to vary continuously along the line of critical points,
whereas in \cite{ak-04a} it turned out to be temperature independent.
A precise determination of this exponents surely deserves further
analytical and numerical study.
Even the determination of $X^\infty$ is problematic: In fact a standard 
FD plot provides the approximate relation
$X^\infty=1/2 (T/T_{KT})$ for $T\leq T_{KT}$ \cite{ak-04a},
whereas  a direct evaluation gives $X^\infty\sim0$, independently 
of $T$ \cite{ak-04a}.
Given this argument has not yet been settled we do 
not report these results in table \ref{tab}. Let us mention that the FDR has
been studied also for the quench between two (critical) points below $T_{KT}$
\cite{bhs-01,ph-04,ak-04a}.

It is worth noting that Schehr and Le Doussal \cite{sl-03,sl-04} obtained a
temperature-dependent FDR (similar to that just mentioned)
$X^\infty=(2(1+e^{\gamma_E} \tau) +O(\tau^2))^{-1}$, where
$\tau = (T_g-T)/T_g$ (being $T_g$ the temperature of the glass
transition in the model) in the glassy 
phase of the two-dimensional Cardy-Ostlund model \cite{co-82}, 
that has a line of fixed points as well.

{\it Three-dimensional $O(N)$ models with $N>1$}:
These models have attracted a modest interest, mainly because
the most relevant experimental realizations of $O(N)$ models belong to
different dynamic universality classes as discussed in section \ref{duc}.

To our knowledge the three-dimensional XY model with purely dissipative 
dynamics has been numerically studied only in \cite{ak-04b}, using the 
approach discussed in \cite{c-03,rt-03}. 
$\lambda/z\simeq 1.34$ was obtained, without providing 
independent determinations of the two exponents.
A direct determination of the FDR $X(t,s)$ in the long-time limit turns out 
to be unreliable, since the numerical data are still affected by large 
statistical fluctuations. On the other hand, a linear extrapolation to the 
small-$s/t$ region provides the estimate $X^\infty=0.43(4)$.

Concerning the $O(3)$ universality class, the determination of $z$ 
has been attempted only in \cite{o3}, suggesting values slightly smaller
than $2$, i.e., $z=1.96(6)$ \cite{o3}. 
On the other hand, FT studies unambiguously give $z>2$ \cite{av-84,z1n,2pe}. 
In fact, we point out that the early three-loop FT computation of 
$z$ \cite{dbz-75} is affected by a numerical error, later corrected 
in \cite{av-84}, that led to the erroneous conclusion $z<2$ in three 
dimensions. 
The correct value is $z = 2 + R \eta$, with a value of $R$ that is
independent of $N$ up to three loops and reads 
$R = 0.726 [1-0.1885 \e+O(\e^2)]$ \cite{av-84}.
A more accurate numerical determination of $z$ surely deserves
further analysis.

{\it Check of the universality of $X^\infty$}:
The issue of universality of $X^\infty$ has been addressed 
in \cite{c-04} by using the algorithm of \cite{c-03,rt-03} for a
variety of models, some of which are not related to the $O(N)$ models considered here.
The long time non-equilibrium dynamics of the Ising, $3$-state 
clock \cite{c-04} and $4$-state Potts \cite{c-04,baxter} models 
in two dimensions has been studied on square [sq], triangular [tr] and
honeycomb [hc] lattices. 
The values found for $X^\infty$ are compatible with
each other and with those already reported in the literature.
For the Ising model it has been found $X^\infty[{\rm sq}]=0.330(5)$,
$X^\infty[{\rm tr}]=0.326(7)$ and $X^\infty[{\rm hc}]=0.330(8)$ on square, 
triangular and honeycomb lattice,
respectively. For the $3$-state clock model, instead, 
$X^\infty[{\rm sq}]=0.406(6)$,
$X^\infty[{\rm tr}]=0.403(5)$ and $X^\infty[{\rm hc}]=0.401(6)$, 
whereas for the 
$4$-state Potts model $X^\infty[{\rm sq}]=0.475(10)$, 
$X^\infty[{\rm tr}]=0.470(12)$ and
$X^\infty[{\rm hc}]=0.467(7)$.
These findings support the universality of $X^\infty$. 
Moreover, Chatelain measured $X^\infty$ for other models on the square 
lattice, which at least in equilibrium are expected  to belong  to the same 
universality class as the previously mentioned models.
Namely, he investigated the Ashkin-Teller model \cite{c-04,baxter} 
at the point of its phase diagram where it
belongs to the same universality class as the $3$-state clock model,
finding $X^\infty=0.405(10)$, in perfect agreement with the previous results. 
Then he considered the Multispin Ising model \cite{c-04} and the 
Baxter-Wu model  \cite{c-04,baxter} (defined on the triangular
lattice and belonging to the same
universality class as the $4$-state Potts model) finding
$X^\infty=0.467(10)$ and $X^\infty=0.550(15)$, respectively (note that, to 
our knowledge, the latter value is the first evidence for $X^\infty>1/2$ in 
ferromagnetic models quenched at their critical points from an initially 
uncorrelated state).
Whereas the former value is in perfect agreement with 
the result for the $4$-state Potts model, 
the latter is definitely incompatible with it. 
This can be a signal of a non-universality of $X^\infty$ (according to
\cite{c-04}, even the dynamical exponent $\theta$ seems to depend on
the specific lattice model, within the same dynamic universality class), or of
the existence of a different (non-equilibrium) universality class to which 
the the Baxter-Wu model belongs, or simply of large
cross-over effects that make particularly difficult to reach 
the asymptotic scaling behaviour. 
It is still unclear whether these possibilities can explain
the (somehow) surprising results of
\cite{c-04}. Further investigations and cross-checks are surely
required.

\fulltable{\label{tab} 
Estimates of $X^\infty$ in purely dissipative models quenched from
a completely disordered initial state.
[tr] and [hc] stand for simulations on the triangular and honeycomb
lattices, respectively. When not indicated, the simulation is on the square 
lattice. Models belonging to the same 
universality class are marked with $^\dagger$ and $^\ddagger$.}
\begin{tabular}{llll}
\br
Model        & Reference\phantom{gggg} & Method\phantom{gggggggggg}& $X^\infty$\\
\mr
Random Walk                 &\cite{ckp-94} & Exact   &   $1/2$\\
Gaussian model              &\cite{ckp-94} & Exact   &   $1/2$\\
$d$-dim Spherical Model     &\cite{gl-00c} & Exact   &   $1-2/d$\\
$1$-dim Ising  Model        &\cite{gl-00i,lz-00}  & Exact   &   $1/2$\\
$2$-dim Ising  Model        &\cite{gl-00c} & MC      &   $0.26(1)$\\
                            &\cite{mbgs-03}& MC      &   $0.340(5)$\\
                            &\cite{c-03}   & MC      &   $0.33(2)$\\
                            &\cite{sdc-03} & MC      &   $0.33(1)$\\
                            &\cite{c-04}   & MC      &   $0.330(5)$\\
                            &\cite{c-04}   & MC[tr]  &   $0.326(7)$\\
                            &\cite{c-04}   & MC[hc]  &   $0.330(8)$\\
                            &\cite{cg-02a2}& FT $O(\e^2)$& $0.30(5)$\\
$2$-dim $3$-state clock Model$^\dagger$&\cite{c-04}  & MC  &  $0.406(6)$\\
                            &\cite{c-04}   & MC[tr]    & $0.403(5)$\\
                            &\cite{c-04}   & MC[hc]    & $0.401(6)$\\
$2$-dim Ashkin-Teller Model$^\dagger$ 	    &\cite{c-04}   & MC    &  $0.405(10)$\\
$2$-dim $4$-state Potts Model$^\ddagger$\phantom{gggg}&\cite{c-04}  & MC &   $0.475(10)$\\
                            &\cite{c-04}   & MC[tr]    & $0.470(12)$\\
                            &\cite{c-04}   & MC[hc]    & $0.467(7)$\\
$2$-dim Baxter-Wu Model$^\ddagger$ 	    &\cite{c-04} & MC[tr]   & $0.550(15)$\\
$2$-dim Multispin Ising Model$^\ddagger$\phantom{gggg}  &\cite{c-04}   & MC   & $0.467(10)$\\
$3$-dim Ising  Model        &\cite{gl-00c} & MC        & $\simeq 0.40$\\
                            &\cite{cg-02a2}& FT $O(\e^2)$& $0.429(6)$\\
$3$-dim $O(2)$ Model        &\cite{cg-02a2}& FT $O(\e^2)$& $0.416(8)$\\
		 	    &\cite{ak-04b} & MC & $0.43(4)$\\
$3$-dim $O(3)$ Model        &\cite{cg-02a2}& FT $O(\e^2)$& $0.405(10)$\\
$3$-dim Random Ising Model  &\cite{cg-02rim}& FT $O(\sqrt\e)$& $\simeq
0.416$\\
\br
\end{tabular}
\endfulltable

\fulltable{\label{tab2} 
Estimates of $X^\infty_\o$  for observables different from the magnetization.
The meaning of the symbols [u], [l] and [q] in the FT estimates is
explained in section \ref{secop}}.
\begin{tabular}{lllll}
\br
Model        & Reference        & Observable & Method    & $X^\infty$\\
\mr
Gaussian model              &\cite{cg-04}  & All & Exact     & $1/2$\\
$d$-dim Spherical Model \phantom{g}    &\cite{s04}    & E & Exact & $\neq 1-2/d$\\
                            &\cite{s04}    & B,P & Exact & $ 1-2/d$\\
$1$-dim Ising  Model        &\cite{mbgs-03}& E & Exact     & $0$\\
                            &\cite{mbgs-03}& Other & Exact     & $1/2$\\
$2$-dim Ising  Model        &\cite{mbgs-03}& E & MC        & $0.33(2)$\\
$2$-dim Ising  Model        &\cite{cg-04}  & E & FT & $0.33$[u], $0.20$[l]\\
$3$-dim Ising  Model        &\cite{cg-04}  & E & FT & $0.39$[u], $0.37$[l]\\
$3$-dim $O(2)$  Model       &\cite{cg-04}  & E & FT & $0.37$[u], $0.31$[l], 
$0.22$[q] \\
                            &\cite{cg-04}  & T & FT & $\simeq $0.41[u]\\
$3$-dim $O(3)$  Model       &\cite{cg-04}  & E & FT & $0.35$[u], $0.30$[l], 
$0.21$[q]\\
                            &\cite{cg-04}  & T & FT & $\simeq $0.41[u]\\
\br
\end{tabular}
\endfulltable

\subsection{Field-Theoretical Approach}

The coarse-graining of the lattice models specified by the Hamiltonian
(\ref{HON}) with a Glauber-like dynamics 
leads to the purely dissipative dynamics of a 
$N$-component field $\p({\bf x},t)$~(Model A of \cite{HH}), described by the 
stochastic Langevin equation (\ref{lang}),
where $\cal{H}[\p]$ is the $\p^4$ Landau-Ginzburg Hamiltonian
given in equation (\ref{lgwA}).
The equilibrium correlation functions, generated by the Langevin 
equation~(\ref{lang}) and averaged over the noise $\zeta$, 
can be obtained, as summarized in section~\ref{sec:PathInt}, 
by means of the field-theoretical action (\ref{mrsh}).
The problem is completely specified once equation (\ref{initialH}) is
assumed to provide the statistical distribution of the
initial condition (with $\tau_0^{-1}=0$ given we are interested only
in the leading scaling behaviour, and $a({\bf x}) = 0$).
The propagators~(i.e., the 
Gaussian two-point correlation and response functions) are
obtained in a standard way, by considering the quadratic part of the action:
\bea
\fl R^G_{\bf q}(t,s)=&\Omega
\langle \tilde{\p_i}({\bf q},s) \p_j(-{\bf q},t) \rangle_G =& \Omega
\delta_{ij} \,\theta(t-s) e^{-\Omega ({\bf q}^2+r_0) (t-s)},\label{Rgaux}\\
\fl C^G_{\bf q}(t,s)=&\langle \p_i({\bf q},s) \p_j(-{\bf q},t) \rangle_G =&
 {\delta_{ij} \over {\bf q}^2+r_0}\left[ e^{-\Omega ({\bf q}^2+r_0)|t-s|}
-e^{-\Omega ({\bf q}^2+r_0)(t+s)}\right], \label{Cgaux}
\eea
where the subscript ``$G$'' is used to remind that we are considering 
the Gaussian approximation.
The response function (\ref{Rgaux}) is the same as in 
equilibrium. Equation (\ref{Cgaux}) reduces, instead, to the equilibrium
form when ${\bf q}\neq 0$ and both times $t$ and $s$ go to infinity while
$t-s$ is kept fixed. This ensures that quasi-equilibrium (in the sense 
discussed in section \ref{quasi}) holds for the order parameter, at least within 
the Gaussian approximation.

\subsection{Gaussian Fluctuation-Dissipation Ratios}
\label{secgaux}

In the Gaussian model (equation (\ref{lgwA}) with $g_0 = 0$)
the response and correlations functions are exactly known and
therefore the FDR can be easily evaluated.
From equations~(\ref{Rgaux}), (\ref{Cgaux}) and the definition (\ref{Xq})
one finds
\be
{\cal X}_{\bf q}(t,s)= \frac{ R_{\bf q}}{\dpar_s C_{\bf q}}=
\frac{1}{1+e^{-2 \Omega ({\bf q}^2+r_0) s}}\,,
\ee
where we set $s<t$ and removed the subscript ``$G$'' since in this section
we are only concerned with the Gaussian approximation.
If the theory is non-critical ($r_0 > 0$), the limit of the FDR for 
$s\rightarrow \infty$ is $1$ for all values of ${\bf q}$, in 
agreement with the idea that in the high-temperature phase all the 
fluctuating modes have a finite equilibration time. 
Equilibrium is hence recovered and the FDT applies.
In the critical theory $r_0=0$, if ${\bf q}\neq 0$,
the limit ratio is again equal to $1$, whereas for ${\bf q}=0$ one has
${\cal X}_{{\bf q}=0}(t,s)=\frac{1}{2}$, independently of $s$ and $t$. 
This shows that the only mode that ``does not relax'' to the 
equilibrium is the zero mode in the critical limit. 
This picture has been confirmed by a one-loop computation \cite{cg-02a1}.

In order to understand why measuring observables in momentum space 
is more effective than doing the same in real space
\cite{mbgs-03},
it is instructive to look at the FDR in real space.
Computing the Fourier transform of equations (\ref{Rgaux}) and (\ref{Cgaux})
one finds (with $\Omega=1$)
\bea
\fl R_{\bf x}(t,s)&=&\int (\rmd q) e^{i {\bf q x}}
e^{-({\bf q}^2+r_0)(t-s)}=
\frac{1}{(4\pi)^{d/2}} (t-s)^{-d/2}e^{-r_0(t-s)-{\bf x}^2/[4(t-s)]}\,,
\label{Rx}\\
\fl C_{\bf x}(t,s)&=&\int (\rmd q) e^{i {\bf q x}}
\frac{e^{-({\bf q}^2+r_0)(t-s)}-e^{-({\bf q}^2+r_0)(t+s)}}{{\bf q}^2+r_0} \,.
\label{Cx}
\eea
The derivative of the $C_{\bf x}(t,s)$ with respect to $s$ is
\be
\fl \dpar_s C_{\bf x}(t,s)=
\frac{1}{(4\pi)^{d/2}}\left[(t-s)^{-d/2}e^{-r_0(t-s)-\frac{{\bf x}^2}{[4(t-s)]}}+
(t+s)^{-d/2}e^{-r_0(t+s)-\frac{{\bf x}^2}{[4(t+s)]}}\right].
\ee
Therefore the FDR is
\be
X_{\bf x}^{-1}(t,s)=1+\left(\frac{t-s}{t+s}\right)^{d/2} 
e^{-2r_0 s+\frac{{\bf x}^2 s}{[2(t^2-s^2)]}}\,.
\ee
Again, for fixed ${\bf x}$ and $r_0\neq 0$, in the ageing regime 
($t, s\rightarrow\infty$, fixed $s/t$) 
this quantity reaches the asymptotic value $X_{\bf x}^\infty=1$, and 
the system equilibrates.
Instead for $r_0=0$ and $t, s\rightarrow\infty$ with fixed $s/t$ 
it reduces to 
\be
X_{\bf x}^{-1}(t,s)=1+\left(\frac{t-s}{t+s}\right)^{d/2}\,,
\label{XX}
\ee
explicitly showing the crossover from the equilibrium value $1$ 
at $s/t=1$ to the pure ageing behaviour $X^\infty=1/2$ for $t\gg s$ (apart
from $d=0$ where the model is a random walk).

The above considerations illustrate that $X_{\bf x}^\infty$ displays quite 
a long 
transient, lasting for a time 
of the order of the age of the system, before reaching its
asymptotic value.
This transient is absent  in ${\cal X}_{{\bf q}=0}$, resulting in a more 
effective
determination of $X^\infty$. As we shall see, the 
corrections to ${\cal X}_{\bf q}(t,s)/X^\infty$ due to the interaction
term $g_0$ 
are very small in magnitude (at least for Model A dynamics of $O(N)$ models),
making this remark still valid.

For the Gaussian theory the FDR can easily be computed for a generic
observable. In particular it was shown \cite{cg-04}
that the FDR for a set of local one-point observables  
is always equal to $1/2$ in the long-time limit.
This argument is very simple and we briefly recall it here: 
Let us consider the operators of the form $\o_{i,n}=\dpar^i\p^n$. 
The critical response and correlation functions of the order 
parameter in real space are given by equations~(\ref{Rx}) 
and (\ref{Cx}).
The two-point correlation function of $\o_{0,n}$ is given by the
$n-1$-loop diagram with the two points connected by $n$ correlation 
lines (see \cite{cg-04} for details).
In real space its expression is simply given by the product of the $n$ 
correlators:
\be
C^\o_{\bf x}(t,s)=c_n [C_{\bf x}(t,s)]^n\,,
\ee
($c_n$ is the combinatorical factor associated with the diagram) 
whose derivative is
\be
\dpar_s C^\o_{\bf x}(t,s)=c_n n 
[C_{\bf x}(t,s)]^{n-1} \dpar_s C_{\bf x}(t,s)\,.
\ee
Analogously, the response function is obtained from that one contributing to 
the correlation function by replacing an order-parameter correlator with a 
response function:
\be
R^\o_{\bf x}(t,s)=c_n n[C_{\bf x}(t,s)]^{n-1} R_{\bf x}(t,s)\,,
\ee
where the factor $n$ comes from the response operator 
$\widetilde{\o}_{0,n}=n \o_{0,n-1} \pht$ (see equation (\ref{defot})). 
Note that the 
combinatorial factor $c_n$ is the same as for the correlation function.
The associated FDR is
\be
X_{\bf x}^\o(t,s)\equiv\frac{R^\o_{\bf x}(t,s)}{\dpar_s C^\o_{\bf x}(t,s)}=
\frac{R_{\bf x}(t,s)}{\dpar_s C_{\bf x}(t,s)}\equiv X_{\bf x}(t,s)\,.
\label{relalltimes}
\ee
Therefore the FDR of powers of the order parameter
in real space is equal, for all times, to the FDR of the order parameter itself.

Before considering the effects of the derivatives let us consider the
previous relation in momentum space. 
Remembering that the product of two functions
after a Fourier transformation becomes a convolution, one has 
\bea
R_{\bf q}^\o(t,s)           &=&c_n n (C * \cdots *C 
*R)_{\bf q} \,,\\
\dpar_s C_{\bf q}^\o(t,s)&=& c_n n (C * \cdots *C 
*C')_{\bf q}\,,
\eea 
where $*$ is the convolution, $C$ and $R$ are in momentum
space (with the time dependence understood), $\cdots$ means $n-1$ times and 
$C'_{\bf q}=\dpar_s C_{\bf q}$. For ${\bf q}=0$ the previous relations become
\bea
R_{{\bf q}=0}^\o(t,s)     &=&c_n n \int (\rmd p) (C * \cdots *C)_{\bf p} R_{-{\bf p}} \,,\\
\dpar_s C_{{\bf q}=0}^\o(t,s)&=&c_n n \int (\rmd p) (C * \cdots *C)_{\bf p} C'_{-{\bf p}}\,.
\eea
Thus 
\be
\fl
\left[{\mathcal X}_{{\bf q}=0}^\o(t,s)\right]^{-1}=
\frac{\int (\rmd p) (C * \cdots *C)_{\bf p} C'_{-{\bf p}}} { \int (\rmd p) (C
* \cdots *C)_{\bf p} R_{-{\bf p}} } =
\frac{\int (\rmd p) (C * \cdots *C)_{\bf p} R_{-{\bf p}} {\cal X}^{-1}_{-{\bf p}}}{ \int (\rmd p) (C * \cdots *C)_{\bf p} R_{-{\bf p}} }\,,
\label{XXX}
\ee
i.e., the inverse of ${\mathcal X}_{{\bf q}=0}^\o$ is a weighted average of 
${\cal X}^{-1}_{-{\bf p}}$, 
with weight $(C * \cdots * C)_{\bf p} R_{-{\bf p}}$. In the limit 
$t\rightarrow\infty$, $s\rightarrow\infty$ this weight is
expected to be peaked around ${\bf p} = {\bf 0}$, giving 
${\mathcal X}^\infty_\o\equiv 
\lim_{s\rightarrow\infty} \lim_{t\rightarrow\infty} 
{\mathcal X}^\o_{{\bf q}=0}(t,s)=
\lim_{s\rightarrow\infty} \lim_{t\rightarrow\infty} {\mathcal X}_{{\bf
p}=0}(t,s) = X^\infty$. Thus the relation between the different FDRs
holds also when considering $\o_{0,n>1}$.
Note that in momentum space, at variance with the relation~(\ref{relalltimes}) holding
between FDR in real space, only the 
long-time limit of the FDR reproduces the FDR for the fields.

Let us now account for the effect of the derivative. In momentum space 
these are simply multiplication by ${\bf q}^{i}$ for each insertion, 
modifying the 
weight in equation (\ref{XX}) by a factor ${\bf q}^{2i}$ which does
not change our conclusion about the long-time limit. 
Since all correlation and response functions of all local operators can be 
written in terms of those of $\o_{i,n}$, this concludes the argument.

It is worth mentioning that for observables different from the order parameter,
${\cal X}_{\bf q}^\o(t,s)$ is expected to depend non-trivially on the times 
even in the Gaussian model. 
Hence the previous argument about the advantage of using 
observables in momentum space to determine $X^\infty$ 
does not apply straightforwardly.

\subsection{Two-loop order parameter FDR}
\label{Atwoloop}

The program of calculating the universal two-point functions of the
order parameter, 
the associated non-universal amplitudes and the corresponding
FDR has been worked out up to two loops
in an $\e$ expansion \cite{cg-02a1,cg-02a2}.
The analytic manipulations required for the calculation are quite long
and involved. We summarize here only the results, referring the reader
interested in technical details to the original works \cite{cg-02a1,cg-02a2}.
The obtained exponents $z$ and $\theta$ 
agree with the already known two-loop expressions \cite{zj,jss-89}.
The scaling of the response function (see equation (\ref{scalR})) 
is characterized by the non-universal amplitude 
\be
A_R = 1+\e^2 {3(N+2)\over 8(N+8)^2} \left[f(0)-4\gamma_E \log{4\over3}
 \right] +O(\e^3)\, \, ,
\label{exprAR}
\ee
and the universal function 
\be
F_R(v) = 1+\e^2 {3(N+2)\over 8(N+8)^2} [f(v)-f(0)] + O(\e^3)\; ,
\label{eqGR}
\ee
whereas the scaling function of the derivative with respect to $s$ of the 
correlation function is (c.f. equation (\ref{scaldC}))
\be
F_{{\dpar C}} (v)=
1 +\e^2 {3(N+2) \over 8(N+8)^2}\left[
2 \log {4\over3} \log{1-v\over 1+v}+\Psi(v)\right] + O(\e^3)\, .
\label{eqGdC}
\ee
The correlation function has been computed only up to one loop
\cite{cg-02a1}. We do not report here 
the quite long expression of $A_{\dpar C}$ that
can be worked out from the results in \cite{cg-02a2}.
In equations~(\ref{exprAR}), (\ref{eqGR}) and (\ref{eqGdC}) 
$f(v)$ and $\Psi(v)$ are regular functions for small $v$, 
defined for $0\le v <1$. 
Explicit expressions can be found in \cite{cg-02a2}. 
For the present purposes it is enough to note that they are of order  
$1$ up to quite large $v$, i.e., $v\sim 0.9$. 
In terms of $A_R$, $A_{\dpar C}$ and $\theta$, the 
long-time limit of the critical FDR is obtained via equation 
(\ref{Xinfamplitratio}):
\be
\fl {(X^\infty)^{-1}\over2}=
1+ {N+2\over 4(N+8)} \e+ \e^2 {N+2\over(N+8)^2}\left[
{N+2\over8} +{3(3N+14)\over4 (N+8)}+ c \right]+O(\e^3)\, ,
\label{Xqbis}
\ee
with $c=-0.0415\ldots$ (its analytic expression is given in \cite{cg-02a2}).

Let us comment on these results and compare them 
with those currently available.
The calculation shows that fluctuations give 
corrections to the Gaussian value of $F_R$ and $F_{\dpar C}$
of order $\e^2$.
Unfortunately these are extremely tiny, making the numerical and experimental 
detection very hard.
This is due to the factor $\e^2 [3(N+2)]/[8(N+8)^2]$ in front of $f(v)$ and 
$\Psi(v)$.
For the Ising model, this factor is
$\simeq0.014$ for $d=3$ and $\simeq 0.05$ for $d=2$, i.e., the two-loop 
corrections to the universal scaling functions are about $1\%$ and $5\%$ of 
the leading term in $d=3$ and $d=2$ respectively.
This is not surprising given that even for the equilibrium scaling 
functions, the corrections to the Gaussian results
involve the same prefactor \cite{cmpv-03}.

However the tiny correction in $F_R(v)$ is very important since it disagrees 
with the prediction, based on the theory Local Scale Invariance (LSI)
\cite{henkel-02}, that $F_R(v)=1$ {\it exactly}. 
The LSI is a proposed extension of conformal invariance to system with 
anisotropic scaling, e.g., to dynamics where time and space scales 
differently.
Given the large success of conformal invariance (see, e.g., \cite{cardyb})
for isotropic scaling it is very interesting to provide generalizations 
to the anisotropic case.
For Glauber-Ising model in $d=2,3$, the agreement of LSI predictions for 
the integrated response function with numerical simulation is 
really remarkable \cite{hpgl-01}.
However, measuring a more suitable integrated response 
function, a small deviation from LSI has been recently detected \cite{gp-04}.
This deviation is in qualitative agreement with the FT prediction in
equation (\ref{eqGR}). 
The failure of LSI has probably to be attributed to the limits of 
applicability of the theory \cite{hu-03,ph-04} and 
might be traced back to the fact that the term
$\Omega \, \pht^2$ (generated by a non-vanishing thermal noise) breaks
the Galilei invariance of the action (\ref{mrsh}) \cite{ph-04}. 
However, it has been argued that LSI should give exact predictions 
whenever $z=2$ \cite{ph-04}, even in the presence of the thermal noise. 
Field-theoretical results agree with this observation
(see also sections \ref{modC} and \ref{RIM}).
In fact, in the $\e$-expansion, $F_R(v)$ is Gaussian at least up to
the order at which $z = 2$ 
(i.e., $O(\e)$ in Model A of $\p^4$ theory, $O(\e^0)$ in Model C,
and $O(\sqrt{\e}^0)$ in Model A of the dilute Ising model).

Let us remark that the time dependence of ${\cal X}_{{\bf q} = {\bf 0}}(t,s)$
shows up only through the two-loop corrections appearing in 
equations (\ref{eqGR}) and (\ref{eqGdC}). 
According to previous discussion, these are expected to be quantitatively 
small, making it difficult to measure the difference between 
${\cal X}_{{\bf q}={\bf 0}}(t,s)$ and its asymptotic value
$X^\infty$, as noted in \cite{mbgs-03r}.

The expression  (\ref{Xqbis}) of $X^\infty$ for general $N$  
provides quantitative predictions for a large class of systems.
To get some numerical estimates out of this two-loop expansion a direct summation
(Pad\'e approximant $[2,0]$) and an ``inverse'' one 
(Pad\'e approximant $[0,2]$) has been performed in \cite{cg-02a2}.
From them some general trends may be understood:
\begin{itemize}
\item 
decreasing the dimensionality, $X^{\infty}$ always decreases (at least up to
$\e=2$);
\item increasing $N$, $X^{\infty}$ decreases, approaching 
the exact result of the spherical model;
\item for $N=\infty$ the curve of the $[0,2]$ approximant reproduces better
the exact result in any dimension with respect to the $[2,0]$ approximant.
\end{itemize}
The last point suggests the use of the $[0,2]$ value as estimate of  
$X^{\infty}$, also for physical $N$. 
As an {\em indicative error bar} the difference between the two
approximants has been used \cite{cg-02a2}.
Accordingly, the numerical predictions for various cases can be 
summarized as follows:

\begin{itemize}
\item[a)]{\it Ising model in two and three dimensions:}
Using the procedure just outlined, the FT estimates are $X^{\infty}=0.30(5)$ for the
two-dimensional Ising model and $X^{\infty}=0.429(6)$ for the 
three-dimensional one \cite{cg-02a2}. 
In two dimensions the agreement is quite satisfactory despite the fact that 
a priori such a low-order perturbative expansion is not expected to provide 
such nice results. 
The estimate $X^{\infty}=0.30(5)$ is marginally
compatible with the first numerical determination 
$X^{\infty}=0.26(1)$ \cite{gl-00c}.
The agreement with the subsequent estimates reviewed in table \ref{tab}, 
and giving $X^{\infty}\simeq 0.33$, is clearly improved.
In three dimensions the preliminary investigation of \cite{gl-00c} gives 
$X^{\infty}\simeq 0.40$ for the Ising model. 
Albeit quite inaccurate, it is in nice agreement
with the FT estimate  $X^{\infty}=0.429(6)$. 
Further simulation of this model in three dimensions would
provide an important test of the actual reliability of 
this quite precise FT estimate.

\item[b)]{\it XY model in three dimensions:}
The result for $X^\infty$ yields also the accurate prediction 
$X^{\infty}=0.416(8)$ for the purely dissipative three-dimensional 
$XY$ universality class. 
Currently, its ageing properties have been investigated only in
\cite{ak-04b} where the value $X^{\infty}=0.43(4)$ is reported. 
On the one hand this result is compatible, within the estimated
error bars, with the FT estimate. On the other hand, its
numerical accuracy is still not sufficient to distinguish the differences
with the Ising and Heisenberg universality classes in three dimensions.

\item[c)]{\it Heisenberg model in three dimensions:}
For the Model A Heisenberg universality class in three 
dimensions, the
estimate of $X^{\infty}$ provided in \cite{cg-02a2} is 
$X^{\infty}=0.405(10)$. 
Currently, no numerical results are available for this case.

\end{itemize}

\subsection{Two-loop FDRs of quadratic operators}
\label{secop}

The two-loop response and correlation function of the energy and the tensor operator
(see equations (\ref{defE}) and (\ref{defT}), respectively) 
were computed for generic times in reference \cite{cg-04}. 
Their general expression is very cumbersome and we refer the interested 
reader to the original reference. 
Here we report and discuss only the value of 
${\cal X}_\o^\infty$ \cite{cg-04}:
\begin{equation}
{\mathcal X}_\o^\infty = \left\{
\begin{array}{cc}
{\displaystyle \frac{1}{2}\left(1-\frac{2}{3}\frac{N+2}{N+8}\e\right) +
O(\e^2)}\;,&\quad \o = E\;,\\ \\
{\displaystyle \frac{1}{2}\left(1-\frac{1}{12}\frac{3N+16}{N+8}\e\right)
+ O(\e^2)}\;, &\quad \o = T \;.
\end{array}
\right.
\label{XinfO}
\end{equation}
Comparing these results with equation (\ref{Xqbis}) one concludes that the long-time 
limit of the fluctuation-dissipation ratio depends on the particular 
observable chosen to compute it.

Let us discuss in more detail some specific cases:
\begin{itemize}
\item {\it $N=\infty$}:
In this case one finds
\be
X^\infty_E=\frac{1}{2}\left(1-\frac{2}{3}\e\right)+O(\e^2)\,, 
\label{XEinf}
\ee
which is clearly different from the expression for the FDR of the order 
parameter, as can be checked expanding $X^\infty_M=1-2/d$
close to $d=4$. We note that equation (\ref{XEinf})  agrees 
with the expansion close to $d=4$ of the result for the total energy in the
spherical model \cite{s04}.
On the basis of combinatorial arguments one expects $X^\infty_T$ to coincide 
with $X^\infty_M$ for $N=\infty$ to all orders in $\e$ \cite{cg-04}.
This expectation is confirmed by the two-loop FT computation equation 
(\ref{XinfO}).

These findings agree with results obtained for the spherical model \cite{s04}.
Unexpectedly, the observable $P$ of the spherical model (see
previous section) can not be naively identified with $\p^2$ for $N=\infty$.
On the other hand $a_B=a_P-2$ agrees with the naive identification of $B$ as 
the Laplacian of $P$ in the continuum limit.
This fact calls for a more rigorous investigation of a
possible correspondence between the two models, beyond the case of
equilibrium dynamics.

\item {\it Ising Model}:
The direct estimate (unconstrained) from the $O(\e)$ series in 
equation (\ref{XinfO}) for $\e=2$ (giving $X^\infty_E{\rm [u]} \sim 0.28$) is 
probably unreliable, as was the case for $X^\infty_M$, i.e.,  
$X^\infty_M{\rm [1loop]}\sim 0.42$ \cite{cg-02a1} whereas
$X^\infty_M{\rm [2loops]}=0.30(5)$ \cite{cg-02a2}. 
To obtain a more reliable result without computing the $O(\e^2)$ term,
it was proposed in \cite{cg-04} to constrain linearly the available one-loop series 
to assume the exactly known value for $d=1$ (i.e., $\e=3$). 
Assuming a smooth behaviour in $\e$ up to $\e=3$, this gives \cite{cg-04}
\be
X^\infty_E{\rm [l]} =\frac{1}{2}\left(1-\frac{\e}{3}\right)\left[1 + \frac{\e}{9}+O(\e^2)\right]\,,
\label{constr}
\ee 
which has the same $\e$-expansion as equation (\ref{XinfO}), but it is expected
to converge more rapidly to the correct result. 
From equation (\ref{constr}) for the two-dimensional Ising model one gets
$X^\infty_E{\rm [l]}\sim 0.20$, that is much lower than the value determined 
so far, $X^\infty_M\simeq 0.33$ \cite{cg-02a2,c-03,mbgs-03,sdc-03,c-04}.
However, a robust field-theoretical prediction of
$X^\infty_E$ for the two-dimensional Ising model requires a difficult
higher-loop computation.
For the three-dimensional Ising model equation (\ref{constr}) gives
$X^\infty_E{\rm [l]}\simeq0.37$, to be compared with the direct estimate 
$X^\infty_E{\rm [u]}\simeq0.39$. 
Note that, as usual in $d=3$, the spreading of the different estimates
is much smaller, signaling a higher reliability of these predictions.

\item {\it $O(N)$ model with $N>1$}:
From equation (\ref{XinfO}) predictions can be obtained for the 
purely dissipative three-dimensional 
$O(N)$ models with arbitrary $N$, again considering constrained analyses 
at the lower-critical dimension, which is $d_{{\rm lcd}}=2$ in this case.
For $N>2$, $X^\infty_{E}(d=2)=0$, since for $d=2$ these 
systems are in the coarsening regime. 
In \cite{cg-04} this was assumed generically for $N\geq 2$.
Within this assumption a linear constraint gives \cite{cg-04}
\be
X^\infty_E{\rm[l]}=\frac{1}{2}\left(1-\frac{\e}{2}\right)\left[1 +\frac{16-N}{6(N+8)} \e+O(\e^2)\right]\,.
\label{lincon}
\ee 
On the other hand the exact results for the spherical model \cite{s04} 
suggest that the approach to $d=2$ is quadratic rather than linear. 
For this reason, also the quadratic constraint \cite{cg-04}
\be
X^\infty_E{\rm[q]}=\frac{1}{2}\left(1-\frac{\e}{2}\right)^2 \left[1 +\frac{20+N}{3(N+8)}\e+O(\e^2)\right]
\label{quadcon}\,,
\ee 
was implemented.
Comparing with the result for the spherical model, it was shown that 
for $N=\infty$ equation (\ref{quadcon}) provides a better approximation of the exact 
result than equation (\ref{lincon}) .
For the three-dimensional $XY$ model ($N=2$) one 
gets $X^\infty_E{\rm [u]}\simeq 0.37$ from direct estimate, 
$X^\infty_E{\rm [l]}\simeq 0.31$ from linear constraint, 
and $X^\infty_E{\rm [q]}\simeq 0.22$ from the quadratic one.
For the three-dimensional Heisenberg model ($N=3$) 
one finds, instead, $X^\infty_E{\rm [u]}\simeq 0.35$, 
$X^\infty_E{\rm [l]}\simeq 0.30$ and 
$X^\infty_E{\rm [q]}\simeq 0.21$.

Even the results with constraints are rather scattered, making it difficult 
to provide firm estimates in $d=3$. However, the conclusion
$X^\infty_E<X^\infty_M$ for all $2<d<4$ is quite robust. 
Furthermore the difference between the two FDRs should be
large enough to be observed in three-dimensional Monte Carlo simulations.
The analysis of the non-equilibrium behaviour within the $\tilde{\e}=d-2$
expansion \cite{zj,2pe} may clarify which, between the linear and the
quadratic constraint, is the proper one close to $d=2$. 

We note that $X^\infty_T$ is very close to $X^\infty_M$
making the numerical detection of such a difference probably very difficult.

\end{itemize}

Let us contrast the FT results of \cite{cg-04} with those obtained for the
one-dimensional Ising model \cite{mbgs-03,ms-04} and for the spherical 
model \cite{s04}, where 
$X^\infty_\o = X^\infty_M$ for all the observables $\o$, 
except for the total energy. 
On the other hand, one could doubt that 
the energy operator is not as suited as 
others to define the effective temperature, being conjugated 
to the temperature 
of the bath but not to the actual one (if any) of the system. 
Nevertheless FT results show that there is at least one further operator,
namely $T_{ij}$, having $X_T^\infty\neq X^\infty_M, X^\infty_E$.
This supports the idea that 
a unique effective temperature can not be defined for this 
class of models \cite{cg-04}.

We recall that in disagreement with FT calculation, the MC simulation of the 
two-dimensional Ising model apparently gives 
$X^\infty_E=X^\infty_M$ \cite{mbgs-03}.
Probably, a more accurate measure of $X^\infty_E$ is required to detect the 
difference (if any in $d=2$) between $X^\infty_E$ and $X^\infty_M$.


\section{Surface behaviour}
\label{secsurf}

It is a well-known experimental fact and theoretical result that the
presence of surfaces influences both the static and dynamic equilibrium
behaviour of critical systems
\cite{diehl-86,diehl-97,p-04a,d-90,dd-83,diehl-94,dj-92}. 
Upon approaching the critical point, local quantities such as the order
parameter may display algebraic singularities at the surface differing from 
those characteristic of the bulk behaviour. 
Accordingly, one usually introduces {\it surface} critical
exponents \cite{diehl-86} in addition to the standard bulk ones. 
In the generic case, the former can not be expressed in terms of
the latter.

\subsection{The semi-infinite $O(N)$ model: statics and dynamics}

As an illustrative example, let us consider the influence of a planar
surface on the model (\ref{HON}) defined on a semi-infinite hypercubic
lattice.  Its Hamiltonian is given by
\be
{\cal H}_{\infty/2}= - \sum_{\langle {\bf ij}\rangle\in {\rm
bulk}} {\bf s_i}\cdot{\bf s_j} - r \sum_{\langle {\bf ij}\rangle\in {\rm
surface}} {\bf s_i}\cdot{\bf s_j} \;,
\label{HONsurb}
\ee
where $r>0$ is the coupling between nearest-neighbouring spins located at 
the surface.
Because of the missing bonds, these 
spins experience an effective local magnetic
field which is smaller than in the bulk, at least for sufficiently small $r$. 
Accordingly, one expects that the surface is not able to order independently
of the bulk. Upon approaching the bulk critical point at the critical
temperature $T_c$, the surface and
the bulk order simultaneously. This defines the so-called {\it
ordinary transition} (OT) and the associated  surface critical behaviour
is characterized by a given set of surface critical exponents
\cite{diehl-86}. On the other hand, upon increasing $r$ at fixed $T>T_c$,
it is possible to overcompensate the missing bonds at the
surface. As a consequence, the surface may order in the presence of a
disordered bulk, provided that $d-1$ is larger than the lower critical
dimension of the model. For $r$ greater than the threshold value 
$r_{\rm sp}$ there is a finite surface
transition temperature $T_s(r)>T_c$ at which the so-called {\it surface
transition} takes place. The surface critical
behaviour of the surface transition is the same as the bulk critical one of the
$d-1$-dimensional system. For $T_s(r)>T>T_c$ the system is
characterized by an ordered surface and a disordered bulk.
Upon further decreasing $T$, the bulk
orders at $T_c$ in the presence of an already ordered surface. This
marks the so-called {\it extraordinary transition}.  
In the $(T,r)$-plane the three lines of phase transitions 
meet at the so-called multicritical {\it special transition} (SpT). 

The same scenario is valid in equilibrium and non-equilibrium dynamics in
semi-infinite systems: The dynamical
behaviour is affected by the presence of the surface and depends on the
surface universality class. On the other hand no new and independent
exponents show up in addition to the static surface and static
and dynamic bulk ones. This is true in the case of 
both equilibrium \cite{dd-83,diehl-94,dj-92} and 
non-equilibrium evolution \cite{nlinrel,rc-95,pi-04}.

The influence of a surface on the dynamical behaviour
following a quench from the high-temperature phase to the critical
point has been investigated in \cite{ms-96}. In particular the
surface autocorrelation function and the associated surface 
autocorrelation exponent
$\lambda_{c,1}$ are determined analytically at the OT and SpT for the 
Model A dynamics of the
$O(N)$ Model up to the first order in the $\epsilon$-expansion and
exactly for the $O(N)$ model with $N\rightarrow\infty$. MC simulations are
instead used to investigate the same quantities at the OT of the
two-dimensional Ising model with Glauber dynamics.  
This problem was reconsidered in \cite{rc-95}, showing that the surface 
autocorrelation exponent is not a new dynamical quantity as claimed 
in \cite{ms-96} but it can be expressed in terms of the bulk
autocorrelation exponent $\lambda_c$ and surface and bulk static 
critical exponents. 
The two-time surface autocorrelation function $C_1(t,s)$ was studied
in \cite{pi-04}. By using scaling arguments, it was shown
that a novel short-time behaviour (referred to as ``cluster
dissolution'') is expected in some cases, among which the
three-dimensional Ising model at OT. In this case a non-algebraic
decay of correlations is expected: $\ln C_1(t,s)\sim
- (t-s)^\kappa$ for $t-s \ll s$ where the exponent $\kappa$ is given in
terms of already known static surface and dynamic bulk exponents. 
This scenario was confirmed by MC simulation \cite{pi-04}.

Previous studies have been extended to the response function and to
the full dynamic scaling regime in \cite{p-04b}. The effects of
the surface on the ageing properties of a critical systems are
investigated via MC simulations at the OT of the two-dimensional Ising model
and at the OT and SpT of the three-dimensional one. In addition 
the Hilhorst-van Leeuwen model is also considered \cite{p-04b}.
The scaling behaviour observed in all these models
agrees with scaling arguments, both for the surface autocorrelation 
function $C_1(t,s)$ and for the integral of the 
surface autoresponse function $R_1(t,s)$. 
In terms of $R_1$ and $C_1$ the surface FDR can be
defined as:
\be
X_1(t,s) = \frac{T\, R_1(t,s)}{\partial_s C_1(t,s)} \;.
\label{Xsurf}
\ee
$X_1$ is analogous to the FDRs previously introduced. 
As in the bulk, it is possible to show that $X_1(t,s)$ is a universal
function and its long-time limit $X_1^\infty$ is a universal amplitude ratio.
As such, it is expected to depend on the
specific surface universality class. In \cite{p-04b} it was found
that $X_1^\infty = 0.37(1)$ for the two-dimensional Ising model (OT)
whereas for the three-dimensional Ising model $X_1^\infty= 0.59(2)$
at OT and $X_1^\infty=0.44(2)$ at SpT. 
  
From these results it is quite clear that $X_1^\infty \neq X^\infty$
(c.f. table \ref{tab})
and in particular that $X_1$ depends on the surface universality
class, as expected. 

It is worth noting that MC simulations suggest that $X_1^\infty$ can be 
greater than $\case{1}{2}$ at the OT point, as a difference with the bulk,
where $X^\infty$ is always $\leq \case{1}{2}$ for $O(N)$ models.

\subsection{The semi-infinite $O(N)$ Universality Class:
Field-Theoretical Approach}

In the previous sections we have discussed in detail how to apply the
field-theoretical approach to study ageing phenomena in the  bulk 
of a critical system. Taking
advantage of previous works on the subject \cite{ms-96,rc-95,p-04b} we
outline here a scaling approach to the non-equilibrium dynamics in 
the presence of spatial surfaces.
Let us go back to equations (\ref{scalfin1}) and (\ref{scalfin2}),
valid for a critical system in the bulk, i.e., far enough from any
surface. From these equations it is possible to compute the two-point
correlation and response functions in real space,  which depend on
the coordinates of the two points $({\bf x}',t')$ and $({\bf
x},t)$. In the following we assume $t>t'$.
Because of the translational invariance in the bulk, this dependence is only 
on ${\bf x}'-{\bf x}$. 
Let us introduce in the system a planar $d-1$-dimensional surface 
${\mathcal S}$. 
As a consequence, the translational invariance is broken
in the direction perpendicular to ${\mathcal S}$. For later
convenience we decompose the position vector as ${\bf x}=({\bf
x}_\|,x_\bot)$ where ${\bf x}_\|$ is the $d-1$-dimensional component
parallel to ${\mathcal S}$ and $x_\bot$ is the one-dimensional one
perpendicular to it. The residual translational symmetry
along ${\mathcal S}$ implies that the
two-point correlation functions depend on
${\bf x}_\|'$ and ${\bf x}_\|$ through 
${\bf x}'_\|-{\bf x}_\|$ whereas they depend on
$x'_\bot$ and $x_\bot$ separately. The dependence on
$x'_\bot - x_\bot$ is recovered only in the bulk, i.e., far away from
${\mathcal S}$. For ${\bf q}_\| = {\bf 0}$ (where ${\bf q}_\|$ is 
the momentum conjugate to
${\bf x}_\|' - {\bf x}_\|$) one can write equations
(\ref{scalfin1}) and (\ref{scalfin2}) as
\bea
\fl R_{{\bf q}_\| = {\bf 0}}(t,t'; x_\bot, x'_\bot) &=& \phantom{t'}
(t-t')^{a-1/z}(t/t')^{\theta} F^{\rm b}_R( x_\bot(t-t')^{-1/z},
x'_\bot(t-t')^{-1/z} ,t'/t)\; ,\label{scalR_bis}\\
\fl C_{{\bf q}_\| = {\bf 0}}(t,t'; x_\bot, x'_\bot) &=&
t'(t-t')^{a-1/z}(t/t')^{\theta} F^{\rm b}_C( x_\bot(t-t')^{-1/z},
x'_\bot(t-t')^{-1/z} ,t'/t)\; ,
\label{scalC_bis}
\eea
where $a = (2-\eta-z)/z$. The superscript ${\rm b}$ 
reminds that these forms are
valid, a priori, in the bulk. As discussed in \cite{rc-95} no
genuinely new divergences arise when discussing the non-equilibrium
dynamics in the presence of a surface. This means that the short-time
expansion $t'\rightarrow 0$ 
of the fields $\p({\bf x}',t')$ and $\pht({\bf x}',t')$ (see
equations (\ref{SDE1}) and (\ref{SDE2})) is not influenced
by the presence of ${\mathcal S}$ and, conversely, that their
short-distance expansion  for $x_\bot' \rightarrow 0$ is not influenced by
the presence of the time surface. Accordingly, the short-distance expansions (SDE)
of $\p({\bf x}',t')$ and $\pht({\bf x}',t')$ 
are
the same (at least in Model A) as in the static case 
and have a SDE coefficient $C_*(x'_\bot)$ 
[$\p({\bf x}',t') \sim C_*(x'_\bot)\p({\bf x}'=({\bf x}'_\|,x'_\bot=0),t')$ and analogous 
for $\pht({\bf x}',t')$]
that in the scaling regime is given by \cite{diehl-86}
\be
C_*(x_\bot') = a_* [x_\bot']^{(\beta_1-\beta)/\nu} \;.
\label{SDEx}
\ee
The exponent $\beta_1$ is defined analogously to the bulk exponent
$\beta$ and describes the behaviour of the surface magnetization $m_1$
upon approaching the critical point. 
For the $O(N)$ universality class field-theoretical estimates of $\beta_1$ 
have been provided using different approaches. 
For recent reviews see \cite{diehl-97,p-04a}. 
Equation (\ref{SDEx}) implies that for $x_\bot \rightarrow 0$,  $R_{{\bf
q}_\| = {\bf 0}}, C_{{\bf
q}_\|  = {\bf 0}} \sim  [x_\bot]^{(\beta_1-\beta)/\nu}$ i.e.,
they have no analytic behaviour in this limit. The previous scaling
forms can be written in a way that singles out this limiting behaviour. 
 As a result the scaling forms in the presence of a surface are given by
(hereafter we omit the subscript ${\bf q}_\| = {\bf 0}$)
\bea
\fl R(t,t'; x_\bot, x'_\bot) = A_R^{\rm s}\phantom{t'}
(t-s)^{a_1}(t/t')^{\theta} \; &
[x_\bot x'_\bot]^{(\beta_1-\beta)/\nu} \nonumber\\
& \quad \times F^{\rm s}_R(x_\bot(t-t')^{-1/z},
x'_\bot(t-t')^{-1/z} ,t'/t)\; ,\label{scalR_s}\\
\fl C(t,t'; x_\bot, x'_\bot) = A_C^{\rm s}
t'(t-t')^{a_1}(t/t')^{\theta} \; &
[x_\bot x'_\bot]^{(\beta_1-\beta)/\nu} \nonumber\\
&\quad\times F^{\rm s}_C(x_\bot(t-t')^{-1/z},
x'_\bot(t-t')^{-1/z} ,t'/t)\; ,
\label{scalC_s}
\eea
where $a_1 = a - 1/z - 2(\beta_1-\beta)/(\nu z)$. The non-universal
amplitudes $A_{R,C}^{\rm s}$ are fixed requiring
$F^{\rm s}_{R,C}(0,0,0) = 1$. The previous equations provide an
extension of the scaling forms presented in \cite{rc-95,p-04b}. As a
consequence of equation (\ref{scalC_s}) one gets
\bea
\fl \partial_{t'} C(t,t'; x_\bot, x'_\bot) = A_{\dpar C}^{\rm s}
(t-t')^{a_1}(t/t')^{\theta} \; &
[x_\bot x'_\bot]^{(\beta_1-\beta)/\nu} \nonumber\\
\fl &\times F^{\rm s}_{\dpar C}(x_\bot(t-t')^{-1/z},
x'_\bot(t-t')^{-1/z} ,t'/t)\; ,
\label{scaldC_s}
\eea
where the non-universal amplitude $A^{\rm s}_{\dpar C}$ is fixed 
requiring $F^{\rm s}_{\dpar C}(0,0,0)=1$. It is easy to show that
$A^{\rm s}_{\dpar C} = (1-\theta) A^{\rm s}_C$. 

It is natural to define the FDR in analogy to what has been done in the bulk:
\be
{\mathcal X^{\rm s}_{{\bf q}_\| = {\bf 0}}(t,t'; x_\bot,
x'_\bot)} 
\equiv \frac{  R(t,t'; x_\bot,
x'_\bot)}{\partial_s C(t,t'; x_\bot, x'_\bot) } \;.
\label{FDRs}
\ee
Using the previous scaling forms ${\mathcal X}^{\rm s}_{{\bf q}_\| =
{\bf 0}}$ can be written as
\be
{\mathcal X^{\rm s}_{{\bf q}_\| = {\bf 0}}(t,t'; x_\bot,
x'_\bot)} =
\frac{A^{\rm s}_R}{A_{\dpar C}^{\rm s}} \frac{F^{\rm s}_R(x_\bot(t-t')^{-1/z},
x'_\bot(t-t')^{-1/z} ,t'/t)}{F^{\rm s}_{\dpar C}(x_\bot(t-t')^{-1/z},
x'_\bot(t-t')^{-1/z} ,t'/t)} \;.
\label{FDRs_bis}
\ee
This explicitly shows that ${\mathcal X}^{\rm s}_{{\bf q}_\| = {\bf
0}}$, just like ${\mathcal X}_{{\bf q}={\bf 0}}$, is a
universal function, being the ratio of two quantities that
have the same scaling dimensions. The specific form of ${\mathcal
X}^{\rm s}_{{\bf q}_\| = {\bf 0}}$  is expected to depend on the
universality class of the transition at the surface.
Moreover, in the long-time limit $t\gg t'$ and sufficiently close to the 
surface $x_\bot, x_\bot' \ll (t-t')^{1/z}$ the leading asymptotic value 
${\mathcal X}^{{\rm s},\infty}$
is approached independently of the distance from the surface, i.e.,
\be
{\mathcal X}^{{\rm s},\infty} \equiv \lim_{t\rightarrow\infty\;,
{\rm fixed}\; t',x_\bot,x_\bot'} {\mathcal X^{\rm s}_{{\bf q}_\| = {\bf 0}}(t,t'; x_\bot,
x'_\bot)} =  \frac{A^{\rm s}_R}{A^{\rm s}_{\dpar C}} =
 \frac{A^{\rm s}_R}{(1-\theta)A^{\rm s}_C}\;.
\label{Xinfs}
\ee
This is a consequence of the fact that at criticality 
the influence of the surface deeply penetrates into the bulk as time goes
by, whereas a bulk-like behaviour can be observed only for
$x_\bot,x_\bot' \gg (t-t')^{1/z}$. Following the argument outlined in section
\ref{secFDRms} one can argue that the asymptotic limit of the FDR just
introduced is the same as that obtained by considering in equation
(\ref{FDRs}) the surface two-point autoresponse and autocorrelation
functions, i.e., quantities in real space. 
On the other hand, as its bulk counterpart ${\mathcal
X}_{{\bf q} = {\bf 0}}$, the expression in the mixed representation
$({\bf q}_\|,x_\bot)$ is more suited to be computed within the
field-theoretical approach and to detect in numerical simulations the
ageing behaviour associated with the slow dynamics of the slow mode
${\bf q}_\| = {\bf 0}$.

Let us check the scaling relations (\ref{scalR_s}) and (\ref{scalC_s})
within the Gaussian model and compute the associated FDR
(\ref{FDRs_bis}) in the case of OT and SpT. 
In the presence of the surface the equilibrium correlation and
response function are given by \cite{diehl-86} 
\bea
\fl R^{\rm (s,e)}({\bf q}_\|, x_\bot, x_\bot',\Delta t) &=& 
R^{\rm (b,e)}({\bf q}_\|, x_\bot- x_\bot',\Delta t) \pm 
R^{\rm (b,e)}({\bf q}_\|, x_\bot + x_\bot',\Delta t) \label{Rse}\\
\fl C^{\rm (s,e)}({\bf q}_\|, x_\bot, x_\bot',\Delta t) &=& 
C^{\rm (b,e)}({\bf q}_\|, x_\bot- x_\bot',\Delta t) \pm 
C^{\rm (b,e)}({\bf q}_\|, x_\bot + x_\bot',\Delta t) \label{Cse}
\eea
for ${{\rm SpT}\atop {\rm OT}}$. 
$R^{\rm (b,e)}({\bf q}_\|,x_\bot,\Delta t)$ and 
$C^{\rm (b,e)}({\bf q}_\|,x_\bot,\Delta t)$ are the bulk equilibrium 
response and correlation functions in the $({\bf
q}_\|,x_\bot)$-representation and $\Delta t = t - t'$. 
Let us recall that in momentum space (we set $\Omega=1$)
they read
\bea
R^{\rm (b,e)}({\bf q},\Delta t) = \theta (\Delta t) \rme^{-({\bf q}^2 +
r_0)\Delta t}\,,\\
C^{\rm (b,e)}({\bf q},\Delta t) = \frac{1}{{\bf q}^2 + r_0}\rme^{
-({\bf q}^2 + r_0)|\Delta t|} 
\eea
[${\bf q} = ({\bf q}_\|,q_\bot)$].
Accordingly,
\bea
\fl R^{\rm (b,e)}({\bf q}_\|, \Delta x_\bot,\Delta t) =& \int \frac{\rmd
q_\bot}{2\pi} R^{\rm (b,e)}({\bf q},\Delta t) \rme^{\rmi q_\bot \Delta
x_\bot} \nonumber\\
\fl &= \theta(\Delta t) (4\pi
 \Delta t)^{-1/2} \exp\left\{- ({\bf q}_\|^2 + r_0)\Delta t -
\frac{(\Delta x_\bot)^2}{4 \Delta t}\right\} \label{Rbe}\\
\fl C^{\rm (b,e)}({\bf q}_\|,\Delta x_\bot, \Delta t) =& 
\int_{|\Delta t|}^\infty\rmd u \;
R^{\rm (b,e)}({\bf q}_\|, \Delta x_\bot,u)\;.\qquad {\rm [FDT]}
\label{Cbe}
\eea
As reviewed in section \ref{FT} the effect of the quench from the
high-temperature disordered phase amounts to the introduction of a time
surface in the field-theoretical description of the model. At the
fixed point $\tau_0 = \infty$ the boundary condition at the time surface
is a Dirichlet one. Accordingly, the correlation function is modified, 
compared to the time-bulk one, by the introduction of an ``image term'' in 
time, i.e., 
$C^{\rm (b,ne)}({\bf q};t,t') = C^{\rm (b,e)}({\bf q},t-t') - C^{\rm (b,e)}({\bf q},t+t')$, whereas
the response function is not modified: $R^{\rm (b,ne)}({\bf q};t,t') =
R^{\rm (b,e)}({\bf q},t-t')$. 
The presence of a spatial surface in the problem does not change the
previous considerations. Accordingly, one finds that for the non-equilibrium
evolution (ne) after a quench 
\bea
\fl R^{\rm (s,ne)}({\bf q}_\|, x_\bot, x_\bot';t,t') &= R^{\rm
(s,e)}({\bf q}_\|, x_\bot, x_\bot',t-t') \;,\\
\fl C^{\rm (s,ne)}({\bf q}_\|, x_\bot, x_\bot';t,t')&= C^{\rm (s,e)}({\bf
q}_\|, x_\bot, x_\bot',t-t') - C^{\rm (s,e)}({\bf q}_\|, x_\bot,
x_\bot',t+t')\;. \label{Csne}
\eea
By means of equations (\ref{Rse}), (\ref{Cse}), (\ref{Rbe}) and
(\ref{Cse}) it is simple to find that $R^{\rm (s,ne)}$ and $C^{\rm
(s,ne)}$ have the expected scaling forms (\ref{scalR_s})
and (\ref{scalC_s}) with the proper (mean-field) values of the
critical exponents. Moreover, the non-universal amplitudes are easily computed:
$A_R^{\rm s}{\rm [SpT]} = \pi^{-1/2}$, 
$A_R^{\rm s}{\rm [OT]} = (4\pi)^{-1/2}$,
$A_C^{\rm s}{\rm [SpT]} = 2 \pi^{-1/2}$ and
$A_C^{\rm s}{\rm [OT]} = \pi^{-1/2}$. Using equation
(\ref{Xinfs}) one concludes that the surface FDR is equal to
$\case{1}{2}$ within the Gaussian theory for both SpT and OT, 
as it is the case for the Gaussian bulk FDR. 

The previous equations can be rewritten in real space. In particular,
the FDT together with equation (\ref{Csne}) yields
\be
\fl
\partial_{t'} C^{\rm (s,ne)}({\bf x}, {\bf x}';t,t') = 
R^{\rm (s,e)}({\bf x}, {\bf x}',t-t') + R^{\rm
(s,e)}({\bf x}, {\bf x}',t+t') \;.
\ee
Accordingly, the FDR in real space can be written as ($t>t'$)
\be
[X^{\rm s}({\bf x}, {\bf x}';t,t')]^{-1} = 1 + \frac{R^{\rm (s,e)}({\bf x}, {\bf x}',t+t')}{R^{\rm (s,e)}({\bf x}, {\bf x}',t-t')}
\ee
where $R^{\rm (s,e)}({\bf x}, {\bf x}',\Delta t)$ is given by the
Fourier transform of equation (\ref{Rse}). Let us investigate the
cross-over between the bulk-like FDR and the surface-like one. For
simplicity we consider the case ${\bf x} = {\bf x}'$. Then it is easy
to find that
\be
[X^{\rm s}({\bf x}, {\bf x};t,t')]^{-1} = 1+
\left(\frac{t-t'}{t+t'}\right)^{d/2} \frac{1 \pm
\exp\{-x_\bot^2/[t+t']\}}{1\pm
\exp\{-x_\bot^2/[t-t']\}} \;.
\ee
For $x_\bot^2 \gg t+t'$ one recovers the bulk value of the FDR 
equation (\ref{XXX}) both for SpT and OT. 
In the opposite limit $x_\bot^2 \ll t-t'$ for the OT one finds
\be
[X^{\rm s}({\bf x}, {\bf x};t,t')]^{-1} = 1 + \left(\frac{t-t'}{t+t'}\right)^{d/2+1}
\label{FDRxs}
\ee
that differs from the corresponding bulk quantity and displays the
same cross-over from the quasi-equilibrium regime $t\gtrsim t'$ to the
ageing regime $t\gg t'$. On the other hand, in the case of SpT, one finds the 
same expression as in the bulk, given in equation~(\ref{XXX}). These results for the 
surface FDR provide the Gaussian theoretical predictions for the quantity 
$X_1$ numerically investigated in \cite{p-04b}. In particular, as far as the asymptotic 
value $X_1^\infty$ is concerned, the Gaussian result is $X_1^\infty=\case{1}{2}$, both 
for OT and SpT. 
At first sight, the non-monotonic behaviour of $X_1^\infty[{\rm OT}]$
as a function of the space 
dimensionality, found in~\cite{p-04b}, can be hardly accounted for within the
field-theoretical approach. 
Further analytical and numerical investigations are required to clarify
this issue.

\section{Dynamics of a conserved order parameter (Model B)}
\label{modB}

In this section we discuss the effect of the conservation law of the order 
parameter. As explained in section \ref{duc} this defines the Model B
universality class.
Compared to the Model A, Model B is much less studied in the literature,
mainly because, due to the
conservation law, its critical slowing-down is much more severe and 
it is quite difficult to reach the asymptotic behaviour in numerical
and experimental investigations.

\subsection{Review of numerical and analytic results}

For Model B dynamics, the critical exponent $z$ is $z=4-\eta$ 
exactly \cite{HH} and
it has been argued that 
$\lambda=d$ \cite{mhl-94}, as found in several numerical 
studies for the one- and two-dimensional Ising 
model \cite{mhl-94,ahj-94,mh-95,gkr-04,sire-04}.

Earlier claims that the one-dimensional FD plot (and hence the FDR) of 
Kawasaki-Ising model reduces asymptotically to the Glauber one \cite{cclz-02},
have been recently revisited in \cite{gkr-04} and 
the apparent equivalence between the two models has been traced back 
to the time range on which the analysis of \cite{cclz-02} was based. 

The scaling form of the
autocorrelation function has attracted particular interest.
For $t,s\gg1$ it has been found to scale 
like \cite{gkr-04}
\be
C_{{\bf x}=0}(t,s)\sim t^{-d/z} {\cal C}(t s^{-\phi})\,,
\label{Cgkr}
\ee
where ${\cal C}(x)$ approaches 
a non-zero constant for $x\rightarrow\infty$ whereas
${\cal C}(x)\sim x^{(d-\lambda')/z}$ for $x\ll1$ and 
$\phi=1+(2-\eta)/(\lambda'-d)$.
This scaling implies the existence of a new relevant length scale
$\sim s^{\phi/z}$
which controls to the crossover between the two regimes.
Based on MC simulations of the one- and two-dimensional 
Ising 
model with Kawasaki dynamics it has been conjectured that 
$\lambda'=d+3/2$ \cite{gkr-04}.
This value has been later revised by Sire, who calculated exactly the 
correlation function for the $O(\infty)$ model, finding 
$\lambda'=d+2$ \cite{sire-04}.
By means of simulations of the Ising model in $d=1,2$ with an 
accelerated dynamics (in the same dynamic universality class),  the relation 
$\lambda'=d+2$ has been checked and it has been conjectured to generally 
hold for 
all systems with conserved order parameter.

So far numerical simulations, both for $d=1$ and $2$, 
have not been able to explore the
long-time regime from which $X^\infty$ is determined. However the conclusion
$X^\infty_{\rm Kawasaki}>X^\infty_{\rm Glauber}$ for the critical FDR
of the magnetization seems quite robust, at least 
in two dimensions \cite{gkr-04}. This fact is somehow surprising if one interprets 
$T/X^\infty$ as an effective temperature. Indeed one would expect that a 
slower dynamics should result in an higher $T_{\rm eff}$.

\subsection{Field-theoretical approach}

The Field-theoretical approach to Model B dynamics starts from the 
effective action (\ref{mrshB}).
The presence of the gradient term (that in momentum space gives a factor 
${\bf q}^2$) makes the difference with Model A dynamics: In fact it leads
to an interaction vertex of the form ${\bf q}^2 g_0$ that reduces the primitive
divergences of all Feynman diagrams so that no new divergences 
are generated by the dynamics \cite{zj,jss-89}.
Hence  the critical exponents are exactly obtained 
on the sole basis of power-counting: 
$\tilde\eta=-\eta$ (i.e., $z=4-\eta$ \cite{zj,tauber}) and 
$\eta_0=\theta=0$ \cite{jss-89}.
Note that the response function in this case is not given by 
$\langle \pht_{\bf -q}(s) \p_{\bf q}(t)\rangle\equiv G_{\bf q}(t,s)$, but
rather by $R_{\bf q}(t,s)=\sigma {\bf q}^2 G_{\bf q}(t,s)$ (see 
equation (\ref{generalresp})).


\subsection{Scaling forms}

The general scaling forms may be obtained solving RG equations with the 
methods of characteristic functions, as outlined in section \ref{secRenScal} for 
Model A dynamics. In particular equations (\ref{genscaling}) and (\ref{genexp})
for the correlation functions 
are unchanged and they provide quite stringent predictions
since one exactly knows that $\tilde\eta=-\eta$ and $\eta_0=0$.

The scaling of the autoresponse function at criticality is simply get from 
these equations with $\delta(1,1,0)=d$:
\be
R_{{\bf x}=0}(t,s)=-\sigma
\left.\nabla_{\bf x}^2 
\langle \p({\bf x},t) \pht ({\bf 0},s)\rangle\right|_{{\bf x}= 0} =
{\cal A}_R (t-s)^{-(d+2)/z}{\cal F}_R (s/t)\,,
\ee
where, as usual, one singles out the non-universal constant ${\cal A}_R$, 
so that ${\cal F}_R(v)$ is a universal function with ${\cal F}_R(0)=1$.

The autocorrelation function at the fixed point $\tau_0^{-1}=0$ is obtained
from $\delta(2,0,0)=d-2+\eta=d+2+z$:
\be
C_{{\bf x}=0}(t,s)={\cal A}_C s(t-s)^{-(d+2)/z}{\cal F}_C (s/t)\,,
\label{CBp}
\ee
where the proportionality to $s$ is explicitly shown to stress the fact that 
the Dirichlet boundary condition in time enforces $C_{\bf x}(t,0)=0$ 
for $\tau_0^{-1}=0$. 
As emphasized in section \ref{sectau0}, the long-time limit of the
correlation function for finite $s$ can be influenced by the
non-vanishing initial correlations, i.e., by a finite $\tau_0$. 
From equation (\ref{at}) we get, in the long-time limit and neglecting
further corrections coming from $\tau_0^{-1}$
\be
A(t)\equiv C_{{\bf x}=0}(t,0)=
\tau_0^{-1} \langle \p ({\bf x},t)  \pht_0({\bf x})\rangle
\simeq \tau_0^{-1} t^{-\delta(1,0,1)/z}=\tau_0^{-1} t^{-d/z},
\label{atb}
\ee
where we used $\delta(1,0,1)=d$ from equation (\ref{genexp}). 
Accordingly, for $t \gg s$, $C_{{\bf x}=0}(t,0) \simeq \tau_0^{-1}
t^{-d/z}$, whereas, from equation (\ref{CBp}), 
$C_{{\bf x}=0}(t,s)\simeq s t^{-(d+2)/z}$.
As a fundamental difference with Model A dynamics, these two terms 
do not scale in the same way with $t$ for $t\gg s$ and in particular
the former is the leading one for large $t$.
Accordingly the autocorrelation exponent
(see equation~(\ref{autocorrexp})) is given by
$\lambda=d$, in agreement with what has been argued in \cite{mhl-94}. 
Furthermore a cross-over behaviour is expected for finite $s$
when the contribution coming from finite initial correlations
is of the same order of magnitude as the contribution with 
$\tau_0^{-1}=0$. Comparing the previously given asymptotic behaviours
one concludes that the former dominates when $\tau_0^{-1}t^{-d/z} \gg
s t^{-(d+2)/z}$, i.e., when $\tau_0 s t^{-2/z} \ll 1$. This agrees
with the prediction  (\ref{Cgkr}), with $\phi=z/2$ 
(equivalently $\lambda'=d+2$),  providing a FT proof of
Sire's conjecture \cite{sire-04}.

From the scaling of the autoresponse function (see the definition in equation~(\ref{autorespexp})) 
we instead get 
$\lambda_R=d+2\neq\lambda$. However, such inequality does not prevent the 
definition of a universal FDR, as one might naively think 
on the basis of the fact that this leads to a different scaling dimensions for
$R$ and $\partial_s C$. Indeed the contribution to $\partial_s C$ 
coming from finite
$\tau_0$, leading in the large-$t$ limit of $C_{{\bf
x}=0}\simeq\tau_0^{-1}t^{-d/z}$, is expected to scale as 
$\tau_0^{-1}t^{-d/z-1}$, turning out to be 
subleading in $\partial_s C_{{\bf x}=0}(t,s)\simeq t^{-(d+2)/z}$.
In fact the derivative is
\be 
\partial_s C_{{\bf x}=0}(t,s)={\cal A}_{\dpar C} (t-s)^{-(d+2)/z}
{\cal F}_{\dpar C} (s/t)\,,
\ee
where, being $\theta=0$, ${\cal A}_{\dpar C}= {\cal A}_C$.
It is possible to define the universal amplitude ratio 
\be
X^{\infty}=\lim_{s\rightarrow \infty} \lim_{t\rightarrow \infty}
\frac{R_{{\bf x}=0}(t,s)}{\partial_s C_{{\bf x}=0}(t,s)}=
\frac{{\cal A}_R}{{\cal A}_{\dpar C}} = \frac{{\cal A}_R}{{\cal A}_C} \,.
\ee

\subsection{Gaussian results}

The Gaussian response and correlation functions read for $t>s$
\bea
R^G_{\bf q}(t,s)&=&\sigma {\bf q}^2
\langle \pht({\bf q},s) \p(-{\bf q},t) \rangle_G = \sigma {\bf q}^2
e^{-\sigma {\bf q}^2 ({\bf q}^2+r_0) (t-s)},\label{RgauxB}\\
 C^G_{\bf q}(t,s)&=&\langle \p({\bf q},s) \p(-{\bf q},t) \rangle_G 
\nonumber\\&=&
\frac{
 e^{-\sigma {\bf q}^2  ({\bf q}^2+r_0)(t-s)}-
e^{-\sigma {\bf q}^2 ({\bf q}^2+r_0)(t+s)}}{{\bf q}^2+r_0}
+\tau_0^{-1} e^{-\sigma {\bf q}^2 ({\bf q}^2+r_0)(t+s)}
\, \label{CgauxB}
\eea
where we keep the correction coming from non-zero $\tau_0^{-1}$.
The  Fourier transforms of the critical functions for  ${\bf x}={\bf 0}$ are 
(setting $\sigma=1$)
\bea
R_{{\bf x}=0}(t,s)&=& r_{1,d} (t-s)^{-(d+2)/4}\,,\\
C_{{\bf x}=0}(t,s)&=& r_{-1,d} s (t-s)^{-(d+2)/4} f_G\left(\frac{s}{t}\right)
+\tau_0^{-1} r_{0,d} (t+s)^{-d/4}\,,
\eea
where $r_{n,d}=\Gamma((d+2n)/4)/[2(4\pi^{d/2})\Gamma(d/2)]$ and 
\be
f_G(x) = (1-x)^{(d+2)/4}\frac{(1-x)^{(2-d)/4} - (1+x)^{(2-d)/4}}{x} \,,
\ee
which is regular
for small $x$:  $f_G(0) = (2-d)/2$.
These forms agree with those obtained from the solution of the RG 
equations (see equations~(\ref{RXX}) and (\ref{CXX})) with $z=4$. 

From the scaling derived in the previous subsection,  
we can define the universal FDRs in real and reciprocal space as in 
equations (\ref{dx}) and (\ref{Xq}).
The argument presented in section~\ref{secFDRms} is easily extended
to the present case, 
concluding $X_{{\bf x}=0}^\infty={\cal X}_{{\bf q}\rightarrow{\bf 0}}^\infty$.
Here we use the notation ${\cal X}_{{\bf q}\rightarrow{\bf 0}}^\infty$ given that
$\p_{{\bf q}={\bf 0}}$ does not change in time, being a conserved quantity. 
Accordingly, $R_{{\bf q}={\bf 0}}$ and 
$\dpar_s C_{{\bf q}={\bf 0}}$ vanish whereas 
the limit ${\bf q}\rightarrow{\bf 0}$ of their ratio is well-defined 
and finite.
On the other hand, in a finite system of linear dimension $L$, the smallest
allowed mode is ${\bf q}_{\rm min}$ with $|{\bf q}_{\rm min}|\propto 1/L$. 
The limit ${\bf q}\rightarrow {\bf 0}$ is explored by studying the mode with
${\bf q}={\bf q}_{\rm min}$ 
in the limit of large $L$. In any case to stay within 
the non-equilibrium regime one has always to consider times much smaller than 
$L^z$, otherwise the system equilibrates.

The Gaussian FDR can be readily computed using the previous
expressions, finding
\be
{\cal X}_{\bf q}^{-1}(t,s)=
1+\left(1-\frac{{\bf q}^2+r_0}{\tau_0}\right) e^{-2\sigma {\bf q}^2({\bf q}^2+r_0)s}.
\ee
As for Model A, all the modes with ${\bf q}\neq {\bf 0}$ 
relax to equilibrium, i.e., ${\cal X}^\infty_{{\bf q}\neq 0}=1$.
At variance with the case of Model A, 
${\cal X}_{{\bf q} \rightarrow {\bf 0}}(t,s) $ reaches a non trivial (but not universal)
value  $1/(2+r_0\tau_0^{-1})$ also for $r_0\neq 0$ (non-critical case).
The conservation law allows arbitrarily slow modes which 
in the case of a purely dissipative dynamics is present only at criticality.
The universal critical zero-momentum FDR is again $\frac{1}{2}$.

Since it was instructive for the Model A dynamics, let us give a look at the 
Gaussian FDR in real space. From the Gaussian response 
and correlation functions reported above one gets
\be
X^{-1}_{\bf x=0}(t,s)=1+\left(\frac{t-s}{t+s}\right)^{(d+2)/4} - \tau_0^{-1} B_d
\left(\frac{t-s}{t+s}\right)^{(d+2)/4}
(t+s)^{-1/2}\,,
\ee
where $B_d = \Gamma(d/4+1)/\Gamma(d/4+1/2)$.
The term proportional to $\tau_0^{-1}$ is only a correction to the asymptotic form for
$t>s\rightarrow\infty$. Anyway,
as in the case of Model A, $X^{-1}_{\bf x=0}$ has a slower 
approach to the asymptotic value $X^\infty=1/2$ than that displayed 
by ${\cal X}_{{\bf q}\rightarrow 0}$.
This faster approach suggests the use of coherent
observables to determine $X^\infty$ also in the case with 
conserved order parameter.
Although this involves the limiting procedure described above, which is expected to be 
more noisy than working in real space, we think that the absence of a long  
transient allows for a more efficient determination of $X^\infty$.

\subsection{Exact solution for $N=\infty$}

The next step in understanding general properties of $X^\infty$ is the 
exact calculation for $N=\infty$. 
Some quantities were already considered in the 
literature \cite{kissner-92,sire-04}.
In this limit the interaction can be self-consistently decoupled as 
$g_0 \p^4({\bf x},t)\rightarrow g/n C_{{\bf x}={\bf 0}}(t,t) \p^2({\bf x},t)$ as 
for Model A dynamics. 
At criticality this leads to the the response function
$R_{\bf q}(t,s) = {\bf q}^2 G_{\bf q}(t,s)$ where
\be
G_{\bf q}(t,s)=e^{-{\bf q}^4(t-s)} e^{2a_d {\bf q}^2(\sqrt{t}-\sqrt{s})}\,.
\ee
The constant $a_d$ is determined by a consistency condition (see 
\cite{kissner-92}, section IV). 
Hence, from the definition of the correlation function, one 
gets \cite{sire-04}
\be 
C_{\bf q}(t,s) = 2 {\bf q}^2 \int_0^s \rmd u G_{\bf q}(t,u) G_{\bf q}(s,u)
+\tau_0^{-1} G_{\bf q}(t,0) G_{\bf q}(s,0)\,,
\ee 
explicitly showing that all the exponents are the
mean-field ones, as expected. 
The FDR ${\cal X}_{{\bf q}\rightarrow{\bf 0}}(t,s)$ turns out to be $1/2$ for all times
and without corrections, as an important difference with model A dynamics, where 
$X^\infty=1-2/d$  for $2\leq d\leq 4$ and $N=\infty$.

\subsection{FDR beyond the Gaussian approximation}

It is easy to realize that the one-loop diagrams do not contribute 
to the critical FDR for ${\bf q}\rightarrow{\bf 0}$. In fact, as in Model A, they are
given by tadpoles \cite{cg-02a1}. Their one-particle irreducible parts
are independent of the external momentum ${\bf q}$. On the other hand,
the vertex carries a factor ${\bf q}^2$. Therefore, in the limit
${\bf q}^2\rightarrow 0$, the contributions of these diagrams vanish.
Possibly non-vanishing corrections (if any) could come from
diagrams whose one-particle irreducible part depends on ${\bf q}$ 
(as the two-loop sunset diagram). So one concludes that
\be 
X^\infty=\frac{1}{2}+O(\e^2)\,.
\ee
A more detailed analysis of this problem is in progress \cite{prep}.

Let us compare this result for $X^\infty_{\rm Kawasaki}$
with that one for Model A ($X^\infty_{\rm Glauber}$). 
By using equation (\ref{Xqbis}) one finds that
$X^\infty_{\rm Kawasaki}/X^\infty_{\rm Glauber} =  1 +
(N+2)/[4(N+8)]\e + O(\e^2)$. Therefore, for small $\e > 0$ one has
$X^\infty_{\rm Kawasaki} > X^\infty_{\rm Glauber}$ a relation that
should not be changed by higher-order terms up to rather large $\e$. 
This expectation is in agreement with the numerical results of
\cite{gkr-04}, where such a relation has been observed for 
the two-dimensional Ising model.

\section{Coupling of a conserved density to the non-conserved 
order parameter (Model C)}
\label{modC}

We now address the problem of when and how the critical behaviour is 
influenced by coupling a non conserved order parameter to a conserved 
density. As discussed in section \ref{duc} such a coupling
characterizes the so-called 
Model C universality class. 
The starting point is the effective dynamic action (\ref{mrshC}) defined 
in terms of the two fields $\p$ and $\en$. 
The propagators~(Gaussian two point correlation and response functions) 
of the resulting theory are the same as Model A (equations (\ref{Rgaux}) and
(\ref{Cgaux})) as far as the order
parameter (and the associated response field) are concerned, whereas,
for $\en$ and $\tilde\en$ one has \cite{oj-93}
\bea
\langle \tilde{\en}({\bf q},s) \en(-{\bf q},t) \rangle_G =& 
R^G_{\en,q}(t,s)=& \theta(t-s) G_\en(t-s),\label{Rgauxe}\\
\langle \en({\bf q},s) \en(-{\bf q},t) \rangle_G =&
 C^G_{\en,q}(t,s)=& G_\en(|t-s|)+ (c_0 - 1) G_\en(t+s), \label{Cgauxe}
\eea
with
$G_\en(t)=\displaystyle{e^{-\rho\Omega q^2 t}} \label{GGe}$.

Let us briefly recall the scenario of fixed points for non-equilibrium
Model C~\cite{bd-75,fm-03,oj-93}.
The fixed-point values for the couplings $g$ and $\gamma$  are determined 
only by the statics.
One has $\tilde{g}^* = \tilde{g}^*_A + 6\tilde{\gamma}^{2*}$, where 
$\tilde{g}^*_A=6\e/(N+8) + O(\e^2)$\ is the fixed-point value
of the coupling constant for Model A \cite{zj}.

The value of $\gamma$ at the stable infrared fixed point depends on the sign
of the specific-heat exponent $\alpha$:
\be
\tilde{\gamma}^{2*}=
\left\{\begin{array}{cll}
0\; , & \mbox{stable for } \alpha <0 \, ,& \mbox{ case I}\, ,\\
\frac{4-N}{N(N+8)}\e + O(\e^2) \;, &  \mbox{stable for }
\alpha >0 \, , & \mbox{ case II}\, ,
\end{array}\right.
\ee
in the case I, the dynamics of the conserved density decouples from that of 
the order parameter and one gets back to Model A (at least asymptotically). 
At the leading order in $\e$-expansion one has, 
for the $O(N)$ model \cite{zj},
\be
\alpha = \frac{4-N}{2(N+8)}\e + O(\e^2) \; ,
\ee   
therefore the truly Model C dynamics fixed point is stable for $N<4+O(\e)$.
In three dimensions, theoretical and experimental investigations showed
that $\alpha$ is negative for $N\geq2$ (see \cite{PV-r} for a
comprehensive review). As a consequence the Model C dynamics
may be realized only for the three-dimensional Ising model ($N=1$) which has
a positive $\alpha$ \cite{PV-r}. 
In two-dimensions, the boundary value between positive and negative $\alpha$
is $N=1$, so no physically relevant genuine Model C dynamics exists in
this case. Albeit for $d=3$ with $N\geq2$ and $d=2$ with
arbitrary $N$ the dynamics 
is asymptotically Model A, a strong crossover is expected, 
especially in models with small $\alpha$ \cite{fm-03}.

As far as $\rho$ is concerned we have two possible stable fixed points 
determined by the equilibrium dynamics \cite{fm-03}
\begin{itemize}
\item[(a)] $\rho^* = \infty$, stable for 
$N>N_1(\e)=4-[15/4+3/2 \log (4/3)] \e+O(\e^2)$;
\item[(b)] $\rho^* = 2/N -1 + O(\e)$, stable for  $N<N_1(\e)$.
\end{itemize}
A third region with $\rho^* = 0$, found in earlier calculation \cite{bd-75}, 
has been recently proved to be only an artifact of the $\e$-expansion
and consequently it does not exist in any dimension \cite{fm-03}.
This (fake) fixed point with $\rho^*=0$ has been mentioned
and discussed in many papers so far \cite{HH,zj,oj-93,cg-03}. Any
reference to it can be ignored.

Finally, as far as the non-equilibrium dynamics is concerned,
it has been shown that, whenever $\alpha>0$, the fixed point value for $c$\ 
is $c^* = 0$~\cite{oj-93}. 

We focus here our attention on the only relevant stable fixed point of the 
model, i.e., II(b).
At this fixed point $z=2+\alpha/\nu$ exactly \cite{zj}, so that $z$ can be 
obtained from the precise estimates of static exponents \cite{PV-r}.
$\theta$ instead is an independent exponent known up to two loops \cite{oj-93}
in $\e$ expansion.
In \cite{cg-03} we obtained the one-loop non-equilibrium universal
scaling function of the response
\be
\fl F_R(v)=1+{4-N\over 4(N+8)(N-1)}
\ln[(1+(N-1)v)^{N-2}(1-(N-1)v)^N]\e+ O(\e^2)\; ,
\ee
that renders, for the physically relevant case of $N=1$, 
\be
F_R(v)=1-\e {v\over 6}+ O(\e^2)\; .
\ee
Also the scaling function $F_C(v)$ has been computed 
\cite{cg-03} but we do not report it here.
For both the universal functions corrections to the Gaussian value
already at one-loop order have been found, at variance with Model A.

In \cite{cg-03} the FDR has been computed, finding
\be
\fl X^\infty = \frac{1}{2}\left\{1 +\frac{4-N}{N+8}\e\left[\frac{N-1}{(4-N)(2-N)} + \frac{N(2-N)}{4(N-1)^2}\ln[N(2-N)]\right] \right\} \ + O(\e^2)\; ,
\ee
that, for $N=1$, is exactly the same as in Model A, 
$X^\infty = 1/2(1-\e/12)+O(\e^2)$.
This is really surprising, in particular looking at the complicate expression
for general $N$. It is probably only a coincidence of one-loop computation,
but only higher-loop calculations may clarify whether this is 
a deeper property or not.

\section{Model A dynamics of a weakly dilute Ising model}
\label{RIM}

A question of theoretical and experimental interest is whether and how 
the critical behaviour is altered by introducing in the systems a small
amount of uncorrelated non-magnetic
impurities leading to models with quenched disorder.

The static critical behaviour of these systems 
with bond or site disorder (the effect of a random field is more complicate
see, e.g., \cite{bel}) is well understood thank to the
Harris criterion~\cite{Harris-74}. It states that the addition of 
impurities to a system which undergoes a second-order 
phase transition does not change the critical behaviour 
if the specific-heat critical exponent $\alpha_{\rm p}$ of the pure 
system is negative. If $\alpha_{\rm p}$ is positive, the transition
is altered. 

For the very important class of the three-dimensional $O(N)$-vector models
it is known that $\alpha_{\rm p}<0$ for $N\geq2$~\cite{PV-r} and the 
critical behaviour is unchanged in the presence of weak quenched disorder
(apart from large crossover effects, see, e.g., \cite{cppv-03,psrd-04}).
Instead, the specific-heat exponent of the three-dimensional Ising model 
is positive~\cite{PV-r}. Therefore the introduction of a small amount
of non-magnetic impurities (dilution) leads to a new universality class
(as confirmed by RG analyses, Monte Carlo simulations, 
and experiments, see \cite{PV-r,ff,cpv-03r} as comprehensive reviews on 
the subject) to which the weakly dilute Random Ising Model (RIM) belongs. 
The RIM is a lattice spin model with nearest-neighbour interaction Hamiltonian 
\be
{\mathcal H}_c = - \sum_{\langle ij\rangle} \rho_i \rho_j\, s_is_j \;,
\label{Hmrim}
\ee
where $s_i$ are the Ising spins at the site $i$, 
and $\rho_i$ are uncorrelated quenched random
variables such that $\rho_i=1$ with probability $c$ ($0<c\le 1$ being the
spin concentration in the lattice), $\rho_i=0$ with probability $1-c$.
Above the percolation threshold $c^*$, the critical properties of the 
model (\ref{Hmrim}) are predicted to be independent of the 
actual impurity concentration \cite{PV-r,ff,cpv-03r}.
However to make this property apparent, the 
corrections to the scaling have to be properly taken into account in the 
analysis of numerical and experimental data (see, e.g., 
\cite{bfm-98,PV-r,cppv-03,psrd-04}).

The purely relaxational equilibrium dynamics (Model A)
of this new universality class is qualitatively well  
understood~\cite{gmm-77,lp-83,jos-95,fd-din,prr-99b,MC,o-95b}. 
The dynamic critical exponent $z$ differs from the mean-field value already
in the one-loop approximation~\cite{gmm-77}, at variance with the pure model.
This exponent is known up to two loops in a 
$\sqrt{\e}$~\cite{jos-95} and up to three loops in
fixed-dimension expansion~\cite{fd-din} in three dimensions. 
Unfortunately the agreement between FT estimates (giving $z\sim 2.2$), 
numerical \cite{MC,prr-99b,sp-05} ($z\sim 2.6$) and experimental \cite{b-88} 
($z\sim1.9 - 2.2$) estimates is still quite poor.

The non-equilibrium dynamics is instead less studied. Within the $\sqrt\e$
expansion, $\theta$ was determined up to two-loop order~\cite{oj-95} and the 
response function up to one loop, both for
conserved and non-conserved order parameter \cite{kissner-92}. 
A careful three-dimensional numerical simulation \cite{sp-05}, 
taking properly into account the corrections to the scaling, 
showed that even $\theta$ is dilution independent and its value 
is (quite surprisingly, compared with $z$) in 
good agreement with the $\sqrt{\e}$ expansion.
In fact the obtained value $\lambda/z=1.05(3)$ \cite{sp-05}, using the 
scaling law $\lambda/z=d-\theta'$, leads to $\theta'=0.100(35)$ that has 
to be compared with the two-loop FT result 
$\theta'_{\rm 2loops}\simeq 0.087$ \cite{oj-95}. 
Also the autocorrelation scaling function obtained by means of numerical
simulation \cite{sp-05} is in nice agreement with the RG 
prediction equation (\ref{CXX}).
The finite-size scaling in the non-equilibrium regime has been also 
investigated \cite{o-95}.

The time evolution of the weakly dilute Ising model 
is described by the stochastic 
Langevin equation (\ref{lang}) with ${\mathcal H}\mapsto {\mathcal
H}_\psi$, where the static Landau-Ginzburg 
Hamiltonian $\cal{H}_\psi[\p]$ has a space-dependent random
temperature \cite{gl-76}
\be
{\cal H}_{\psi}[\p] = \int\!\! \rmd^d x \left[
\frac{1}{2} (\nabla \p )^2 + \frac{1}{2} (r_0 +\psi({\bf x}))\p^2
+\frac{1}{4!} g_0 \p^4 \right] .
\label{lg}
\ee
Here $\psi({\bf x})$ is a spatially uncorrelated random field with 
Gaussian distribution
\be
P(\psi) = {1\over \sqrt{4\pi w} } \exp\left[ - {\psi^2\over 4 w}\right].
\ee
Dynamical correlation functions, generated by Langevin equation 
and averaged over the noise $\zeta$, are given by equation (\ref{mrsh}) 
with $\mathcal{H}$ replaced by $\mathcal{H}_\psi$ (we denote $S_\psi$
the resulting dynamical functional).

In the analysis of static critical behaviour, the average over the quenched
disorder $\psi$ is usually performed by means of the replica 
trick~\cite{ea-75,gl-76}.
If instead one is interested in dynamic processes it is simpler to perform 
directly the average at the beginning of the calculation \cite{dedominicis-78}
\be
\int [\rmd \psi] P(\psi) \exp(-S_{\psi}[\p,\tilde{\p}])=
\exp(-S [\p,\tilde{\p}])
\ee
obtaining the $\psi$-independent action~\cite{oj-95} (with $v_0\propto w$)
\bea
S [\p,\tilde{\p}]&=\int\!\! \rmd^dx \Bigg\{
\int_0^\infty\!\! \rmd t \, &\tilde{\p}\left[ \dpar_t \p+
\Omega( r_0 -\Delta)\p
-\Omega\tilde{\p}\right] \nonumber\\
&&+{\Omega g_0\over 3!} \int_0^\infty\!\! \rmd t\, \tilde{\p}\p^3 -{\Omega^2 v_0 \over2}
\left(\int_0^\infty\!\! \rmd t\,\tilde{\p}\p\right)^2\Bigg\}\, .
\label{ranaction}
\eea
Note that a non-local (in time) effective interaction term has been generated.

The perturbative expansion is performed in terms of the two fourth-order
couplings $g_0$ and $v_0$ and using the propagators given in
equations (\ref{Rgaux}) and (\ref{Cgaux}).
The scaling of the response function is characterized by \cite{cg-02rim}
\be
A_R=1-{1\over2}\sqrt{6\e\over 53}\gamma_E +O(\e)\; ,
\mbox{ and } \;
F_R(x)=1 + O(\e),
\ee
and the correlation (see equation (\ref{scalC})) by \cite{cg-02rim}
\bea
{A_C\over2}&=&1+{1\over2} \sqrt{6\e\over 53} (2-\gamma_E)+ O(\e)\,,\\
F_C(x)&=&1+{1\over2}\sqrt{6\e\over 53}
\left[1+{1\over2} \left(1+{1\over x}\right) \log {1-x\over1+x} \right]+ O(\e) \,.
\eea
Note that, at variance with the pure model, the function $F_C(x)$ gets 
a contribution already at one-loop order. $F_R(v)$ remains $1$ in apparent
agreement with LSI although $z\neq 2$, already at this order.

The FDR for generic times is
\be
{\cal X}_{\bf q=0}(s/t)=\frac{1}{2}\left(
1-{1\over2}\sqrt{6\e\over 53}
\left[1+{1\over2} \log {1-s/t\over 1+s/t}\right]\right) + O(\e) \;.
\label{XgenRIM}
\ee
Therefore, assuming even in this case the validity of equation (\ref{equiv}),
one finds
\be
X^\infty = {1\over2}-{1\over4}\sqrt{6\e\over 53} + O(\e)\;,
\label{XRIM}
\ee
that for $\e=1$ (that is the only physical relevant case, given that 
$\alpha_{\rm p}=0$ in the two-dimensional Ising model) leads to 
$X^\infty\sim 0.416$.
To this order it is not clear whether randomness really changes
$X^\infty$ in a sensible way or not. In any case this could not be safely stated
from low-order computations
since the $\sqrt{\e}$ is known to be not well-behaved at $d=3$ \cite{PV-r,ff}.
However, a qualitative conclusion can be drawn from equation (\ref{XgenRIM}): 
${\cal X}_{{\bf q}={\bf 0}}$ has a quite strong 
dependence on the time ratio $s/t$,
that, at variance with the pure model, should be easily identified in 
simulations. 

The relation between $X_{{\bf x}=0}^\infty$ and 
${\cal X}_{{\bf q}=0}^\infty$ in the RIM 
has been studied in more detail by Schehr and Paul
\cite{sp-05}. Due to the quench disorder,
the response function $R_{\bf q}$ has a power-law decay for 
$q^zt\gg1$ \cite{kissner-92}. Therefore
the argument leading to 
$X_{{\bf x}=0}^\infty = {\cal X}_{{\bf q}=0}^\infty$ does not apply 
straightforwardly in the form outlined in  section \ref{secFDRms}. 
To clarify this point $R_{\bf q}(t,s)$ and $C_{\bf q}(t,s)$
were calculated for generic momenta up to the first order in the 
$\sqrt\epsilon$-expansion.
Then, via the Fourier
transform, one can compute explicitly $R_{{\bf x}=0}(t,s)$ and
$C_{{\bf x}=0}(t,s)$ and therefore $X_{{\bf x}=0}(t,s)$.
In doing that one faces the problem that
$R_{\bf q}$ and $C_{\bf q}$ \cite{sp-05} have a power-law decay for
large $q$ and fixed times, in contrast to the exponential one of the
pure model. As a consequence one has to introduce an ultraviolet cut-off
$\Lambda_0$, finding \cite{sp-05}:
\bea
R_{{\bf x}=0}(t,s)&=&[A_R^0+A_R^1 \log(t-s)] (t-s)^{a-d/z} 
\left(\frac{t}{s}\right)^\theta\,, \label{RxRIM}\\
C_{{\bf x}=0}(t,s)&=&[A_C^0+A_C^1 \log(t-s)] (t-s)^{a+1-d/z} 
\left(\frac{t}{s}\right)^\theta F_C(s/t)\,,\label{CxRIM}
\eea
where $A_{R,C}^{0,1}$ are non-universal amplitudes (with $A_R^1=A_C^1$).
In particular $A_{R,C}^0$ explicitly depend
on $\Lambda_0$ and are not finite for $\Lambda_0 \rightarrow \infty$.
Note that these expressions do {\it not}
agree with the general RG predictions equations (\ref{RXX}) and (\ref{CXX}),
because of the presence of the logarithmic term.
However this fact is not surprising: Logarithmic corrections to
the scaling behaviour are generically expected in the presence of 
quenched disorder, as shown by Cardy using conformal invariance in
generic dimension \cite{c-99}.

We point out (as already noticed in \cite{sp-05}) that, 
up to one-loop order, the logarithms can 
be absorbed in the exponent of $(t-s)$, giving rise to non-universal,
cut-off dependent exponents $a^{\rm (eff)}_{C,R} = a + A^1_{C,R}/A^0_{C,R}$. 
This possibility seems rather unlikely, but only a two-loop 
calculation can rule it out.
We mention that the same kind of logarithmic behaviour has been found in 
pinned elastic interfaces near the depinning transition \cite{ls-05}. 

From equation (\ref{RxRIM}) and (\ref{CxRIM}) one easily derives the FDR
in the real space \cite{sp-05}
\be
\fl
X_{{\bf x}=0}^{-1}=F_X(t/s);\qquad {\rm with} \qquad
F_X(u)= 2\frac{u^2+1}{(u+1)^2}+\sqrt{\frac{6\epsilon}{53}}
\left(\frac{u-1}{u+1}\right)^2+O(\e)\,, 
\ee
where $F_X(u)$ interpolates between $1$ in the quasi-equilibrium regime
for $u\rightarrow1$, and its asymptotic value for $u\rightarrow\infty$
given by $\lim_{u\rightarrow\infty} F_X(u)^{-1}=X^\infty$ of 
equation (\ref{XRIM}).
This result explicitly shows, at order $\sqrt\e$, that the asymptotic FDR 
for total and local magnetization are indeed equal, despite the 
power-law behaviour of $R_{\bf q}$ and $C_{\bf q}$ for large $q$.
This fact calls for a deeper understanding of the relation between 
$X_{{\bf x}=0}^\infty$ and ${\cal X}_{{\bf q}=0}^\infty$ and of the
fact that equation (\ref{equiv}) seems to hold independently of 
the argument presented in section \ref{secFDRms}.

\section{Purely dissipative dynamics of a $\p^3$ Landau-Ginzburg theory}
\label{phi3}

So far we have  only  considered models whose static critical properties are 
described by the effective $\p^4$ Landau-Ginzburg Hamiltonian given in
equation~(\ref{lgwA}). 
In all these cases $\theta\geq 0$ and $0\leq X^\infty\leq1/2$.
To understand whether these bounds on such universal 
quantities are general properties or they are accidentally
connected with the diagrammatic 
properties of $\p^4$ theories, it is worth studying different 
effective Hamiltonians.
In particular we consider a model whose static critical properties are 
described by the $\p^3$ Landau-Ginzburg Hamiltonian
\be
{\cal H}_3[\p]= \int \rmd^d x 
\left[\frac{1}{2}(\nabla \p)^2 +\frac{1}{2}r_0\p^2
+{g_0\over3!} \sum_{i,j,k=1}^N \,d_{ijk}\p_i\p_j\p_k \right]\,,
\label{Hamphi3}
\ee
where the coefficients $d_{ijk}$ specify the model and $\p$ is a 
$N$-component field. A variety of critical phenomena belongs to the
universality class of equation (\ref{Hamphi3}).
Examples are provided by the isotropic-to-nematic phase transition in liquid 
crystals \cite{d-69}, 
the $N+1$-state Potts model \cite{hamp}, its percolation transition
($q=N+1\rightarrow 1$) \cite{fk-72} and its limit $q \rightarrow 0$,
representing  electrical resistor networks  \cite{fk-72}.
The model with one component $N=1$ and $d_{111}=i$ 
describes the Yang-Lee edge  singularity \cite{f-78}. 
Moreover a $\p^3$-like interaction is one of the most relevant in 
the effective field-theoretical action of the Edwards-Anderson spin glass 
\cite{spinglass}.

The upper critical dimension for the universality class specified by
equation~(\ref{Hamphi3}) is $d_c=6$.
Therefore 
it is possible to compute critical exponents and all other universal 
quantities in a $\e$-expansion with $\e=6-d$
\cite{hlhd-75,pl-76,amit-76,3loop}. Fixed-dimension results have also
been provided \cite{fd}. 

However the field-theoretical study of the Hamiltonian (\ref{Hamphi3}) 
presents several potential pitfalls. 
For example in the $\e$-expansion of the Potts-symmetric Hamiltonian 
a stable fixed point of the renormalized coupling constant $g$ is 
found only for $N<7/3$ and its value $g^*$ diverges as 
$N$ approaches $7/3$, whereas it takes imaginary values for $N>7/3$. On
the other hand the $\e$ expansion is not reliable 
for all values of $N$ for which $g^*$ is not 
sufficiently small \cite{amit-76}. 
This condition naively leads to the conclusion that 
the $\e$-expansion makes sense only for $N\leq 1$. 
In spite of these problems the universality class of the percolation 
transition ($N=0$) has to be considered free from pitfalls.
This fact is corroborated by the remarkable agreement of field-theoretical 
estimates with numerical and experimental results \cite{percbook} in $d>2$ 
(in $d=2$ it was claimed that the $\e$ expansion breaks down due to the 
relevance of $\p^4$ interactions \cite{fg-81}).

The simplest dynamics that one can think of realizing for some of the model
just described is the  
purely dissipative one, i.e., the Model A of 
section \ref{duc}, defined by the Langevin equation (\ref{lang}). The
dynamic critical properties can be worked out from the dynamical
functional (\ref{mrsh}) with ${\cal H}$ replaced by ${\cal H}_3$.
We point out that the dynamics just introduced is {\it not} the actual 
dynamics of isotropic percolation (reviewed, e.g., in \cite{jt-04}).

The calculation of the one-loop non-equilibrium 
response and correlation function following a quench from the
disordered state (introduced by means of ${\mathcal H}_0$, as in
section~\ref{secRenScal})
to the critical point is 
straightforward and we do not provide any details but only 
the final results. 
Zero-momentum response and correlation functions satisfy the expected 
scaling laws (\ref{scalR}) and (\ref{scalC}) with exponents
\be
a = -\frac{\alpha}{4(\alpha-4\beta)} \e+O(\e^2)\,, \qquad
\theta =
-\frac{2\alpha-\gamma}{8(\alpha-4\beta)}\e+O(\e^2)\,,
\ee
($a$ agrees with the known exponents 
\cite{amit-76,bj-81} through the scaling relation $a=(2-\eta-z)/z$) and
scaling functions
\bea
F_R(x)&=&1- \e \frac{\alpha}{4(\alpha-4\beta)}  x +O(\e^2)\,,\\
F_C(x)&=&1- \e\frac{\alpha}{8(\alpha-4\beta)}
\left[x - \left(1+\frac{1}{x}\right)\log(1-x^2)+O(\e^2)
\right]\,.
\eea
The resulting FDR is given by
\be
X^\infty=
\frac{1}{2}\left[1+\frac{3\alpha-\gamma}{8(\alpha-4\beta)}\e\right]+O(\e^2)\,.
\ee
In the previous equations we introduced the three contractions
\be 
d_{ijk}d_{jkl}= \alpha \delta_{il}\,, \qquad
d_{ilm}d_{jmn}d_{knl}= \beta d_{ijk}\, ,\qquad
d_{ijk}d_{kll}= \gamma \delta_{ij}\, ,
\ee
that completely characterize the model at one-loop level. 
Accordingly, for the models we are interested in, 
we do not report the coefficients $d_{ijk}$ \cite{hamp,f-78}, but only
the contractions:
\bea
\fl
\mbox{Yang-Lee edge singularity:} &\,\, \alpha=-1, &\,\, \beta=-1,\, \qquad\qquad\quad\,  \gamma=-1,\\  \fl
N+1-\mbox{state Potts Model:}  &\,\, \alpha= {(N+1)^2 (N-1)},&\,\, \beta={(N+1)^2 (N-2)},\,\, \gamma=0.
\eea
The obtained $F_R(x)$ disagrees at one-loop level with the prediction of 
Local Scale Invariance $F_R(x)=1$ \cite{henkel-02}.
This fact is not surprising, since at one-loop level, we have $z\neq2$
(see the remarks in section \ref{Atwoloop}).

Let us discuss these results for some specific model.  
For the Yang-Lee edge singularity we found
\be
\theta=\frac{1}{56}\e+O(\e^2)\,, \qquad
X^\infty=\frac{1}{2}\left[1-\frac{1}{28}\e+O(\e^2)\right]\,,
\ee
i.e., $\theta>0$ and $X^\infty<1/2$.
For the models with the symmetry of the $q=N+1$-state Potts model, instead
\be
\theta=\frac{1-N}{8(7-3N)}\e+O(\e^2)\,, \qquad
X^\infty=\frac{1}{2}\left[1-\frac{3}{8}\frac{1-N}{7-3N}\e+O(\e^2)\right]\,.
\ee
For $N<1$, where the perturbative expansion of such models is considered
under control, we get $\theta>0$ and $X^\infty<1/2$. 
Conversely, for $N>1$ we apparently get $X^\infty>1/2$ and $\theta<0$.
However, such a result is not completely under 
control, given that one finds  $\nu<1/2$ \cite{amit-76}  for $N>1$, 
in disagreement with the results for the $q=N+1$-state Potts 
model \cite{wu-82}.

\section{Conclusions and Perspectives}
\label{disc}

We reviewed the ageing properties of systems which are quenched from the 
high-temperature phase exactly to their critical points.
It clearly emerges that the theoretical understanding of such phenomena is 
quite complete in comparison with the richer ageing effects in glassy
phenomelogy.
Nevertheless there are still some points that require further investigation
in order to be completely understood:

First, we mention the equivalence between the asymptotic FDR 
$X^\infty$ defined in momentum and in real space, see equation (\ref{equiv}). 
To our knowledge, the only general argument supporting this result is 
the one outlined in section \ref{secFDRms} (originally proposed 
in \cite{cg-02a1}).
However, a recent analysis of the dynamics of the weakly dilute Ising 
model \cite{sp-05}, suggest that probably such an equivalence has a wider
range of applicability. It would be desirable to have a clear proof of it
and to know the conditions under which one can (or can not) expect equation~(\ref{equiv}) to be valid, especially in connection 
the possible definition of the  effective temperature 
(see section \ref{secFDRms}).

Another interesting issue is the behaviour of these critical systems
just at their lower critical dimensions (i.e., the dimension where the 
critical temperature vanishes, $d=1$ for Ising-like systems and 
$d=2$ for systems with continuous symmetries).
Indeed, in passing, we mentioned \cite{detlg,lz-00,gl-00i}
that different one-dimensional systems with Ising symmetry may
have different non-equilibrium critical exponents (i.e., $\theta$). 
In this context it has been 
recently argued \cite{clz-04} that a quench to zero-temperature
at the lower critical dimension is not
the dimensional continuation of a line of critical quenches in 
the $(T,d)$ plane (as often implicitly assumed), but it is the continuation
of a line of zero-temperature quenches, i.e., the system behaves as in the 
coarsening regime, although $X^\infty\neq0$ (see for details \cite{clz-04}).
Moreover the critical quench at the lower critical dimension has a very
interesting aspect: Being the autocorrelation exponent $a-d/z+1$ (see 
equation (\ref{RXX}))
equal to zero, it is possible to write the FDR $X_{{\bf x}=0}$ as a function 
of the autocorrelation function $C_{{\bf x}=0}$ 
as it happens in more complex instances of glassy 
behaviour (see, e.g., \cite{cugl-02,cr-03}).

Field-theoretical RG provides methods going well-beyond the expansion close 
to the upper critical dimension that has been used so far and that we
have reviewed. 
It will be interesting to consider different approaches like 
expansions close to the lower critical dimensions, 
perturbative expansions in fixed-dimensions and
non-perturbative approaches like the so-called 
Exact RG and $1/N$-expansion. These methods might lead to more accurate
theoretical estimates for the universal quantities of interest and
hopefully to a more general understanding of the phenomena.

It is natural to wonder whether some of the results that have
been reviewed carry over to glassy systems, which was the original
motivation of the study of ageing behaviour in critical models.  
Recent numerical simulations of finite-dimensional ($d=3,4$)
spin-glass models exactly 
at the critical point \cite{hp-05} show 
that even in this case the response and correlation functions scale
according to equations (\ref{RXX}) and (\ref{CXX}), as in critical
non-disordered system, 
with a non-trivial value of $X^\infty$. On the other hand, for a
quench in the  
low-temperature phase a completely different behaviour 
is expected \cite{cr-03}.

Last but not least, we remark that the ultimate check of all the 
physical theories is provided by  experiments.
Unfortunately, to our knowledge, only quasi-one-dimensional systems with 
Ising symmetry have been considered so far \cite{1dquench}.
Although very interesting, the physics of one-dimensional systems is too 
peculiar ($d=1$ is the lower critical dimension of Ising 
systems) to provide a test of the theoretical results currently available.
Experiments in two and three dimensions are required to check the relevant 
aspects of ageing dynamics {\it at the critical point}.
In principle, this could be done by 
considering the same experimental systems as those considered
when studying  
phase ordering---magnetic systems, nematic-liquid crystals, binary liquids,
etc. \cite{bray}--- exactly at their critical points.


\ack

It is a pleasure to thank L. Berthier, S. Caracciolo, F. Corberi, 
L. Cugliandolo, H.~W. Diehl, S. Dietrich, M. Henkel, F. Krzakala, A. Lefevre, 
E. Lippiello, M. Pleimling, 
P. Sibani, A. Pelissetto, F. Ricci-Tersenghi, P. Sollich, E. Vicari,
and M. Zannetti for stimulating discussions, correspondence and suggestions about ageing
phenomena during the last two years. The authors acknowledge 
S. Andergassen and P. Sibani for a careful reading of a first version of this 
manuscript.
PC acknowledges financial support from EPSRC Grant No. GR/R83712/01.

\end{document}